\documentclass[twocolumn,english,superscriptaddress,prl,nofootinbib]{revtex4-1}
\usepackage{lmodern}
\usepackage{lmodern}
\usepackage[T1]{fontenc}
\usepackage[latin9]{inputenc}
\usepackage{color}
\usepackage{babel}
\usepackage{amsmath}
\usepackage{amssymb}
\usepackage{graphicx}
\usepackage[unicode=true,pdfusetitle,
bookmarks=true,bookmarksnumbered=false,bookmarksopen=false,
breaklinks=false,pdfborder={0 0 1},backref=false,colorlinks=true]
{hyperref}
\hypersetup{pdfborderstyle=,urlcolor=orange,linkcolor=blue,citecolor=red}

\makeatletter
\usepackage{color}
\usepackage{babel}
\usepackage{units}

\usepackage{tikz}
\usepackage{color}
\usepackage{xcolor}
\usepackage{dsfont}

\usepackage[cal=boondoxo]{mathalfa}
\usepackage{grffile}

\catcode`,\active

\catcode`\,12

\newsavebox{\@brx}
\newcommand{\llangle}[1][]{\savebox{\@brx}{\(\m@th{#1\langle}\)}%
	\mathopen{\copy\@brx\kern-0.5\wd\@brx\usebox{\@brx}}}
\newcommand{\rrangle}[1][]{\savebox{\@brx}{\(\m@th{#1\rangle}\)}%
	\mathclose{\copy\@brx\kern-0.5\wd\@brx\usebox{\@brx}}}

\makeatletter
\def\maketitle{
	\@author@finish
	\title@column\titleblock@produce
	\suppressfloats[t]}
\makeatother

\newcommand{\kp}{\kappa_{\scriptscriptstyle E}}

\newcommand{\R}{\mathds{R}}

\newcommand{\bgamma}{\boldsymbol{\gamma}}
\newcommand{\bx}{\boldsymbol{x}}
\newcommand{\bxi}{\boldsymbol{\xi}}
\newcommand{\bW}{\boldsymbol{W}}
\allowdisplaybreaks

\makeatother

\begin{document}
	\title[]{The star-shaped space of solutions of the spherical negative perceptron}
	\author{Brandon Livio Annesi}
	\affiliation{Department of Computing Sciences, Bocconi University, Milano, Italy}
	\author{Clarissa Lauditi}
	\affiliation{Department of Applied Science and Technology, Politecnico di Torino, Torino, Italy}
	\author{Carlo Lucibello}
	\affiliation{Department of Computing Sciences, Bocconi University, Milano, Italy}
	\affiliation{Bocconi Institute for Data Science and Analytics, Milano, Italy}
	\author{Enrico M. Malatesta}
	\affiliation{Department of Computing Sciences, Bocconi University, Milano, Italy}
	\affiliation{Bocconi Institute for Data Science and Analytics, Milano, Italy}
	\author{Gabriele Perugini}
	\affiliation{Department of Computing Sciences, Bocconi University, Milano, Italy}
	\author{Fabrizio Pittorino}
	\affiliation{Department of Electronics, Information and Bioengineering, Politecnico di Milano, Milano, Italy}
	\affiliation{Bocconi Institute for Data Science and Analytics, Milano, Italy}
	\author{Luca Saglietti}
	\affiliation{Department of Computing Sciences, Bocconi University, Milano, Italy}
	\affiliation{Bocconi Institute for Data Science and Analytics, Milano, Italy}
	
\begin{abstract}
	Empirical studies on the landscape of neural networks have shown that low-energy configurations are often found in complex connected structures, where zero-energy paths between pairs of distant solutions can be constructed. 
	Here we consider the spherical negative perceptron, a prototypical non-convex neural network model framed as a
	continuous constraint satisfaction problem. 
	We introduce a general analytical method
	for computing energy barriers in the simplex with vertex configurations sampled from the equilibrium.
	We find that in the over-parameterized 
	regime the solution manifold displays simple connectivity properties. There exists a large geodesically convex component that is attractive for a wide range of optimization dynamics. Inside this region we identify a subset of 
	atypical high-margin solutions that are geodesically connected with most other solutions, giving rise to a star-shaped geometry.
	We analytically characterize the organization of the connected space of solutions and show numerical evidence of a transition, at larger constraint densities, where the aforementioned simple geodesic connectivity breaks down. 
	
\end{abstract}
\maketitle

In constraint satisfaction problems (CSPs), the goal is to find a configuration of the $N$ variables that satisfies a system of constraints. 
In the case of random instances~\cite{MPZ2002} and for large size $N$, one can typically identify sharp ``structural'' phase transitions in the geometrical organization of 
the solution space~\cite{MMZ2001,MMZ2005,ZK2007}. In the past decades, statistical physics methods from spin glass theory~\cite{mezard1987spin} have been successfully employed to investigate the impact of these landscape features on the performance of solution-sampling algorithms~\cite{monasson1999determining}. A deeper understanding of this interplay in the case of continuous variables \cite{cavaliere2023biased} is becoming a crucial prerequisite for the study of learning dynamics in neural networks.

The characterization of the manifold of low-energy lying states in neural networks has become one of the central theoretical questions of the field~\cite{epfl_workshop}. In typical setups, the high degree of over-parameterization of the models guarantees the existence of multiple zero-energy configurations, but different local geometries induce vastly different accessibility and generalization properties~\cite{keskar,LiVisualizing2018}. 
Growing theoretical~\cite{baldassi2015subdominant, unreasoanable,relu_locent, baldassi2020shaping, baldassi2020wide,baldassi2021learning} and empirical~\cite{sagun2016eigenvalues,sagun2017empirical,pittorino2021, foret2021sharpnessaware} evidence seems to show that there exist flat degenerate areas in the landscape of neural networks. The dynamics of common stochastic gradient descent (SGD) based algorithms, seem to be quickly attracted to the borders of these regions and then drift toward their core~\cite{FengTu2021,Chen2022,kunin2021rethinking}.

Linear paths between two minimizers (e.g. as given by SGD with different random initial conditions) display energy barriers. Nonetheless, zero-energy paths can be systematically constructed between them~\cite{draxler2018,garipov2018}. This surprising finding on \emph{mode connectivity} is compatible with the hypothesis of the existence of a single connected component of zero-energy configurations, organized in an intricate network of tunnels and plateaus \cite{entezari2022, pittorino22, jordan2023repair}. 
Understanding the extent of linear mode connectivity may unlock progress in some of the most debated topics in deep learning, from the ``lottery ticket'' hypothesis and pruning \cite{frankle2018, frankle20} to multitask/continual learning and ensemble methods \cite{mirzadeh2021linear, FastEnsembling}.

\begin{figure}[h]
		\begin{centering}
			\includegraphics[width=0.90\columnwidth]{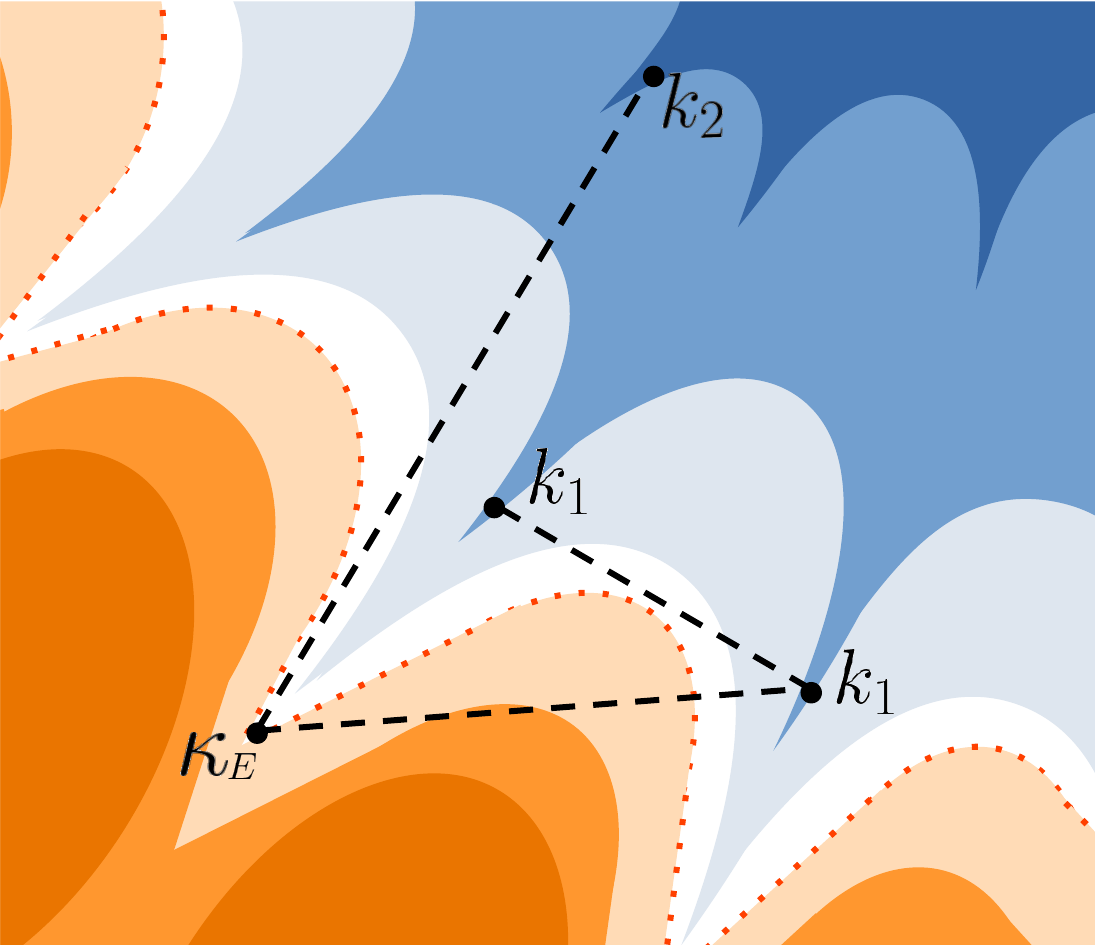}
		\end{centering} 
		\caption{\label{fig:sketch} \textbf{Sketch of the solution space} of the negative perceptron in the RS phase. The red dotted line represents the border of the connected manifold of solutions for a given margin $\kp$ (white-blue region). In the orange regions, the configurations have non-zero energy. The solutions that satisfy margins larger than the one of the problem, $\kp<k_1<k_2$, are organized in a nested structure (darker shades of blue). When a typical solution with margin $\kp$ is geodesically connected with a solution with margin $k_1$, an energy barrier (a crossing of the orange region) is observed. However, the $k_1$ solutions belong to a \emph{geodesically convex} sub-manifold (the \emph{geodesic path} falls within the white-blue region). Solutions with an even higher margin $k_2$, located in the \emph{kernel}, are connected to almost any other solution. Definitions are provided in Appendix \ref{app:def}. 
		}
	\end{figure}

Here, we consider the simplest non-convex neural network model, the negative spherical  perceptron~\cite{gardner1988optimal,engel-vandenbroek, scipost2017}, and characterize the connectivity properties of its solution space via \emph{geodesic} (minimum length) paths on the high-dimensional sphere. 
In particular, we introduce a novel analytical method, based on the replica analysis of the model \cite{mezard1987spin}, that yields the typical energy barriers in the convex hull of a group of $y$ solutions sampled from the zero-energy manifold. 
We find that, in the low constraint density regime, the domain of solutions is \emph{star-shaped}~\cite{Hansen2020}: almost all solutions are geodesically connected through zero energy paths to a subset of atypical high-margin solutions. This subset, which we call the \emph{kernel}, is nested in the core of the largest geodesically convex component of the solution manifold. 
	A sketch of this geometrical organization is shown in Fig.~\ref{fig:sketch}. 
	
	We empirically investigate the behavior of different classes of solvers and their bias towards different regions of the solution manifold. In particular, we find that the dynamics of SGD with cross-entropy loss naturally flows toward the large convex component of the solution manifold. Moreover, for any solver, one can identify a phase at low constraint densities and up to a certain threshold, where the sampled solutions are geodesically connected to the most robust solutions (see below) of the problem. Above this threshold and up to the limit density for satisfiability, energy barriers are encountered, signaling the breakdown of the star-shaped organization of solutions. 
	We compare these thresholds with the known structural transitions identified through the statistical physics analysis of this model~\cite{baldassi2023typical}.
	

	\emph{The model.} 
The spherical perceptron is defined by $N$ weights $W_i\in\mathbb{R}$, trained to satisfy an extensive number $P=\alpha N$ of constraints 
\begin{equation}
	\label{eq::constraints}
	\Delta^\mu \equiv \boldsymbol{W}\cdot\boldsymbol{\xi}^\mu \ge \kp \sqrt{N}\,, \quad \mu \in [P]
\end{equation}
where $\kp$ is the margin of the problem, the $\Delta^\mu$ are called stabilities, and $\boldsymbol{\xi}^\mu \sim \mathcal{N}(0, I_N)$. The weights are also subject to the spherical constraint $\|\boldsymbol{W}\|^2 = N$. We analyze the large $N$ limit at constant $\alpha$. When a negative margin $\kp<0$ is considered, the so-called negative perceptron, linear-separability of the dataset is not a necessary condition for satisfiability (SAT) and the problem is non-convex. 
The negative perceptron has recently received attention in both the physics~\cite{franz2016,scipost2017,baldassi2023typical} and mathematics communities~\cite{elAlaoui2022algorithmic,montanari2021tractability}. 
A detailed analysis of the different structural transitions affecting the solution space, as $\kp$ and $\alpha$ are increased, shows that the model enters a non-ergodic phase with Replica Symmetry Breaking (RSB) (details on the phase diagram in the SM). 
In this work, we further investigate the geometric properties of the ground states in the region below the critical line $\alpha_{\text{dAT}}(\kp)$ where Replica Symmetry (RS) holds.

\emph{Organization of the solutions.} In the RS phase, the problem is SAT 
w.h.p. for large $N$. We consider the uniform probability density over the solutions,
\begin{equation}
	p_{\boldsymbol{\xi}, \kp}(\bW) = \frac{1}{{Z_{\bxi,\kp}}}\delta(\lVert \bW\rVert^2 - N) \prod_{\mu=1}^{P} \Theta(\bW\cdot \bxi^\mu-\kp\sqrt{N}),
	\label{eq:prob}
\end{equation} 
where the partition function $Z_{\bxi,\kp}$ plays the role of a normalization factor. 
%
%
While the typical solutions obtained by sampling from \eqref{eq:prob} have minimum stabilities exactly equal to $\kp$, the solution space also contains an exponential number of atypical solutions that satisfy the constraints \eqref{eq::constraints}) with a larger margin $k$ \cite{baldassi2021unveiling}, with $\kp<k \le \kappa_\text{max}(\alpha)$, where $\kappa_\text{max}(\alpha)$ represents the SAT-UNSAT transition line. 
In order to characterize the geometry of the solution space, for a given sample $\bxi$ we consider two configurations independently sampled from $p_{\boldsymbol{\xi}, k_1}$ and $p_{\boldsymbol{\xi}, k_2}$ respectively and employ the replica method \cite{Huang_pairs} to compute their overlaps
$q_1=\frac{1}{N}\mathbb{E}\langle\bW^1\cdot\bW^2\rangle_{k_1,k_1}$, $p=\frac{1}{N}\mathbb{E}\langle\bW^1 \cdot \bW^2\rangle_{k_1,k_2}$ and $q_2=\frac{1}{N}\mathbb{E}\langle\bW^1 \cdot \bW^2\rangle_{k_2,k_2}$. 
Here $\langle\,\cdot \, \rangle_{k,k'}$ represents the average over the Cartesian product of densities \eqref{eq:prob} with the corresponding margins and $\mathbb{E}$ is the expectation over disorder $\bxi$ (see SM). For $k_1 < k_2$, we find that they satisfy the simple inequality $q_1 < p < q_2$. This ordering is compatible with a nested organization of solutions with different margins. 
The degree of anisotropy of this structure can be evaluated analytically (details in the SM). 
\begin{figure*}[t]
	\begin{centering}
		\includegraphics[width=2\columnwidth]{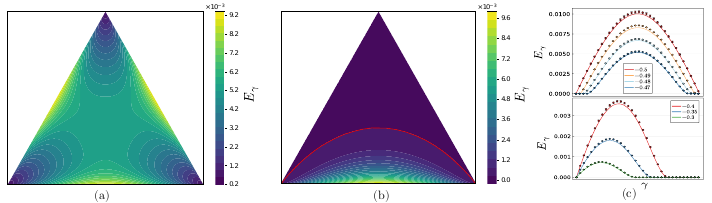}
	\end{centering}
	\caption{\label{fig:y3}\textbf{Interpolating manifold} between triplets of solutions for $\kp = -0.5$, $\alpha=1$.  (a) Solutions with the same margin $k = \kp$.  (b) Solutions with different margins. The two bottom vertices are two typical solutions to the problem, i.e. $k_{1} = \kp$; the top vertex is sampled with $k_2 = -0.1 >\kappa_{\text{krn}}(\kp, \alpha) \simeq -0.171$. The red level curve separates the zero from the non-zero energy region on the simplex. 
		(c-top) Energy along the geodesic connecting two solutions sampled with margin $k$ for $k=\kp, -0.49, -0.48, -0.47$. 
		(c-bottom) Same as the top panel but with the left endpoint margin fixed to  $k_1 = \kp$ and the one on the right having margin $k_2 = -0.4, -0.35, -0.3$. Lines are the theoretical predictions, dots are large-$N$ extrapolations of the numerical simulations.} 
\end{figure*}


\emph{Interpolating between solutions.} Our main analytic result is a formula for studying the typical energy landscape between groups of $y$ solutions. 
In particular, we consider the projection on the $N$-sphere of the $(y-1)-$simplex;
\begin{equation}\label{eq:interpolating_vector}
	\boldsymbol{W}_{\boldsymbol{\gamma}} = \frac{\sqrt{N}  \sum_{r=1}^y \gamma_r \bW^r}{\|\sum_{r=1}^y \gamma_r \boldsymbol{W}^r \|} ,
\end{equation}
with $\gamma_r \ge 0$  and $\sum_{r=1}^y \gamma_r = 1$. 
By varying the margins $\{k_r\}_{r=1}^y$ of the solutions $\{\boldsymbol{W}^r\}_{r=1}^y$ on the vertices, different regions of the solution manifold can be explored. 
To obtain the asymptotic energy of the interpolating configurations with respect to the margin $\kp$ of the problem, we evaluate the probability of a constraint violation:
\begin{equation}\label{eq:general_energybarrier}
	E_{\boldsymbol{\gamma}} = \lim_{N\to+\infty}\ \,\mathbb{E}_{\bxi}\,\big\langle \Theta\big( - \boldsymbol{W}_{\boldsymbol{\gamma}} \cdot \boldsymbol{\xi}^\mu + \kp \sqrt{N} \big)\big\rangle_{k_1,\dots,k_y}.
\end{equation}
In the high dimensional limit, this quantity only depends on the typical overlaps between pairs of solutions with different margins $q_{rs}$, $r, s \in [y]$. Analytic details are reported in Appendix \ref{app:computation}.
With a similar approach, one can also derive the stability distribution $\Delta^\mu$ in Eq.~(\ref{eq::constraints}) for the interpolating configurations 
(details in the SM).

\begin{figure*}[t]
	\begin{centering}
		\includegraphics[width=.99\columnwidth]{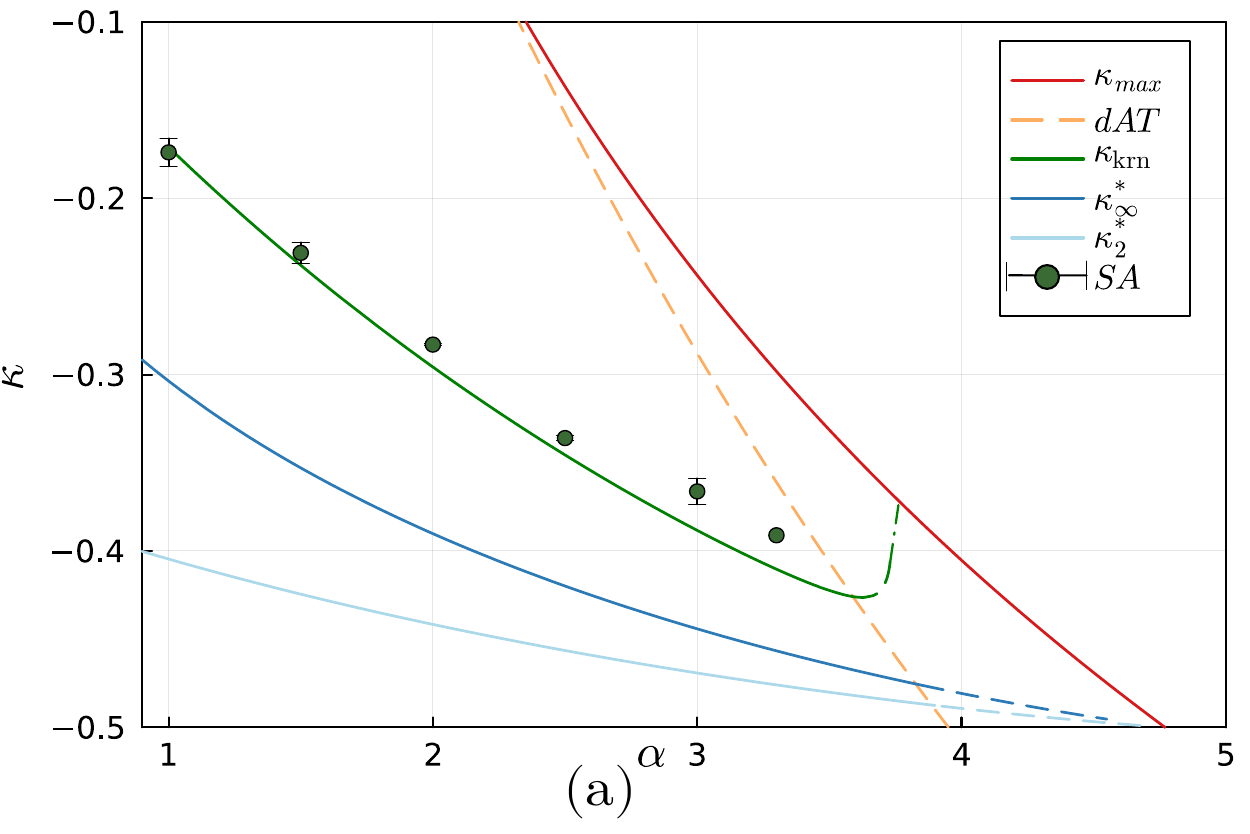}
		\includegraphics[width=.99\columnwidth]{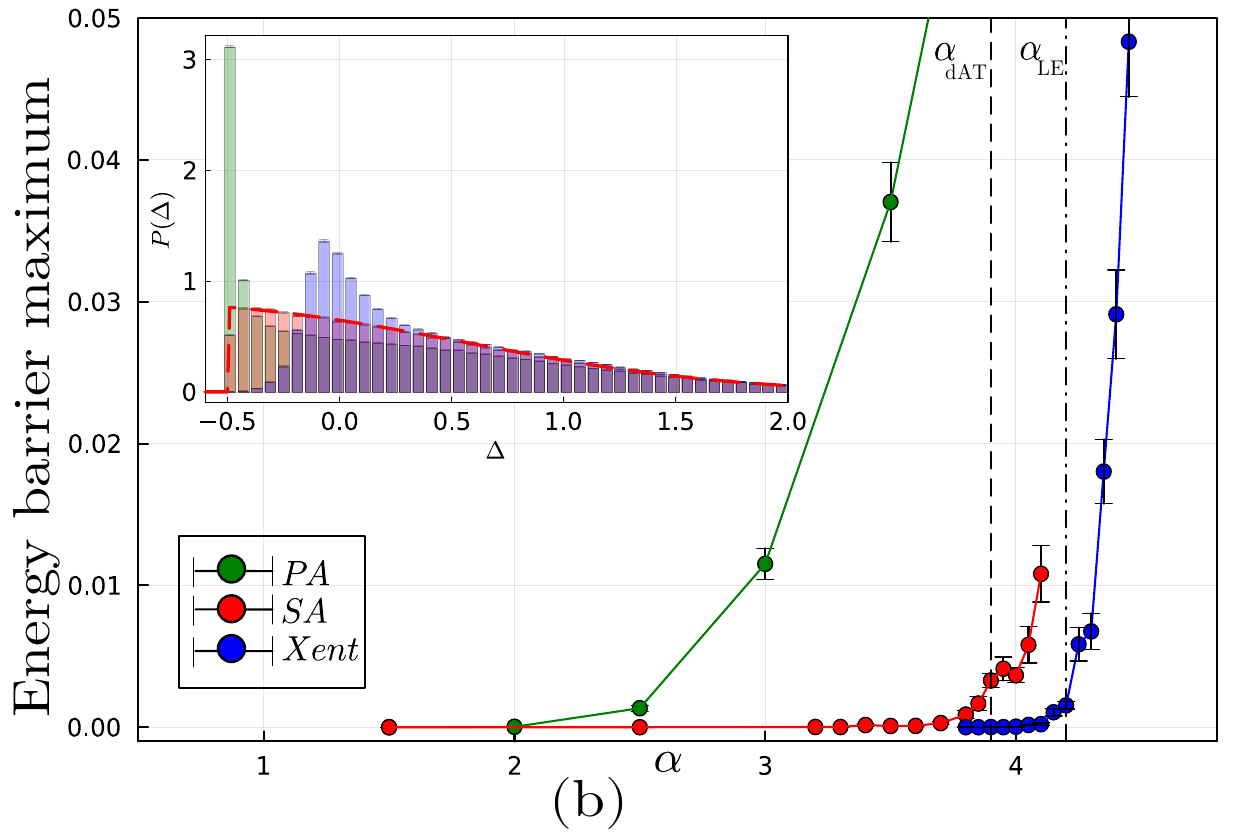}
	\end{centering}
	\caption{\label{fig:phase_diagram}              
		(a) \textbf{Coalescence threshold lines}  at $\kp=-0.5$ (blue and cyan), and the $\kappa_{\text{krn}}$ threshold (green line) as a function of $\alpha$. In orange the dAT transition line, delimiting the RS-stable region; in red the RS estimate of the $\kappa_{max}$. Above the dAT line, dashed lines indicate the continuation of the (unstable) RS predictions. Points are numerical extrapolations from SA samples (see SM for details). (b) \textbf{Maximum error along the geodesic path} ($y=2$) connecting numerical solutions found with different algorithms (PA, SA and Xent) with the fBP max-margin solutions. Non-zero energies along the path indicate disconnection in the solution space. The vertical dashed lines denote the values of $\alpha_{dAT}$ and $\alpha_{LE}$ at $\kp=-0.5$. The inset shows stability distributions for PA, SA and Xent at $\alpha=2$ compared with the theoretical stability distribution of typical solutions (red dashed line).
	} 
\end{figure*}

\emph{The largest geodesically convex component.} We first consider the case where the vertices of the simplex are all sampled with identical margin $k_r = k$, with $\kp\le k<\kappa_{dAT}(\alpha)$. 
One finds that the energy on the projected $(y-1)-$simplex is always strictly greater than zero when $k = \kp$,
while extended regions around each vertex fall to zero energy for $k>\kp$. 
For each value of $y$ one can identify a \emph{coalescence threshold}, 
$\kappa^\star_y(\kp, \alpha)$, 
corresponding to the value of the margin above which the entire $(y-1)-$simplex lies at zero energy. 
In particular, we find the minimum margin $\kappa^\star_2(\kp, \alpha)$ that ensures \emph{linear mode connectivity}.
These thresholds are displayed in Fig.\,\ref{fig:phase_diagram} as a function of $\alpha$ for $\kp=-0.5$, and satisfy $\kappa^\star_2 < \kappa^\star_3< ...<\kappa^\star_\infty$. Above the last \emph{coalescence threshold}, $\kappa^\star_\infty$, the projected convex hull of the entire ensemble of solutions 
lies at zero energy: this is what we call geodesically convex component of the manifold of solutions (see also~\ref{app:def}). The size of this region can be bounded by the typical overlap $q$ between $\kappa_{\infty}^\star$ solutions. 
By inspecting the distribution of stabilities across the zero-energy manifold, one finds that the geodesic paths encounter different solutions from those of the equilibrium description at the corresponding margin (details in the SM). 


In Fig.~\ref{fig:y3} (a), we plot $E_{\boldsymbol{\gamma}}$ on the two-dimensional simplex, with all vertices at $k = \kp$. Since the maximum energy barriers are located on the edges of the projected simplex, the minimum energy path connecting the corner solutions needs to deviate through its barycenter. Notice that, since $\kappa^\star_2<\kappa^\star_3$, 
as $k$ is increased from $k = \kp$, the energy barriers along the edges go to zero faster than the energy at the center of the simplex.
At the top of panel (c), we show how the barrier on the edges goes to zero as the margin of the vertices is increased.

\emph{The kernel of the solution space.}
We now focus on the connectivity of solutions with different margins. Specifically, we start by considering the geodesic path between a typical solution, $k_1 = \kp$, and a higher margin solution, $k_2>\kp$. 
One can show that, for any $(\kp, \alpha)$ below the dAT transition, there exists a threshold $\kappa_{\text{krn}}$ such that, w.h.p., no energy barrier is encountered along the geodesic path between any solution with $k_2 \ge \kappa_{\text{krn}}$ and any other typical solution with margin $k_1\ge\kp$. 
These findings imply that the solution space is \emph{star-shaped}, and allow us to identify the \emph{kernel} of solution space, i.e. a subset of solutions
that are ``visible'' -- through geodesic paths -- from any typical point of the solution manifold. 
In the bottom of Fig.~\ref{fig:y3} (c), we show the decrease of the energy barrier with $k_2$. 

In Fig.~\ref{fig:phase_diagram} (a), we display the line $\kappa_{\text{krn}}(\alpha)$ for a problem with margin $\kp=-0.5$. 
Notice that its continuation above the dAT line (dashed), where the RS results are incorrect, would predict an intersection between the $\kappa_{\text{krn}}$ and the $\kappa_{\text{max}}$ lines, implying a break-down of the star-shaped property. We revert to numerical experiments, in the next section, to understand what happens to the connectivity of typical solutions in this phase.

The analysis of the connectivity of solutions with different margins can be carried out also with $y>2$. In Fig.~\ref{fig:y3} (b), we display the $y=3$ case, with the bottom vertices being typical solutions with margin $\kp$ and the upper vertex having a margin larger than $\kappa_{\text{krn}}$. As expected from the $y=2$ analysis, by deviating through the core of the solution manifold, one can construct a piece-wise geodesic path between \emph{any} pair of solutions lying exactly at zero energy. 


\emph{Sampling bias and disconnection.} 
We compare the properties of solutions found with different solvers on instances of the negative perceptron with $\kp=-0.5$. In particular, in Fig.~\ref{fig:phase_diagram} (b), we characterize their geodesic connectivity to solutions located in the kernel region, as a function of $\alpha$. Note that, because of the nested overlap structure, we expect the maximum-margin solutions of the problem to be located in the kernel. Therefore, for obtaining them we employ the focusing-BP (fBP) algorithm, which was shown in \cite{baldassi2023typical} to yield good proxies of the $\kappa_{\text{max}}$ solutions.

Typical solutions instead are approximated by carefully applying Simulated Annealing (SA) on the square hinge loss with margin $\kp$ and are found to be in good agreement with the theory (cf. the points in Fig.~\ref{fig:y3} (a)-(b) and \ref{fig:phase_diagram} (a), and further experiments in the SM). Non-zero energy barriers with the fBP solutions seem to appear in close proximity of the dAT transition line, confirming the star-shapedness of typical solutions in the RS stable region.

SGD on the cross-entropy loss -- the most common optimization objective for this class of problems -- yields \emph{robust} solutions \cite{baldassi2020shaping} with higher average stability than typical, as shown in the inset in Fig.~\ref{fig:phase_diagram} (b). The geodesic path between independent optimization trajectories, starting from random initialization, shows no energy barriers as soon as the zero-energy region is accessed, revealing an algorithmic bias towards the geodesically convex component of the solution manifold (details in Appendix \ref{app:simulations} and in the SM). The disconnection transition with the core solutions (Xent in Fig.~\ref{fig:phase_diagram} (b)) is delayed with respect to SA, and seems to happen in close proximity of the $\alpha_{LE}$ transition characterized in \cite{baldassi2023typical}.


Finally, we implement the classic Perceptron Algorithm (PA). When the learning rate is sufficiently small, this algorithm is able to sample solutions with a large mass of stabilities at threshold (inset of Fig.~\ref{fig:phase_diagram} (b)), and therefore less robust than typical ones. The disconnection with the core region of the solution manifold is in this case anticipated before the dAT line. Notice that this result is not incompatible with our predictions, since these solutions seem to be sub-dominant in the flat measure over solutions, and cannot be seen through an equilibrium analysis.

These numerical results are consistent with our theoretical picture of a star-shaped space of solutions in the over-parameterized regime, and reveal a progressive disconnection transition that affects different types of solutions according to their degree of robustness. 

{\em Discussion and conclusions.} In the present work, we characterized the connectivity properties of a prototypical model of non-convex neural networks. The theoretical analysis unveiled the presence, in the overparameterized regime, of a connected manifold of solutions 
organized in a star-shaped structure. 
Similar types of structures have been shown to appear in completely unrelated high-dimensional problems~\cite{ZhangStrogatz2021,Martiniani2023}. We conjecture that simple mode connectivity may be a universal property of non-convex optimization problems in the over-parameterized regime. 
A promising future research direction would be to investigate analytically whether the star-shaped geometry, or a generalization thereof, holds in more complex \cite{multilayerjamming} and more realistic models of neural networks \cite{spigler2019, jacot2022}. 

With a precise picture in hand, we were also able to characterize where different solvers end up in the solution space. Understanding how algorithmic bias can be exploited to enhance learning performance is a central question in the field of deep learning. At the same time, probing the landscape with different dynamics could help characterize the solution space in more complex settings, following up on~\cite{entezari2022, pittorino22}.

\appendix

\section{Definitions} \label{app:def}



\paragraph{\textbf{Typical solutions.}}

In high dimensions, due to entropic factors, independent sampling from a given measure will return w.h.p. configurations with specific shared properties. For instance, almost all the exponentially many solutions of the perceptron problem have the same stability distribution profile.
In this work, we define i.i.d. samples from measure (\ref{eq:prob}) as  \emph{typical} solutions. The measure also contains different types of solutions that we call \emph{atypical} and that are exponentially sub-dominant in number and therefore become statistically irrelevant when considering the high-dimensional limit.
Some atypical solutions satisfy a higher margin constraint than the one of the problem, or more generally have a different stability profile (e.g., \emph{robust} solutions sampled with SGD have higher average stability than typical).  




\paragraph{\textbf{Projection of the Euclidean $y$-symplex on the $N$-dimensional sphere: geodesic paths and polytopes}}

\begin{figure}[h]
	\begin{centering}
		\includegraphics[height=2.2cm]{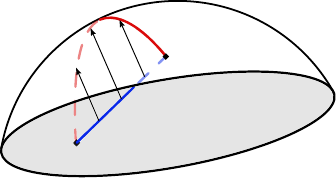}
		\includegraphics[height=3.cm]{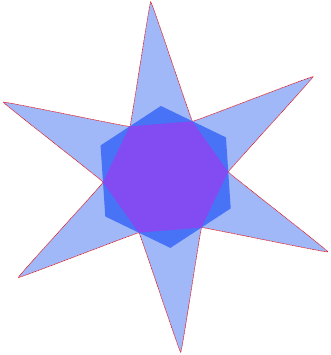}
	\end{centering} 
	\caption{\label{fig::geodesic_proj} Left: illustrative example of the projection of a straight path (blue) onto the geodesic (red) in the case of a one-dimensional path, i.e. $y=2$. All points along the geodesic (or that can be connected by a geodesic) are said to be \emph{geodesically connected}.
		Right: Simple 2-dimensional visualization of the shape of the manifold of solutions. The large geodesically convex component, (darker blue) and its kernel (purple) are also shown.
	}
\end{figure}

All the results presented in the paper for the interpolation energy Eq.~\eqref{eq:interpolating_vector} are obtained first on the Euclidean simplex by interpolating configurations among $y$ normalized solutions, and then by normalizing all interpolated configurations, i.e. projecting them on the $N$-dimensional sphere. This gives rise, on the $N$-dimensional sphere, to geodesic paths for $y=2$ and geodesic polytopes of dimension $y-1$ for $y>2$ (their edges are themselves geodesic polytopes of dimension $y-2$). See Fig.~\ref{fig::geodesic_proj} for an illustration of the projection of the straight path onto the geodesic for $y=2$.

\paragraph{\textbf{Geodesic connectivity and geodesic convexity}}

Given two points $\bx_1, \bx_2$ of a set $S$ in a Euclidean space, we say that $\bx_1$ is connected to $\bx_2$ via $S$ if the straight path $[\bx_1,\bx_2]=\{\gamma \bx_1 + (1-\gamma) \bx_2\,: \,\gamma\in [0,1]\}$ is such that $[\bx_1,\bx_2] \subset S$. A set $C\subset\R^d$ is convex if all $\bx_1,\bx_2 \in C$ are connected via $C$ \cite{Hansen2020}. 
Keeping in mind the projection operation of the $y-1$-simplex on the $N$-dimensional sphere illustrated in Fig.~\ref{fig::geodesic_proj} for a geodesic path ($y=2$), we can notice that \emph{all notions of connectivity and convexity usually defined in the Euclidean space straight-forwardly generalize to the $N$-sphere by substituting the concept of straight path with the one of geodesic}. Since our model is defined on the $N$-sphere, we thus mention along the paper the concepts of \emph{geodesic} connectivity and \emph{geodesic} convexity. 
Fig.~\ref{fig::geodesic_proj} also illustrates the notion of geodesic connectivity.

\paragraph{\textbf{Star-shaped set its kernel and the geodesically convex component}}
As a natural 
relaxation of the convexity notion, a set $S\subset \R^d$ is star-shaped if $\exists \, \bx_1 \in S$ such that $\forall \bx_2 \in S$ we have $[\bx_1, \bx_2] \subset S$.
The \emph{kernel} of a star-shaped set $S$ is defined as the set of all $\bx_1 \in S$ such that $[\bx_1,\bx_2] \subset S$ $\forall \bx_2 \in S$.
Its elements are called the star centers of $S$. The kernel of a star-shaped set is a convex set.

The kernel of the star-shaped set of solutions of the negative perceptron considered in this paper is represented by the set of solutions with $\kappa>\kappa_{\text{krn}}$, see Fig.~\ref{fig:phase_diagram}.  In Fig.~\ref{fig::geodesic_proj} we represent a simple 2-dimensional star-shaped set, its kernel, the possible existence inside the star-shaped set of a convex set containing the kernel and distinct from it. This is called geodesically convex component in the main text. 



\section{Details on the analytical computation for the connectivity thresholds} \label{app:computation}

To compute the asymptotic training error $E_{\boldsymbol{\gamma}}$ in Eq.~\eqref{eq:general_energybarrier} on the manifold spanned by $y$ independently sampled solutions, we resort to the replica trick~\cite{mezard1987spin}. 
The computation for the general case in which each sampled solution has a distinct margin $\{k_r\}_{r=1}^y \geq \kappa_E$ involves standard steps (see SM for details). 
The final expression of $E_{\boldsymbol{\gamma}}$ can be factored into the product of two terms: 
\begin{equation}
	E_{\boldsymbol{\gamma}} =\Theta\left(f_{\boldsymbol{\gamma}}(\kp, \left\{ k_r \right\}_r) \right)  I_{\boldsymbol{\gamma}}\left( \kp, \left\{ k_r \right\}_r\right) \,,
\end{equation}
where $f_{\boldsymbol{\gamma}}(\kp, \left\{ k_r \right\}_r) = \kp c_{\boldsymbol{\gamma}} - \sum_r \gamma_r k_r$ is a linear function of the margin, with $c_{\boldsymbol{\gamma}} = \sqrt{\sum_{rs} q_{rs} \gamma_r \gamma_s}$, and $I_{\boldsymbol{\gamma}}$ is a non-negative function involving $2y$-dimensional Gaussian integral, whose expression is fully reported in the SM. Since the expression evaluates to zero only if the argument of the Heaviside $\Theta$ function is negative, the sign of $f_{\boldsymbol{\gamma}}$ is sufficient to analytically derive the connectivity thresholds reported in Fig.~\ref{fig:phase_diagram}. It is useful to distinguish between cases:
\begin{enumerate}
	\item  All sampled solutions corresponding to the vertices of the $y-1$ dimensional simplex have the same margin $k_r = k \geq \kappa_E$, 
	giving $f_{\boldsymbol{\gamma}}=\kp c_{\boldsymbol{\gamma}}-k$. In this case, since all solutions are statistically equivalent, the only order parameter is the typical overlap between them $q=\frac{1}{N}\mathbb{E}\langle\bW^1\cdot\bW^2\rangle_{k,k}$. Consequently the norm of the interpolating vector is $c_{\boldsymbol{\gamma}} = \left(1-q\right)\sum_r \gamma_r^{2}+q\ \leq 1$. We consider two subcases:
	\begin{enumerate}
		\item In the case where all vertices are typical solutions with $\kappa \equiv \kp$, since $c_{\boldsymbol{\gamma}} \le 1$ then $f_{\boldsymbol{\gamma}}=\kp c_{\boldsymbol{\gamma}} - \kp >0$ when $\kp < 0$ and therefore every interpolated configuration has non-vanishing training error (Fig.~\ref{fig:y3}). When $\kp \ge 0$ instead, we have that $f_{\boldsymbol{\gamma}} \le 0$ and therefore $E_{\boldsymbol{\gamma}}=0$ for all $\boldsymbol{\gamma}$, consistently with the convexity of the solution space. 
		\item When $\kp < \kappa < 0$, 
		the minimum value of $c_{\boldsymbol{\gamma}}$ is attained on the barycenter $\gamma_r = \frac{1}{y} \,\,\, \forall r$. Hence, if the inequality $\kp c_{\boldsymbol{\gamma}} - k < 0$ holds for the barycenter, all interpolated configurations have zero training error. The condition
		$
		\label{eq::kappaystar}
		\kappa > \kp c_{\text{barycenter}} = \kp \sqrt{(1-q) \frac{1}{y} + q } \equiv \kappa_y^\star
		$
		defines the coalescence thresholds $\kappa_2^\star < \kappa_3^\star < \dots < \kappa_\infty^\star = \kp \sqrt{q}$ (see Fig.~\ref{fig:phase_diagram}) indicating that for $\kappa>\kappa_y^\star$ the normalized $y-1$-simplex lies at zero error. Their ordering in $y$ is due to the fact that the overlap $q$ is an increasing monotonic function of the margin $\kappa$. 
	\end{enumerate}
	\item Solutions sampled with different margins. If $y-1$ vertices have margins $k_1$ and one vertex has margin $k_2$, with $k_2 > k_1 \geq \kp$, then the overlap matrix will depend on $q_1=\frac{1}{N}\mathbb{E}\langle\bW^1\cdot\bW^2\rangle_{k_1,k_1}$, $q_2=\frac{1}{N}\mathbb{E}\langle\bW^1\cdot\bW^2\rangle_{k_2,k_2}$ and $p=\frac{1}{N}\mathbb{E}\langle\bW^1\cdot\bW^2\rangle_{k_1,k_2}$. In this case we can rewrite $f_{\boldsymbol{\gamma}}=\kp c_{\boldsymbol{\gamma}} - k_1 \sum_{r=1}^{y-1} \gamma_r - k_2 \gamma_y$ and study it analytically when an explicit algebraic relation can be derived. For $y=2$ (and by using the explicit expression of the norm $c_{\boldsymbol{\gamma}}$, see SM for details) we can find 
	the minimum margin $k_2$ that should be imposed on $\bW^2$ given the margin $k_1$ on $\bW^1$ such that the two solutions are geodesically connected. We call it $\kappa_{\text{krn}}(k_1)$ and define it as:
	$
	\kappa_{\text{krn}}(k_1) = k_1 p - \sqrt{(1-p^2)(\kp^2 - k_1^2)}.
	$
	The maximum value of $\kappa_{\text{krn}}$ is obtained for $k_1 = \kp$, for which $ \kappa_{\text{krn}} = k_1 p$ which is reported in Fig.~\ref{fig:phase_diagram}.
\end{enumerate}

\section{Numerical simulations on the attractiveness of the convex core for SGD} \label{app:simulations}

\begin{figure}[h]
	\begin{centering}
		\includegraphics[width=0.9\columnwidth]{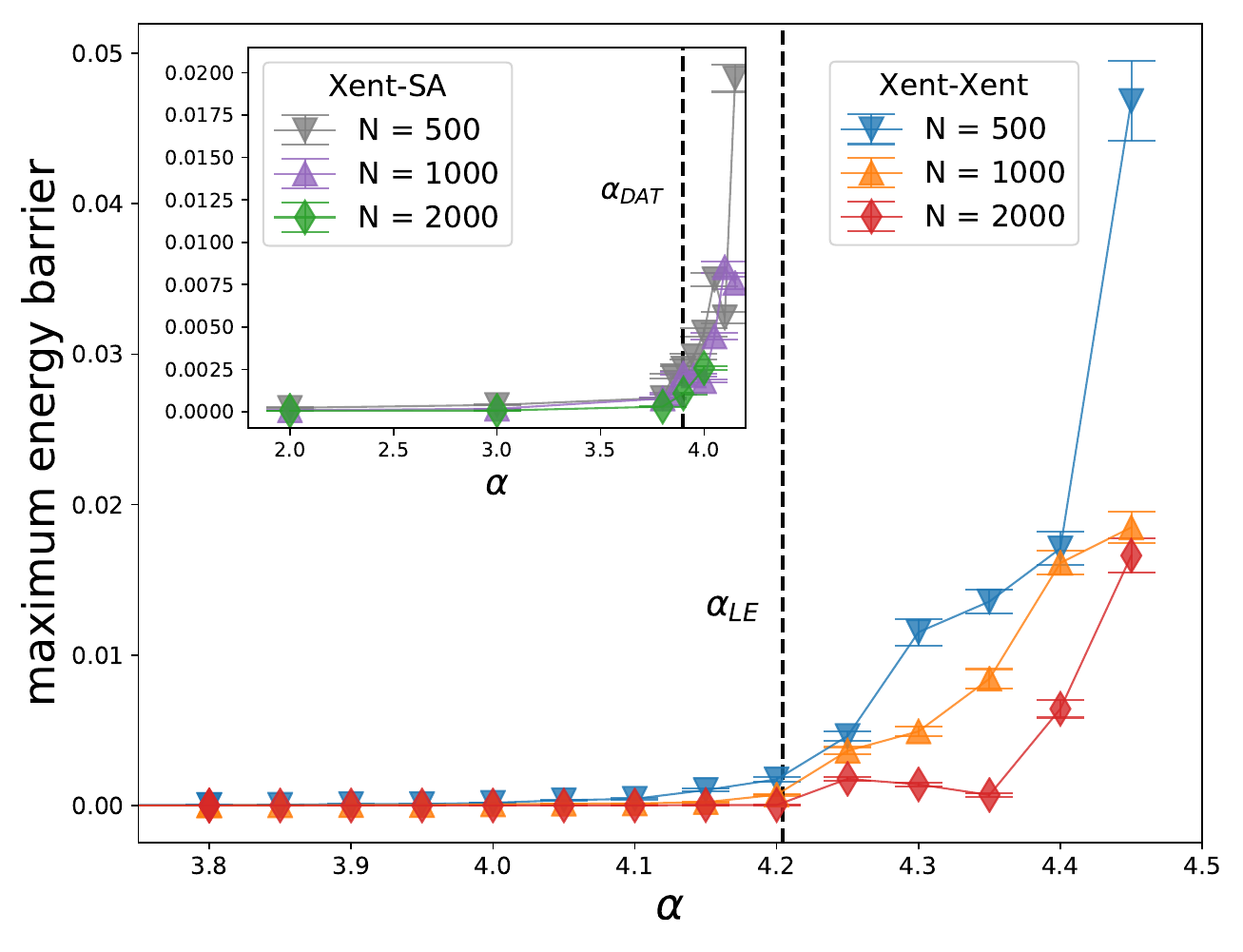}
		\includegraphics[width=0.9\columnwidth]{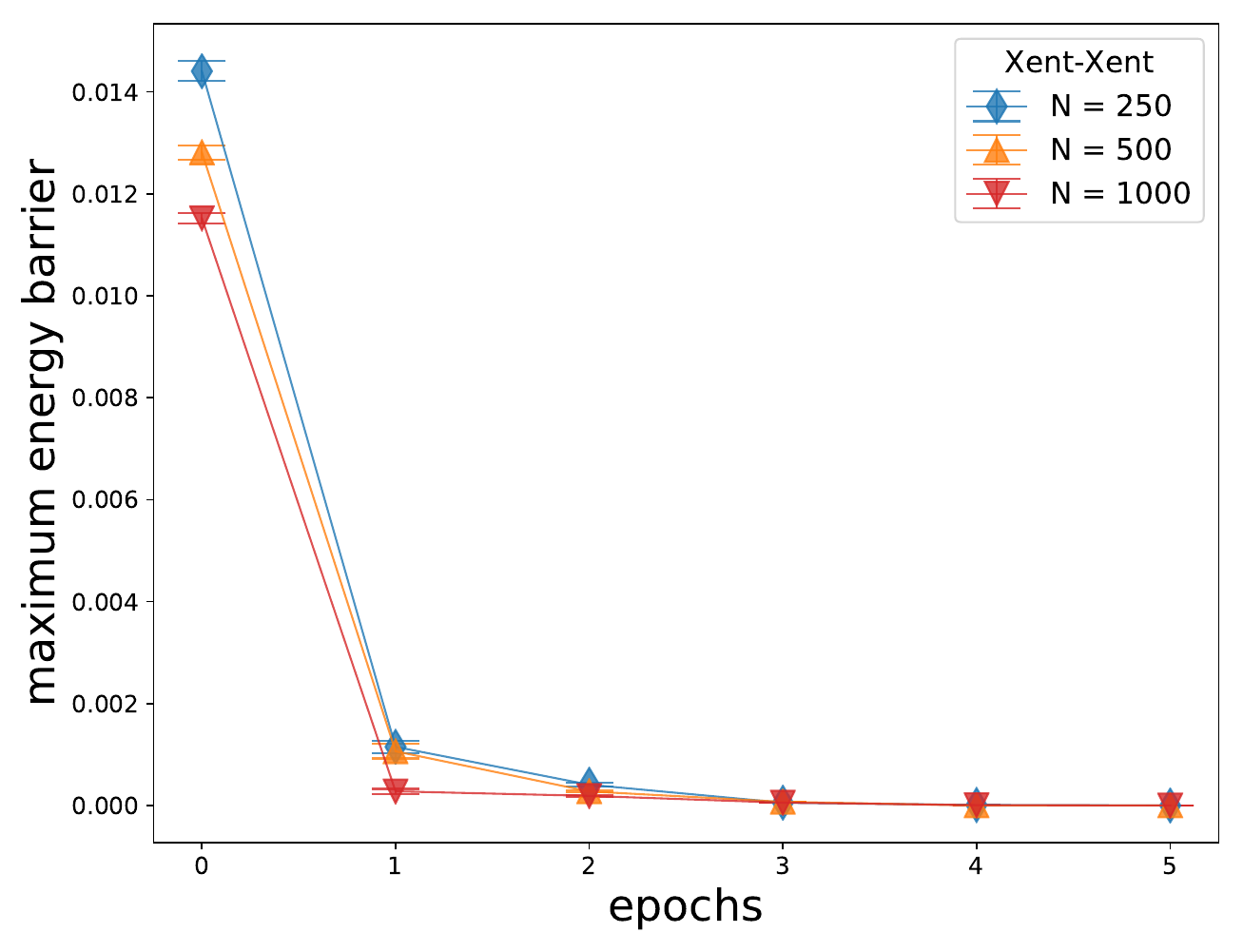}
	\end{centering}
	\vspace{-0.5cm}
	\caption{\label{fig:sgd_barr} 
		(Top) Maximum barrier height along the geodesic path connecting two SGD solutions obtained with the cross-entropy loss in function of $\alpha$. The training dynamics is stopped as soon as a solution is found. (Inset) Maximum barrier height along the geodesic path connecting SGD solutions with SA solutions.
		(Bottom) Maximum barrier height between two configurations initialized on typical solutions (obtained with SA) as a function of the number of SGD epochs on the cross-entropy loss. After a few epochs, the gradient drives the solutions towards the convex core. 
	} 
\end{figure}

We study the inductive bias of SGD dynamics on the cross-entropy loss by evaluating the average maximum energy barrier's height between pairs of independent solutions (means are over different initial conditions). 

We observe that randomly initialized  trajectories on the $N$-sphere access the zero-energy star-shaped manifold of solutions by entering directly in the geodesically convex component: as soon as the solutions are obtained, they are not only connected to fBP solutions in the core of the solution space (see \emph{Xent} curve in Fig.~\ref{fig:phase_diagram}~(b)), but they are also connected between each-other with zero-energy geodesic paths, as shown in the upper panel of Fig.~\ref{fig:sgd_barr}, where the optimization is stopped as soon as the solutions are found. The disconnection transition of the SGD-SGD barrier is located close to the $\alpha_{LE}$ transition \cite{baldassi2023typical}.

To further investigate the dynamical bias of SGD towards the convex manifold, we consider pairs of trajectories initialized inside the zero-energy region in correspondence to typical solutions (provided by SA). We find that the initial non-zero energy barrier between them quickly drops to zero in a few epochs, showing that SGD naturally drifts to the inner and convex region of the solution space, see lower panel of Fig.~\ref{fig:sgd_barr}.

\bibliography{references.bib}

\begin{thebibliography}{50}%
\makeatletter
\providecommand \@ifxundefined [1]{%
 \@ifx{#1\undefined}
}%
\providecommand \@ifnum [1]{%
 \ifnum #1\expandafter \@firstoftwo
 \else \expandafter \@secondoftwo
 \fi
}%
\providecommand \@ifx [1]{%
 \ifx #1\expandafter \@firstoftwo
 \else \expandafter \@secondoftwo
 \fi
}%
\providecommand \natexlab [1]{#1}%
\providecommand \enquote  [1]{``#1''}%
\providecommand \bibnamefont  [1]{#1}%
\providecommand \bibfnamefont [1]{#1}%
\providecommand \citenamefont [1]{#1}%
\providecommand \href@noop [0]{\@secondoftwo}%
\providecommand \href [0]{\begingroup \@sanitize@url \@href}%
\providecommand \@href[1]{\@@startlink{#1}\@@href}%
\providecommand \@@href[1]{\endgroup#1\@@endlink}%
\providecommand \@sanitize@url [0]{\catcode `\\12\catcode `\$12\catcode
  `\&12\catcode `\#12\catcode `\^12\catcode `\_12\catcode `\%12\relax}%
\providecommand \@@startlink[1]{}%
\providecommand \@@endlink[0]{}%
\providecommand \url  [0]{\begingroup\@sanitize@url \@url }%
\providecommand \@url [1]{\endgroup\@href {#1}{\urlprefix }}%
\providecommand \urlprefix  [0]{URL }%
\providecommand \Eprint [0]{\href }%
\providecommand \doibase [0]{http://dx.doi.org/}%
\providecommand \selectlanguage [0]{\@gobble}%
\providecommand \bibinfo  [0]{\@secondoftwo}%
\providecommand \bibfield  [0]{\@secondoftwo}%
\providecommand \translation [1]{[#1]}%
\providecommand \BibitemOpen [0]{}%
\providecommand \bibitemStop [0]{}%
\providecommand \bibitemNoStop [0]{.\EOS\space}%
\providecommand \EOS [0]{\spacefactor3000\relax}%
\providecommand \BibitemShut  [1]{\csname bibitem#1\endcsname}%
\let\auto@bib@innerbib\@empty
\bibitem [{\citenamefont {M\'ezard}\ \emph {et~al.}(2002)\citenamefont
  {M\'ezard}, \citenamefont {Parisi},\ and\ \citenamefont
  {Zecchina}}]{MPZ2002}%
  \BibitemOpen
  \bibfield  {author} {\bibinfo {author} {\bibfnamefont {M.}~\bibnamefont
  {M\'ezard}}, \bibinfo {author} {\bibfnamefont {G.}~\bibnamefont {Parisi}}, \
  and\ \bibinfo {author} {\bibfnamefont {R.}~\bibnamefont {Zecchina}},\ }\href
  {\doibase 10.1126/science.1073287} {\bibfield  {journal} {\bibinfo  {journal}
  {Science}\ }\textbf {\bibinfo {volume} {297}},\ \bibinfo {pages} {812}
  (\bibinfo {year} {2002})},\ \Eprint
  {http://arxiv.org/abs/https://www.science.org/doi/pdf/10.1126/science.1073287}
  {https://www.science.org/doi/pdf/10.1126/science.1073287} \BibitemShut
  {NoStop}%
\bibitem [{\citenamefont {Martin}\ \emph {et~al.}(2001)\citenamefont {Martin},
  \citenamefont {Monasson},\ and\ \citenamefont {Zecchina}}]{MMZ2001}%
  \BibitemOpen
  \bibfield  {author} {\bibinfo {author} {\bibfnamefont {O.~C.}\ \bibnamefont
  {Martin}}, \bibinfo {author} {\bibfnamefont {R.}~\bibnamefont {Monasson}}, \
  and\ \bibinfo {author} {\bibfnamefont {R.}~\bibnamefont {Zecchina}},\ }\href
  {\doibase https://doi.org/10.1016/S0304-3975(01)00149-9} {\bibfield
  {journal} {\bibinfo  {journal} {Theoretical Computer Science}\ }\textbf
  {\bibinfo {volume} {265}},\ \bibinfo {pages} {3} (\bibinfo {year} {2001})},\
  \bibinfo {note} {phase Transitions in Combinatorial Problems}\BibitemShut
  {NoStop}%
\bibitem [{\citenamefont {M\'ezard}\ \emph {et~al.}(2005)\citenamefont
  {M\'ezard}, \citenamefont {Mora},\ and\ \citenamefont {Zecchina}}]{MMZ2005}%
  \BibitemOpen
  \bibfield  {author} {\bibinfo {author} {\bibfnamefont {M.}~\bibnamefont
  {M\'ezard}}, \bibinfo {author} {\bibfnamefont {T.}~\bibnamefont {Mora}}, \
  and\ \bibinfo {author} {\bibfnamefont {R.}~\bibnamefont {Zecchina}},\ }\href
  {\doibase 10.1103/PhysRevLett.94.197205} {\bibfield  {journal} {\bibinfo
  {journal} {Phys. Rev. Lett.}\ }\textbf {\bibinfo {volume} {94}},\ \bibinfo
  {pages} {197205} (\bibinfo {year} {2005})}\BibitemShut {NoStop}%
\bibitem [{\citenamefont {Zdeborov\'a}\ and\ \citenamefont
  {Krz\k{a}ka\l{}a}(2007)}]{ZK2007}%
  \BibitemOpen
  \bibfield  {author} {\bibinfo {author} {\bibfnamefont {L.}~\bibnamefont
  {Zdeborov\'a}}\ and\ \bibinfo {author} {\bibfnamefont {F.}~\bibnamefont
  {Krz\k{a}ka\l{}a}},\ }\href {\doibase 10.1103/PhysRevE.76.031131} {\bibfield
  {journal} {\bibinfo  {journal} {Phys. Rev. E}\ }\textbf {\bibinfo {volume}
  {76}},\ \bibinfo {pages} {031131} (\bibinfo {year} {2007})}\BibitemShut
  {NoStop}%
\bibitem [{\citenamefont {M{\'e}zard}\ \emph {et~al.}(1987)\citenamefont
  {M{\'e}zard}, \citenamefont {Parisi},\ and\ \citenamefont
  {Virasoro}}]{mezard1987spin}%
  \BibitemOpen
  \bibfield  {author} {\bibinfo {author} {\bibfnamefont {M.}~\bibnamefont
  {M{\'e}zard}}, \bibinfo {author} {\bibfnamefont {G.}~\bibnamefont {Parisi}},
  \ and\ \bibinfo {author} {\bibfnamefont {M.}~\bibnamefont {Virasoro}},\
  }\href@noop {} {\emph {\bibinfo {title} {Spin glass theory and beyond: An
  Introduction to the Replica Method and Its Applications}}},\ Vol.~\bibinfo
  {volume} {9}\ (\bibinfo  {publisher} {World Scientific Publishing Company},\
  \bibinfo {year} {1987})\BibitemShut {NoStop}%
\bibitem [{\citenamefont {Monasson}\ \emph {et~al.}(1999)\citenamefont
  {Monasson}, \citenamefont {Zecchina}, \citenamefont {Kirkpatrick},
  \citenamefont {Selman},\ and\ \citenamefont
  {Troyansky}}]{monasson1999determining}%
  \BibitemOpen
  \bibfield  {author} {\bibinfo {author} {\bibfnamefont {R.}~\bibnamefont
  {Monasson}}, \bibinfo {author} {\bibfnamefont {R.}~\bibnamefont {Zecchina}},
  \bibinfo {author} {\bibfnamefont {S.}~\bibnamefont {Kirkpatrick}}, \bibinfo
  {author} {\bibfnamefont {B.}~\bibnamefont {Selman}}, \ and\ \bibinfo {author}
  {\bibfnamefont {L.}~\bibnamefont {Troyansky}},\ }\href {\doibase
  10.1038/22055} {\bibfield  {journal} {\bibinfo  {journal} {Nature}\ }\textbf
  {\bibinfo {volume} {400}},\ \bibinfo {pages} {133} (\bibinfo {year}
  {1999})}\BibitemShut {NoStop}%
\bibitem [{\citenamefont {Cavaliere}\ and\ \citenamefont
  {Ricci-Tersenghi}(2023)}]{cavaliere2023biased}%
  \BibitemOpen
  \bibfield  {author} {\bibinfo {author} {\bibfnamefont {A.~G.}\ \bibnamefont
  {Cavaliere}}\ and\ \bibinfo {author} {\bibfnamefont {F.}~\bibnamefont
  {Ricci-Tersenghi}},\ }\href@noop {} {\bibfield  {journal} {\bibinfo
  {journal} {arXiv preprint arXiv:2303.14879}\ } (\bibinfo {year}
  {2023})}\BibitemShut {NoStop}%
\bibitem [{epf()}]{epfl_workshop}%
  \BibitemOpen
  \href
  {https://www.epfl.ch/labs/lcn/epfl-virtual-symposium-loss-landscape-of-neural-networks-theoretical-insights-and-practical-implications-15-16-february-2022/}
  {\bibinfo  {journal} {Loss Landscape of Neural Networks: theoretical insights
  and practical implications, EPFL Virtual Symposium}\ }\BibitemShut {NoStop}%
\bibitem [{\citenamefont {Keskar}\ \emph {et~al.}(2016)\citenamefont {Keskar},
  \citenamefont {Mudigere}, \citenamefont {Nocedal}, \citenamefont
  {Smelyanskiy},\ and\ \citenamefont {Tang}}]{keskar}%
  \BibitemOpen
\bibfield  {journal} {  }\bibfield  {author} {\bibinfo {author} {\bibfnamefont
  {N.~S.}\ \bibnamefont {Keskar}}, \bibinfo {author} {\bibfnamefont
  {D.}~\bibnamefont {Mudigere}}, \bibinfo {author} {\bibfnamefont
  {J.}~\bibnamefont {Nocedal}}, \bibinfo {author} {\bibfnamefont
  {M.}~\bibnamefont {Smelyanskiy}}, \ and\ \bibinfo {author} {\bibfnamefont
  {P.~T.~P.}\ \bibnamefont {Tang}},\ }\href {http://arxiv.org/abs/1609.04836}
  {\bibfield  {journal} {\bibinfo  {journal} {CoRR}\ }\textbf {\bibinfo
  {volume} {abs/1609.04836}} (\bibinfo {year} {2016})},\ \Eprint
  {http://arxiv.org/abs/1609.04836} {arXiv:1609.04836} \BibitemShut {NoStop}%
\bibitem [{\citenamefont {Li}\ \emph {et~al.}(2018)\citenamefont {Li},
  \citenamefont {Xu}, \citenamefont {Taylor}, \citenamefont {Studer},\ and\
  \citenamefont {Goldstein}}]{LiVisualizing2018}%
  \BibitemOpen
  \bibfield  {author} {\bibinfo {author} {\bibfnamefont {H.}~\bibnamefont
  {Li}}, \bibinfo {author} {\bibfnamefont {Z.}~\bibnamefont {Xu}}, \bibinfo
  {author} {\bibfnamefont {G.}~\bibnamefont {Taylor}}, \bibinfo {author}
  {\bibfnamefont {C.}~\bibnamefont {Studer}}, \ and\ \bibinfo {author}
  {\bibfnamefont {T.}~\bibnamefont {Goldstein}},\ }in\ \href
  {https://proceedings.neurips.cc/paper/2018/file/a41b3bb3e6b050b6c9067c67f663b915-Paper.pdf}
  {\emph {\bibinfo {booktitle} {Advances in Neural Information Processing
  Systems}}},\ Vol.~\bibinfo {volume} {31},\ \bibinfo {editor} {edited by\
  \bibinfo {editor} {\bibfnamefont {S.}~\bibnamefont {Bengio}}, \bibinfo
  {editor} {\bibfnamefont {H.}~\bibnamefont {Wallach}}, \bibinfo {editor}
  {\bibfnamefont {H.}~\bibnamefont {Larochelle}}, \bibinfo {editor}
  {\bibfnamefont {K.}~\bibnamefont {Grauman}}, \bibinfo {editor} {\bibfnamefont
  {N.}~\bibnamefont {Cesa-Bianchi}}, \ and\ \bibinfo {editor} {\bibfnamefont
  {R.}~\bibnamefont {Garnett}}}\ (\bibinfo  {publisher} {Curran Associates,
  Inc.},\ \bibinfo {year} {2018})\BibitemShut {NoStop}%
\bibitem [{\citenamefont {Baldassi}\ \emph {et~al.}(2015)\citenamefont
  {Baldassi}, \citenamefont {Ingrosso}, \citenamefont {Lucibello},
  \citenamefont {Saglietti},\ and\ \citenamefont
  {Zecchina}}]{baldassi2015subdominant}%
  \BibitemOpen
  \bibfield  {author} {\bibinfo {author} {\bibfnamefont {C.}~\bibnamefont
  {Baldassi}}, \bibinfo {author} {\bibfnamefont {A.}~\bibnamefont {Ingrosso}},
  \bibinfo {author} {\bibfnamefont {C.}~\bibnamefont {Lucibello}}, \bibinfo
  {author} {\bibfnamefont {L.}~\bibnamefont {Saglietti}}, \ and\ \bibinfo
  {author} {\bibfnamefont {R.}~\bibnamefont {Zecchina}},\ }\href {\doibase
  10.1103/PhysRevLett.115.128101} {\bibfield  {journal} {\bibinfo  {journal}
  {Phys. Rev. Lett.}\ }\textbf {\bibinfo {volume} {115}},\ \bibinfo {pages}
  {128101} (\bibinfo {year} {2015})}\BibitemShut {NoStop}%
\bibitem [{\citenamefont {Baldassi}\ \emph {et~al.}(2016)\citenamefont
  {Baldassi}, \citenamefont {Borgs}, \citenamefont {Chayes}, \citenamefont
  {Ingrosso}, \citenamefont {Lucibello}, \citenamefont {Saglietti},\ and\
  \citenamefont {Zecchina}}]{unreasoanable}%
  \BibitemOpen
  \bibfield  {author} {\bibinfo {author} {\bibfnamefont {C.}~\bibnamefont
  {Baldassi}}, \bibinfo {author} {\bibfnamefont {C.}~\bibnamefont {Borgs}},
  \bibinfo {author} {\bibfnamefont {J.~T.}\ \bibnamefont {Chayes}}, \bibinfo
  {author} {\bibfnamefont {A.}~\bibnamefont {Ingrosso}}, \bibinfo {author}
  {\bibfnamefont {C.}~\bibnamefont {Lucibello}}, \bibinfo {author}
  {\bibfnamefont {L.}~\bibnamefont {Saglietti}}, \ and\ \bibinfo {author}
  {\bibfnamefont {R.}~\bibnamefont {Zecchina}},\ }\href {\doibase
  10.1073/pnas.1608103113} {\bibfield  {journal} {\bibinfo  {journal}
  {Proceedings of the National Academy of Sciences}\ }\textbf {\bibinfo
  {volume} {113}},\ \bibinfo {pages} {E7655} (\bibinfo {year}
  {2016})}\BibitemShut {NoStop}%
\bibitem [{\citenamefont {Baldassi}\ \emph {et~al.}(2019)\citenamefont
  {Baldassi}, \citenamefont {Malatesta},\ and\ \citenamefont
  {Zecchina}}]{relu_locent}%
  \BibitemOpen
  \bibfield  {author} {\bibinfo {author} {\bibfnamefont {C.}~\bibnamefont
  {Baldassi}}, \bibinfo {author} {\bibfnamefont {E.~M.}\ \bibnamefont
  {Malatesta}}, \ and\ \bibinfo {author} {\bibfnamefont {R.}~\bibnamefont
  {Zecchina}},\ }\href {\doibase 10.1103/PhysRevLett.123.170602} {\bibfield
  {journal} {\bibinfo  {journal} {Phys. Rev. Lett.}\ }\textbf {\bibinfo
  {volume} {123}},\ \bibinfo {pages} {170602} (\bibinfo {year}
  {2019})}\BibitemShut {NoStop}%
\bibitem [{\citenamefont {Baldassi}\ \emph
  {et~al.}(2020{\natexlab{a}})\citenamefont {Baldassi}, \citenamefont
  {Pittorino},\ and\ \citenamefont {Zecchina}}]{baldassi2020shaping}%
  \BibitemOpen
  \bibfield  {author} {\bibinfo {author} {\bibfnamefont {C.}~\bibnamefont
  {Baldassi}}, \bibinfo {author} {\bibfnamefont {F.}~\bibnamefont {Pittorino}},
  \ and\ \bibinfo {author} {\bibfnamefont {R.}~\bibnamefont {Zecchina}},\
  }\href {\doibase 10.1073/pnas.1908636117} {\bibfield  {journal} {\bibinfo
  {journal} {Proceedings of the National Academy of Sciences}\ }\textbf
  {\bibinfo {volume} {117}},\ \bibinfo {pages} {161} (\bibinfo {year}
  {2020}{\natexlab{a}})}\BibitemShut {NoStop}%
\bibitem [{\citenamefont {Baldassi}\ \emph
  {et~al.}(2020{\natexlab{b}})\citenamefont {Baldassi}, \citenamefont
  {Malatesta}, \citenamefont {Negri},\ and\ \citenamefont
  {Zecchina}}]{baldassi2020wide}%
  \BibitemOpen
  \bibfield  {author} {\bibinfo {author} {\bibfnamefont {C.}~\bibnamefont
  {Baldassi}}, \bibinfo {author} {\bibfnamefont {E.~M.}\ \bibnamefont
  {Malatesta}}, \bibinfo {author} {\bibfnamefont {M.}~\bibnamefont {Negri}}, \
  and\ \bibinfo {author} {\bibfnamefont {R.}~\bibnamefont {Zecchina}},\ }\href
  {\doibase 10.1088/1742-5468/abcd31} {\bibfield  {journal} {\bibinfo
  {journal} {Journal of Statistical Mechanics: Theory and Experiment}\ }\textbf
  {\bibinfo {volume} {2020}},\ \bibinfo {pages} {124012} (\bibinfo {year}
  {2020}{\natexlab{b}})}\BibitemShut {NoStop}%
\bibitem [{\citenamefont {Baldassi}\ \emph {et~al.}(2022)\citenamefont
  {Baldassi}, \citenamefont {Lauditi}, \citenamefont {Malatesta}, \citenamefont
  {Pacelli}, \citenamefont {Perugini},\ and\ \citenamefont
  {Zecchina}}]{baldassi2021learning}%
  \BibitemOpen
  \bibfield  {author} {\bibinfo {author} {\bibfnamefont {C.}~\bibnamefont
  {Baldassi}}, \bibinfo {author} {\bibfnamefont {C.}~\bibnamefont {Lauditi}},
  \bibinfo {author} {\bibfnamefont {E.~M.}\ \bibnamefont {Malatesta}}, \bibinfo
  {author} {\bibfnamefont {R.}~\bibnamefont {Pacelli}}, \bibinfo {author}
  {\bibfnamefont {G.}~\bibnamefont {Perugini}}, \ and\ \bibinfo {author}
  {\bibfnamefont {R.}~\bibnamefont {Zecchina}},\ }\href {\doibase
  10.1103/PhysRevE.106.014116} {\bibfield  {journal} {\bibinfo  {journal}
  {Phys. Rev. E}\ }\textbf {\bibinfo {volume} {106}},\ \bibinfo {pages}
  {014116} (\bibinfo {year} {2022})}\BibitemShut {NoStop}%
\bibitem [{\citenamefont {Sagun}\ \emph {et~al.}(2016)\citenamefont {Sagun},
  \citenamefont {Bottou},\ and\ \citenamefont {LeCun}}]{sagun2016eigenvalues}%
  \BibitemOpen
  \bibfield  {author} {\bibinfo {author} {\bibfnamefont {L.}~\bibnamefont
  {Sagun}}, \bibinfo {author} {\bibfnamefont {L.}~\bibnamefont {Bottou}}, \
  and\ \bibinfo {author} {\bibfnamefont {Y.}~\bibnamefont {LeCun}},\
  }\href@noop {} {\bibfield  {journal} {\bibinfo  {journal} {arXiv preprint
  arXiv:1611.07476}\ } (\bibinfo {year} {2016})}\BibitemShut {NoStop}%
\bibitem [{\citenamefont {Sagun}\ \emph {et~al.}(2017)\citenamefont {Sagun},
  \citenamefont {Evci}, \citenamefont {Guney}, \citenamefont {Dauphin},\ and\
  \citenamefont {Bottou}}]{sagun2017empirical}%
  \BibitemOpen
  \bibfield  {author} {\bibinfo {author} {\bibfnamefont {L.}~\bibnamefont
  {Sagun}}, \bibinfo {author} {\bibfnamefont {U.}~\bibnamefont {Evci}},
  \bibinfo {author} {\bibfnamefont {V.~U.}\ \bibnamefont {Guney}}, \bibinfo
  {author} {\bibfnamefont {Y.}~\bibnamefont {Dauphin}}, \ and\ \bibinfo
  {author} {\bibfnamefont {L.}~\bibnamefont {Bottou}},\ }\href@noop {}
  {\bibfield  {journal} {\bibinfo  {journal} {arXiv preprint arXiv:1706.04454}\
  } (\bibinfo {year} {2017})}\BibitemShut {NoStop}%
\bibitem [{\citenamefont {Pittorino}\ \emph {et~al.}(2021)\citenamefont
  {Pittorino}, \citenamefont {Lucibello}, \citenamefont {Feinauer},
  \citenamefont {Perugini}, \citenamefont {Baldassi}, \citenamefont
  {Demyanenko},\ and\ \citenamefont {Zecchina}}]{pittorino2021}%
  \BibitemOpen
  \bibfield  {author} {\bibinfo {author} {\bibfnamefont {F.}~\bibnamefont
  {Pittorino}}, \bibinfo {author} {\bibfnamefont {C.}~\bibnamefont
  {Lucibello}}, \bibinfo {author} {\bibfnamefont {C.}~\bibnamefont {Feinauer}},
  \bibinfo {author} {\bibfnamefont {G.}~\bibnamefont {Perugini}}, \bibinfo
  {author} {\bibfnamefont {C.}~\bibnamefont {Baldassi}}, \bibinfo {author}
  {\bibfnamefont {E.}~\bibnamefont {Demyanenko}}, \ and\ \bibinfo {author}
  {\bibfnamefont {R.}~\bibnamefont {Zecchina}},\ }in\ \href
  {https://openreview.net/forum?id=xjXg0bnoDmS} {\emph {\bibinfo {booktitle}
  {International Conference on Learning Representations}}}\ (\bibinfo {year}
  {2021})\BibitemShut {NoStop}%
\bibitem [{\citenamefont {Foret}\ \emph {et~al.}(2021)\citenamefont {Foret},
  \citenamefont {Kleiner}, \citenamefont {Mobahi},\ and\ \citenamefont
  {Neyshabur}}]{foret2021sharpnessaware}%
  \BibitemOpen
  \bibfield  {author} {\bibinfo {author} {\bibfnamefont {P.}~\bibnamefont
  {Foret}}, \bibinfo {author} {\bibfnamefont {A.}~\bibnamefont {Kleiner}},
  \bibinfo {author} {\bibfnamefont {H.}~\bibnamefont {Mobahi}}, \ and\ \bibinfo
  {author} {\bibfnamefont {B.}~\bibnamefont {Neyshabur}},\ }in\ \href
  {https://openreview.net/forum?id=6Tm1mposlrM} {\emph {\bibinfo {booktitle}
  {International Conference on Learning Representations}}}\ (\bibinfo {year}
  {2021})\BibitemShut {NoStop}%
\bibitem [{\citenamefont {Feng}\ and\ \citenamefont {Tu}(2021)}]{FengTu2021}%
  \BibitemOpen
  \bibfield  {author} {\bibinfo {author} {\bibfnamefont {Y.}~\bibnamefont
  {Feng}}\ and\ \bibinfo {author} {\bibfnamefont {Y.}~\bibnamefont {Tu}},\
  }\href {\doibase 10.1073/pnas.2015617118} {\bibfield  {journal} {\bibinfo
  {journal} {Proceedings of the National Academy of Sciences}\ }\textbf
  {\bibinfo {volume} {118}} (\bibinfo {year} {2021}),\
  10.1073/pnas.2015617118},\ \Eprint
  {http://arxiv.org/abs/https://www.pnas.org/content/118/9/e2015617118.full.pdf}
  {https://www.pnas.org/content/118/9/e2015617118.full.pdf} \BibitemShut
  {NoStop}%
\bibitem [{\citenamefont {Chen}\ \emph {et~al.}(2022)\citenamefont {Chen},
  \citenamefont {Qu},\ and\ \citenamefont {Gong}}]{Chen2022}%
  \BibitemOpen
  \bibfield  {author} {\bibinfo {author} {\bibfnamefont {G.}~\bibnamefont
  {Chen}}, \bibinfo {author} {\bibfnamefont {C.~K.}\ \bibnamefont {Qu}}, \ and\
  \bibinfo {author} {\bibfnamefont {P.}~\bibnamefont {Gong}},\ }\href {\doibase
  https://doi.org/10.1016/j.neunet.2022.01.019} {\bibfield  {journal} {\bibinfo
   {journal} {Neural Networks}\ }\textbf {\bibinfo {volume} {149}},\ \bibinfo
  {pages} {18} (\bibinfo {year} {2022})}\BibitemShut {NoStop}%
\bibitem [{\citenamefont {Kunin}\ \emph {et~al.}(2021)\citenamefont {Kunin},
  \citenamefont {Sagastuy-Brena}, \citenamefont {Gillespie}, \citenamefont
  {Margalit}, \citenamefont {Tanaka}, \citenamefont {Ganguli},\ and\
  \citenamefont {Yamins}}]{kunin2021rethinking}%
  \BibitemOpen
  \bibfield  {author} {\bibinfo {author} {\bibfnamefont {D.}~\bibnamefont
  {Kunin}}, \bibinfo {author} {\bibfnamefont {J.}~\bibnamefont
  {Sagastuy-Brena}}, \bibinfo {author} {\bibfnamefont {L.}~\bibnamefont
  {Gillespie}}, \bibinfo {author} {\bibfnamefont {E.}~\bibnamefont {Margalit}},
  \bibinfo {author} {\bibfnamefont {H.}~\bibnamefont {Tanaka}}, \bibinfo
  {author} {\bibfnamefont {S.}~\bibnamefont {Ganguli}}, \ and\ \bibinfo
  {author} {\bibfnamefont {D.~L.}\ \bibnamefont {Yamins}},\ }\href@noop {} {\
  (\bibinfo {year} {2021})}\BibitemShut {NoStop}%
\bibitem [{\citenamefont {Draxler}\ \emph {et~al.}(2018)\citenamefont
  {Draxler}, \citenamefont {Veschgini}, \citenamefont {Salmhofer},\ and\
  \citenamefont {Hamprecht}}]{draxler2018}%
  \BibitemOpen
  \bibfield  {author} {\bibinfo {author} {\bibfnamefont {F.}~\bibnamefont
  {Draxler}}, \bibinfo {author} {\bibfnamefont {K.}~\bibnamefont {Veschgini}},
  \bibinfo {author} {\bibfnamefont {M.}~\bibnamefont {Salmhofer}}, \ and\
  \bibinfo {author} {\bibfnamefont {F.}~\bibnamefont {Hamprecht}},\ }in\ \href
  {http://proceedings.mlr.press/v80/draxler18a.html} {\emph {\bibinfo
  {booktitle} {Proceedings of the 35th International Conference on Machine
  Learning}}},\ \bibinfo {series} {Proceedings of Machine Learning Research},
  Vol.~\bibinfo {volume} {80},\ \bibinfo {editor} {edited by\ \bibinfo {editor}
  {\bibfnamefont {J.}~\bibnamefont {Dy}}\ and\ \bibinfo {editor} {\bibfnamefont
  {A.}~\bibnamefont {Krause}}}\ (\bibinfo  {publisher} {PMLR},\ \bibinfo {year}
  {2018})\ pp.\ \bibinfo {pages} {1309--1318}\BibitemShut {NoStop}%
\bibitem [{\citenamefont {Garipov}\ \emph {et~al.}(2018)\citenamefont
  {Garipov}, \citenamefont {Izmailov}, \citenamefont {Podoprikhin},
  \citenamefont {Vetrov},\ and\ \citenamefont {Wilson}}]{garipov2018}%
  \BibitemOpen
  \bibfield  {author} {\bibinfo {author} {\bibfnamefont {T.}~\bibnamefont
  {Garipov}}, \bibinfo {author} {\bibfnamefont {P.}~\bibnamefont {Izmailov}},
  \bibinfo {author} {\bibfnamefont {D.}~\bibnamefont {Podoprikhin}}, \bibinfo
  {author} {\bibfnamefont {D.~P.}\ \bibnamefont {Vetrov}}, \ and\ \bibinfo
  {author} {\bibfnamefont {A.~G.}\ \bibnamefont {Wilson}},\ }in\ \href
  {https://proceedings.neurips.cc/paper/2018/file/be3087e74e9100d4bc4c6268cdbe8456-Paper.pdf}
  {\emph {\bibinfo {booktitle} {Advances in Neural Information Processing
  Systems}}},\ Vol.~\bibinfo {volume} {31},\ \bibinfo {editor} {edited by\
  \bibinfo {editor} {\bibfnamefont {S.}~\bibnamefont {Bengio}}, \bibinfo
  {editor} {\bibfnamefont {H.}~\bibnamefont {Wallach}}, \bibinfo {editor}
  {\bibfnamefont {H.}~\bibnamefont {Larochelle}}, \bibinfo {editor}
  {\bibfnamefont {K.}~\bibnamefont {Grauman}}, \bibinfo {editor} {\bibfnamefont
  {N.}~\bibnamefont {Cesa-Bianchi}}, \ and\ \bibinfo {editor} {\bibfnamefont
  {R.}~\bibnamefont {Garnett}}}\ (\bibinfo  {publisher} {Curran Associates,
  Inc.},\ \bibinfo {year} {2018})\BibitemShut {NoStop}%
\bibitem [{\citenamefont {Entezari}\ \emph {et~al.}(2022)\citenamefont
  {Entezari}, \citenamefont {Sedghi}, \citenamefont {Saukh},\ and\
  \citenamefont {Neyshabur}}]{entezari2022}%
  \BibitemOpen
  \bibfield  {author} {\bibinfo {author} {\bibfnamefont {R.}~\bibnamefont
  {Entezari}}, \bibinfo {author} {\bibfnamefont {H.}~\bibnamefont {Sedghi}},
  \bibinfo {author} {\bibfnamefont {O.}~\bibnamefont {Saukh}}, \ and\ \bibinfo
  {author} {\bibfnamefont {B.}~\bibnamefont {Neyshabur}},\ }in\ \href
  {https://openreview.net/forum?id=dNigytemkL} {\emph {\bibinfo {booktitle}
  {International Conference on Learning Representations}}}\ (\bibinfo {year}
  {2022})\BibitemShut {NoStop}%
\bibitem [{\citenamefont {Pittorino}\ \emph {et~al.}(2022)\citenamefont
  {Pittorino}, \citenamefont {Ferraro}, \citenamefont {Perugini}, \citenamefont
  {Feinauer}, \citenamefont {Baldassi},\ and\ \citenamefont
  {Zecchina}}]{pittorino22}%
  \BibitemOpen
  \bibfield  {author} {\bibinfo {author} {\bibfnamefont {F.}~\bibnamefont
  {Pittorino}}, \bibinfo {author} {\bibfnamefont {A.}~\bibnamefont {Ferraro}},
  \bibinfo {author} {\bibfnamefont {G.}~\bibnamefont {Perugini}}, \bibinfo
  {author} {\bibfnamefont {C.}~\bibnamefont {Feinauer}}, \bibinfo {author}
  {\bibfnamefont {C.}~\bibnamefont {Baldassi}}, \ and\ \bibinfo {author}
  {\bibfnamefont {R.}~\bibnamefont {Zecchina}},\ }in\ \href
  {https://proceedings.mlr.press/v162/pittorino22a.html} {\emph {\bibinfo
  {booktitle} {Proceedings of the 39th International Conference on Machine
  Learning}}},\ \bibinfo {series} {Proceedings of Machine Learning Research},
  Vol.\ \bibinfo {volume} {162},\ \bibinfo {editor} {edited by\ \bibinfo
  {editor} {\bibfnamefont {K.}~\bibnamefont {Chaudhuri}}, \bibinfo {editor}
  {\bibfnamefont {S.}~\bibnamefont {Jegelka}}, \bibinfo {editor} {\bibfnamefont
  {L.}~\bibnamefont {Song}}, \bibinfo {editor} {\bibfnamefont {C.}~\bibnamefont
  {Szepesvari}}, \bibinfo {editor} {\bibfnamefont {G.}~\bibnamefont {Niu}}, \
  and\ \bibinfo {editor} {\bibfnamefont {S.}~\bibnamefont {Sabato}}}\ (\bibinfo
   {publisher} {PMLR},\ \bibinfo {year} {2022})\ pp.\ \bibinfo {pages}
  {17759--17781}\BibitemShut {NoStop}%
\bibitem [{\citenamefont {Jordan}\ \emph {et~al.}(2023)\citenamefont {Jordan},
  \citenamefont {Sedghi}, \citenamefont {Saukh}, \citenamefont {Entezari},\
  and\ \citenamefont {Neyshabur}}]{jordan2023repair}%
  \BibitemOpen
  \bibfield  {author} {\bibinfo {author} {\bibfnamefont {K.}~\bibnamefont
  {Jordan}}, \bibinfo {author} {\bibfnamefont {H.}~\bibnamefont {Sedghi}},
  \bibinfo {author} {\bibfnamefont {O.}~\bibnamefont {Saukh}}, \bibinfo
  {author} {\bibfnamefont {R.}~\bibnamefont {Entezari}}, \ and\ \bibinfo
  {author} {\bibfnamefont {B.}~\bibnamefont {Neyshabur}},\ }in\ \href
  {https://openreview.net/forum?id=gU5sJ6ZggcX} {\emph {\bibinfo {booktitle}
  {The Eleventh International Conference on Learning Representations}}}\
  (\bibinfo {year} {2023})\BibitemShut {NoStop}%
\bibitem [{\citenamefont {Frankle}\ and\ \citenamefont
  {Carbin}(2019)}]{frankle2018}%
  \BibitemOpen
  \bibfield  {author} {\bibinfo {author} {\bibfnamefont {J.}~\bibnamefont
  {Frankle}}\ and\ \bibinfo {author} {\bibfnamefont {M.}~\bibnamefont
  {Carbin}},\ }in\ \href {https://openreview.net/forum?id=rJl-b3RcF7} {\emph
  {\bibinfo {booktitle} {International Conference on Learning
  Representations}}}\ (\bibinfo {year} {2019})\BibitemShut {NoStop}%
\bibitem [{\citenamefont {Frankle}\ \emph {et~al.}(2020)\citenamefont
  {Frankle}, \citenamefont {Dziugaite}, \citenamefont {Roy},\ and\
  \citenamefont {Carbin}}]{frankle20}%
  \BibitemOpen
  \bibfield  {author} {\bibinfo {author} {\bibfnamefont {J.}~\bibnamefont
  {Frankle}}, \bibinfo {author} {\bibfnamefont {G.~K.}\ \bibnamefont
  {Dziugaite}}, \bibinfo {author} {\bibfnamefont {D.}~\bibnamefont {Roy}}, \
  and\ \bibinfo {author} {\bibfnamefont {M.}~\bibnamefont {Carbin}},\ }in\
  \href {https://proceedings.mlr.press/v119/frankle20a.html} {\emph {\bibinfo
  {booktitle} {Proceedings of the 37th International Conference on Machine
  Learning}}},\ \bibinfo {series} {Proceedings of Machine Learning Research},
  Vol.\ \bibinfo {volume} {119},\ \bibinfo {editor} {edited by\ \bibinfo
  {editor} {\bibfnamefont {H.~D.}\ \bibnamefont {III}}\ and\ \bibinfo {editor}
  {\bibfnamefont {A.}~\bibnamefont {Singh}}}\ (\bibinfo  {publisher} {PMLR},\
  \bibinfo {year} {2020})\ pp.\ \bibinfo {pages} {3259--3269}\BibitemShut
  {NoStop}%
\bibitem [{\citenamefont {Mirzadeh}\ \emph {et~al.}(2021)\citenamefont
  {Mirzadeh}, \citenamefont {Farajtabar}, \citenamefont {Gorur}, \citenamefont
  {Pascanu},\ and\ \citenamefont {Ghasemzadeh}}]{mirzadeh2021linear}%
  \BibitemOpen
  \bibfield  {author} {\bibinfo {author} {\bibfnamefont {S.~I.}\ \bibnamefont
  {Mirzadeh}}, \bibinfo {author} {\bibfnamefont {M.}~\bibnamefont
  {Farajtabar}}, \bibinfo {author} {\bibfnamefont {D.}~\bibnamefont {Gorur}},
  \bibinfo {author} {\bibfnamefont {R.}~\bibnamefont {Pascanu}}, \ and\
  \bibinfo {author} {\bibfnamefont {H.}~\bibnamefont {Ghasemzadeh}},\ }in\
  \href {https://openreview.net/forum?id=Fmg_fQYUejf} {\emph {\bibinfo
  {booktitle} {International Conference on Learning Representations}}}\
  (\bibinfo {year} {2021})\BibitemShut {NoStop}%
\bibitem [{\citenamefont {Benton}\ \emph {et~al.}(2021)\citenamefont {Benton},
  \citenamefont {Maddox}, \citenamefont {Lotfi},\ and\ \citenamefont
  {Wilson}}]{FastEnsembling}%
  \BibitemOpen
  \bibfield  {author} {\bibinfo {author} {\bibfnamefont {G.}~\bibnamefont
  {Benton}}, \bibinfo {author} {\bibfnamefont {W.}~\bibnamefont {Maddox}},
  \bibinfo {author} {\bibfnamefont {S.}~\bibnamefont {Lotfi}}, \ and\ \bibinfo
  {author} {\bibfnamefont {A.~G.~G.}\ \bibnamefont {Wilson}},\ }in\ \href
  {https://proceedings.mlr.press/v139/benton21a.html} {\emph {\bibinfo
  {booktitle} {Proceedings of the 38th International Conference on Machine
  Learning}}},\ \bibinfo {series} {Proceedings of Machine Learning Research},
  Vol.\ \bibinfo {volume} {139},\ \bibinfo {editor} {edited by\ \bibinfo
  {editor} {\bibfnamefont {M.}~\bibnamefont {Meila}}\ and\ \bibinfo {editor}
  {\bibfnamefont {T.}~\bibnamefont {Zhang}}}\ (\bibinfo  {publisher} {PMLR},\
  \bibinfo {year} {2021})\ pp.\ \bibinfo {pages} {769--779}\BibitemShut
  {NoStop}%
\bibitem [{\citenamefont {Gardner}\ and\ \citenamefont
  {Derrida}(1988)}]{gardner1988optimal}%
  \BibitemOpen
  \bibfield  {author} {\bibinfo {author} {\bibfnamefont {E.}~\bibnamefont
  {Gardner}}\ and\ \bibinfo {author} {\bibfnamefont {B.}~\bibnamefont
  {Derrida}},\ }\href {\doibase 10.1088/0305-4470/21/1/031} {\bibfield
  {journal} {\bibinfo  {journal} {Journal of Physics A: Mathematical and
  General}\ }\textbf {\bibinfo {volume} {21}},\ \bibinfo {pages} {271}
  (\bibinfo {year} {1988})}\BibitemShut {NoStop}%
\bibitem [{\citenamefont {Engel}\ and\ \citenamefont {Van~den
  Broeck}(2001)}]{engel-vandenbroek}%
  \BibitemOpen
  \bibfield  {author} {\bibinfo {author} {\bibfnamefont {A.}~\bibnamefont
  {Engel}}\ and\ \bibinfo {author} {\bibfnamefont {C.}~\bibnamefont {Van~den
  Broeck}},\ }\href@noop {} {\emph {\bibinfo {title} {Statistical mechanics of
  learning}}}\ (\bibinfo  {publisher} {Cambridge University Press},\ \bibinfo
  {year} {2001})\BibitemShut {NoStop}%
\bibitem [{\citenamefont {Franz}\ \emph {et~al.}(2017)\citenamefont {Franz},
  \citenamefont {Parisi}, \citenamefont {Sevelev}, \citenamefont {Urbani},\
  and\ \citenamefont {Zamponi}}]{scipost2017}%
  \BibitemOpen
  \bibfield  {author} {\bibinfo {author} {\bibfnamefont {S.}~\bibnamefont
  {Franz}}, \bibinfo {author} {\bibfnamefont {G.}~\bibnamefont {Parisi}},
  \bibinfo {author} {\bibfnamefont {M.}~\bibnamefont {Sevelev}}, \bibinfo
  {author} {\bibfnamefont {P.}~\bibnamefont {Urbani}}, \ and\ \bibinfo {author}
  {\bibfnamefont {F.}~\bibnamefont {Zamponi}},\ }\href {\doibase
  10.21468/SciPostPhys.2.3.019} {\bibfield  {journal} {\bibinfo  {journal}
  {SciPost Phys.}\ }\textbf {\bibinfo {volume} {2}},\ \bibinfo {pages} {019}
  (\bibinfo {year} {2017})}\BibitemShut {NoStop}%
\bibitem [{\citenamefont {Hansen}\ \emph {et~al.}(2020)\citenamefont {Hansen},
  \citenamefont {Herburt}, \citenamefont {Martini},\ and\ \citenamefont
  {Moszy{\'{n}}ska}}]{Hansen2020}%
  \BibitemOpen
  \bibfield  {author} {\bibinfo {author} {\bibfnamefont {G.}~\bibnamefont
  {Hansen}}, \bibinfo {author} {\bibfnamefont {I.}~\bibnamefont {Herburt}},
  \bibinfo {author} {\bibfnamefont {H.}~\bibnamefont {Martini}}, \ and\
  \bibinfo {author} {\bibfnamefont {M.}~\bibnamefont {Moszy{\'{n}}ska}},\
  }\href {\doibase 10.1007/s00010-020-00720-7} {\bibfield  {journal} {\bibinfo
  {journal} {Aequationes mathematicae}\ }\textbf {\bibinfo {volume} {94}},\
  \bibinfo {pages} {1001} (\bibinfo {year} {2020})}\BibitemShut {NoStop}%
\bibitem [{\citenamefont {Baldassi}\ \emph {et~al.}(2023)\citenamefont
  {Baldassi}, \citenamefont {Malatesta}, \citenamefont {Perugini},\ and\
  \citenamefont {Zecchina}}]{baldassi2023typical}%
  \BibitemOpen
  \bibfield  {author} {\bibinfo {author} {\bibfnamefont {C.}~\bibnamefont
  {Baldassi}}, \bibinfo {author} {\bibfnamefont {E.~M.}\ \bibnamefont
  {Malatesta}}, \bibinfo {author} {\bibfnamefont {G.}~\bibnamefont {Perugini}},
  \ and\ \bibinfo {author} {\bibfnamefont {R.}~\bibnamefont {Zecchina}},\
  }\href {\doibase 10.1103/PhysRevE.108.024310} {\bibfield  {journal} {\bibinfo
   {journal} {Phys. Rev. E}\ }\textbf {\bibinfo {volume} {108}},\ \bibinfo
  {pages} {024310} (\bibinfo {year} {2023})}\BibitemShut {NoStop}%
\bibitem [{\citenamefont {Franz}\ and\ \citenamefont
  {Parisi}(2016)}]{franz2016}%
  \BibitemOpen
  \bibfield  {author} {\bibinfo {author} {\bibfnamefont {S.}~\bibnamefont
  {Franz}}\ and\ \bibinfo {author} {\bibfnamefont {G.}~\bibnamefont {Parisi}},\
  }\href {\doibase 10.1088/1751-8113/49/14/145001} {\bibfield  {journal}
  {\bibinfo  {journal} {Journal of Physics A: Mathematical and Theoretical}\
  }\textbf {\bibinfo {volume} {49}},\ \bibinfo {pages} {145001} (\bibinfo
  {year} {2016})}\BibitemShut {NoStop}%
\bibitem [{\citenamefont {El~Alaoui}\ and\ \citenamefont
  {Sellke}(2022)}]{elAlaoui2022algorithmic}%
  \BibitemOpen
  \bibfield  {author} {\bibinfo {author} {\bibfnamefont {A.}~\bibnamefont
  {El~Alaoui}}\ and\ \bibinfo {author} {\bibfnamefont {M.}~\bibnamefont
  {Sellke}},\ }\href {\doibase 10.1007/s10955-022-02976-6} {\bibfield
  {journal} {\bibinfo  {journal} {Journal of Statistical Physics}\ }\textbf
  {\bibinfo {volume} {189}},\ \bibinfo {pages} {27} (\bibinfo {year}
  {2022})}\BibitemShut {NoStop}%
\bibitem [{\citenamefont {Montanari}\ \emph {et~al.}(2021)\citenamefont
  {Montanari}, \citenamefont {Zhong},\ and\ \citenamefont
  {Zhou}}]{montanari2021tractability}%
  \BibitemOpen
  \bibfield  {author} {\bibinfo {author} {\bibfnamefont {A.}~\bibnamefont
  {Montanari}}, \bibinfo {author} {\bibfnamefont {Y.}~\bibnamefont {Zhong}}, \
  and\ \bibinfo {author} {\bibfnamefont {K.}~\bibnamefont {Zhou}},\ }\href
  {https://arxiv.org/abs/2110.15824} {\bibfield  {journal} {\bibinfo  {journal}
  {arXiv preprint arXiv:2110.15824}\ } (\bibinfo {year} {2021})}\BibitemShut
  {NoStop}%
\bibitem [{\citenamefont {Baldassi}\ \emph {et~al.}(2021)\citenamefont
  {Baldassi}, \citenamefont {Lauditi}, \citenamefont {Malatesta}, \citenamefont
  {Perugini},\ and\ \citenamefont {Zecchina}}]{baldassi2021unveiling}%
  \BibitemOpen
  \bibfield  {author} {\bibinfo {author} {\bibfnamefont {C.}~\bibnamefont
  {Baldassi}}, \bibinfo {author} {\bibfnamefont {C.}~\bibnamefont {Lauditi}},
  \bibinfo {author} {\bibfnamefont {E.~M.}\ \bibnamefont {Malatesta}}, \bibinfo
  {author} {\bibfnamefont {G.}~\bibnamefont {Perugini}}, \ and\ \bibinfo
  {author} {\bibfnamefont {R.}~\bibnamefont {Zecchina}},\ }\href {\doibase
  10.1103/PhysRevLett.127.278301} {\bibfield  {journal} {\bibinfo  {journal}
  {Phys. Rev. Lett.}\ }\textbf {\bibinfo {volume} {127}},\ \bibinfo {pages}
  {278301} (\bibinfo {year} {2021})}\BibitemShut {NoStop}%
\bibitem [{\citenamefont {Huang}\ \emph {et~al.}(2013)\citenamefont {Huang},
  \citenamefont {Wong},\ and\ \citenamefont {Kabashima}}]{Huang_pairs}%
  \BibitemOpen
  \bibfield  {author} {\bibinfo {author} {\bibfnamefont {H.}~\bibnamefont
  {Huang}}, \bibinfo {author} {\bibfnamefont {K.~Y.~M.}\ \bibnamefont {Wong}},
  \ and\ \bibinfo {author} {\bibfnamefont {Y.}~\bibnamefont {Kabashima}},\
  }\href {\doibase 10.1088/1751-8113/46/37/375002} {\bibfield  {journal}
  {\bibinfo  {journal} {Journal of Physics A: Mathematical and Theoretical}\
  }\textbf {\bibinfo {volume} {46}},\ \bibinfo {pages} {375002} (\bibinfo
  {year} {2013})}\BibitemShut {NoStop}%
\bibitem [{\citenamefont {Zhang}\ and\ \citenamefont
  {Strogatz}(2021)}]{ZhangStrogatz2021}%
  \BibitemOpen
  \bibfield  {author} {\bibinfo {author} {\bibfnamefont {Y.}~\bibnamefont
  {Zhang}}\ and\ \bibinfo {author} {\bibfnamefont {S.~H.}\ \bibnamefont
  {Strogatz}},\ }\href {\doibase 10.1103/PhysRevLett.127.194101} {\bibfield
  {journal} {\bibinfo  {journal} {Phys. Rev. Lett.}\ }\textbf {\bibinfo
  {volume} {127}},\ \bibinfo {pages} {194101} (\bibinfo {year}
  {2021})}\BibitemShut {NoStop}%
\bibitem [{\citenamefont {Martiniani}\ and\ \citenamefont
  {Casiulis}(2023)}]{Martiniani2023}%
  \BibitemOpen
  \bibfield  {author} {\bibinfo {author} {\bibfnamefont {S.}~\bibnamefont
  {Martiniani}}\ and\ \bibinfo {author} {\bibfnamefont {M.}~\bibnamefont
  {Casiulis}},\ }\href {\doibase 10.4279/pip.150001} {\bibfield  {journal}
  {\bibinfo  {journal} {Papers in Physics}\ }\textbf {\bibinfo {volume} {15}},\
  \bibinfo {pages} {150001} (\bibinfo {year} {2023})}\BibitemShut {NoStop}%
\bibitem [{\citenamefont {Franz}\ \emph {et~al.}(2019)\citenamefont {Franz},
  \citenamefont {Hwang},\ and\ \citenamefont {Urbani}}]{multilayerjamming}%
  \BibitemOpen
  \bibfield  {author} {\bibinfo {author} {\bibfnamefont {S.}~\bibnamefont
  {Franz}}, \bibinfo {author} {\bibfnamefont {S.}~\bibnamefont {Hwang}}, \ and\
  \bibinfo {author} {\bibfnamefont {P.}~\bibnamefont {Urbani}},\ }\href
  {\doibase 10.1103/PhysRevLett.123.160602} {\bibfield  {journal} {\bibinfo
  {journal} {Phys. Rev. Lett.}\ }\textbf {\bibinfo {volume} {123}},\ \bibinfo
  {pages} {160602} (\bibinfo {year} {2019})}\BibitemShut {NoStop}%
\bibitem [{\citenamefont {Spigler}\ \emph {et~al.}(2019)\citenamefont
  {Spigler}, \citenamefont {Geiger}, \citenamefont {d~Ascoli}, \citenamefont
  {Sagun}, \citenamefont {Biroli},\ and\ \citenamefont {Wyart}}]{spigler2019}%
  \BibitemOpen
  \bibfield  {author} {\bibinfo {author} {\bibfnamefont {S.}~\bibnamefont
  {Spigler}}, \bibinfo {author} {\bibfnamefont {M.}~\bibnamefont {Geiger}},
  \bibinfo {author} {\bibfnamefont {S.}~\bibnamefont {d~Ascoli}}, \bibinfo
  {author} {\bibfnamefont {L.}~\bibnamefont {Sagun}}, \bibinfo {author}
  {\bibfnamefont {G.}~\bibnamefont {Biroli}}, \ and\ \bibinfo {author}
  {\bibfnamefont {M.}~\bibnamefont {Wyart}},\ }\href {\doibase
  10.1088/1751-8121/ab4c8b} {\bibfield  {journal} {\bibinfo  {journal} {Journal
  of Physics A: Mathematical and Theoretical}\ }\textbf {\bibinfo {volume}
  {52}},\ \bibinfo {pages} {474001} (\bibinfo {year} {2019})}\BibitemShut
  {NoStop}%
\bibitem [{\citenamefont {Jacot}\ \emph {et~al.}(2022)\citenamefont {Jacot},
  \citenamefont {Ged}, \citenamefont {Simsek}, \citenamefont {Hongler},\ and\
  \citenamefont {Gabriel}}]{jacot2022}%
  \BibitemOpen
  \bibfield  {author} {\bibinfo {author} {\bibfnamefont {A.}~\bibnamefont
  {Jacot}}, \bibinfo {author} {\bibfnamefont {F.}~\bibnamefont {Ged}}, \bibinfo
  {author} {\bibfnamefont {B.}~\bibnamefont {Simsek}}, \bibinfo {author}
  {\bibfnamefont {C.}~\bibnamefont {Hongler}}, \ and\ \bibinfo {author}
  {\bibfnamefont {F.}~\bibnamefont {Gabriel}},\ }\href@noop {} {\enquote
  {\bibinfo {title} {Saddle-to-saddle dynamics in deep linear networks: Small
  initialization training, symmetry, and sparsity},}\ } (\bibinfo {year}
  {2022}),\ \Eprint {http://arxiv.org/abs/2106.15933} {arXiv:2106.15933
  [stat.ML]} \BibitemShut {NoStop}%
\bibitem [{\citenamefont {Parisi}\ and\ \citenamefont
  {Zamponi}(2010)}]{Parisi_2010}%
  \BibitemOpen
  \bibfield  {author} {\bibinfo {author} {\bibfnamefont {G.}~\bibnamefont
  {Parisi}}\ and\ \bibinfo {author} {\bibfnamefont {F.}~\bibnamefont
  {Zamponi}},\ }\href {\doibase 10.1103/revmodphys.82.789} {\bibfield
  {journal} {\bibinfo  {journal} {Reviews of Modern Physics}\ }\textbf
  {\bibinfo {volume} {82}},\ \bibinfo {pages} {789} (\bibinfo {year}
  {2010})}\BibitemShut {NoStop}%
\bibitem [{\citenamefont {Berthier}\ and\ \citenamefont
  {Biroli}(2011)}]{Berthier_2011}%
  \BibitemOpen
  \bibfield  {author} {\bibinfo {author} {\bibfnamefont {L.}~\bibnamefont
  {Berthier}}\ and\ \bibinfo {author} {\bibfnamefont {G.}~\bibnamefont
  {Biroli}},\ }\href {\doibase 10.1103/revmodphys.83.587} {\bibfield  {journal}
  {\bibinfo  {journal} {Reviews of Modern Physics}\ }\textbf {\bibinfo {volume}
  {83}},\ \bibinfo {pages} {587} (\bibinfo {year} {2011})}\BibitemShut
  {NoStop}%
\bibitem [{\citenamefont {Kirkpatrick}\ and\ \citenamefont
  {Thirumalai}(2015)}]{Kirkpatrick2015}%
  \BibitemOpen
  \bibfield  {author} {\bibinfo {author} {\bibfnamefont {T.~R.}\ \bibnamefont
  {Kirkpatrick}}\ and\ \bibinfo {author} {\bibfnamefont {D.}~\bibnamefont
  {Thirumalai}},\ }\href {\doibase 10.1103/RevModPhys.87.183} {\bibfield
  {journal} {\bibinfo  {journal} {Rev. Mod. Phys.}\ }\textbf {\bibinfo {volume}
  {87}},\ \bibinfo {pages} {183} (\bibinfo {year} {2015})}\BibitemShut
  {NoStop}%
\end{thebibliography}%

		\clearpage

	\title{The star-shaped set of solutions of the spherical negative perceptron \\ SUPPLEMENTAL MATERIAL}
	
	\maketitle
	
	\onecolumngrid 
	
	\tableofcontents{}

\section{Typical phase diagram of the negative perceptron}\label{sec1}

We denote by $\boldsymbol{W}$ the $N$-dimensional weight vector of the perceptron, whose output on a certain input $\boldsymbol{\xi}$ is given by $\sigma=\text{sign}\Big(\frac{\boldsymbol{W}\cdot\boldsymbol{\xi}}{\sqrt{N}}\Big)$. We consider a dataset composed of $P=\alpha N$ input-output pairs $(\boldsymbol{\xi}^{\mu},\sigma^{\mu})_{\mu=1}^{P}$, where both $\sigma^{\mu}$ and each component of $\boldsymbol{\xi}^{\mu}$ are i.i.d. Rademacher random variables. We are interested in characterizing the space of solutions with a certain margin $\kp$, that is the set of $\boldsymbol{W}$ such that $\forall\mu\in\{1,...,P\}$, $\sigma^{\mu}\frac{\boldsymbol{W}\cdot \boldsymbol{\xi}^{\mu}}{\sqrt{N}}-\kp>0$ and that are normalized on the $N$-dimensional sphere of radius $\sqrt{N}$, i.e. $\lVert \boldsymbol{W} \rVert^2 = N$. For $\kp<0$, this is known to be an example of a non-convex Constraint Satisfaction Problem (CSP), where $\alpha=\frac{P}{N}$ is the constraint density. 
Notice that we can safely impose from now on $\sigma^\mu = 1$ for every $\mu = 1, \ldots, P$  without loss of generality, since one can always gauge-transform $\xi^\mu_i \to \sigma^\mu \,\xi^\mu_i$ without affecting the probability measure over the patterns.

\begin{figure}[h]
	\begin{centering}
		\includegraphics[width=0.70\columnwidth]{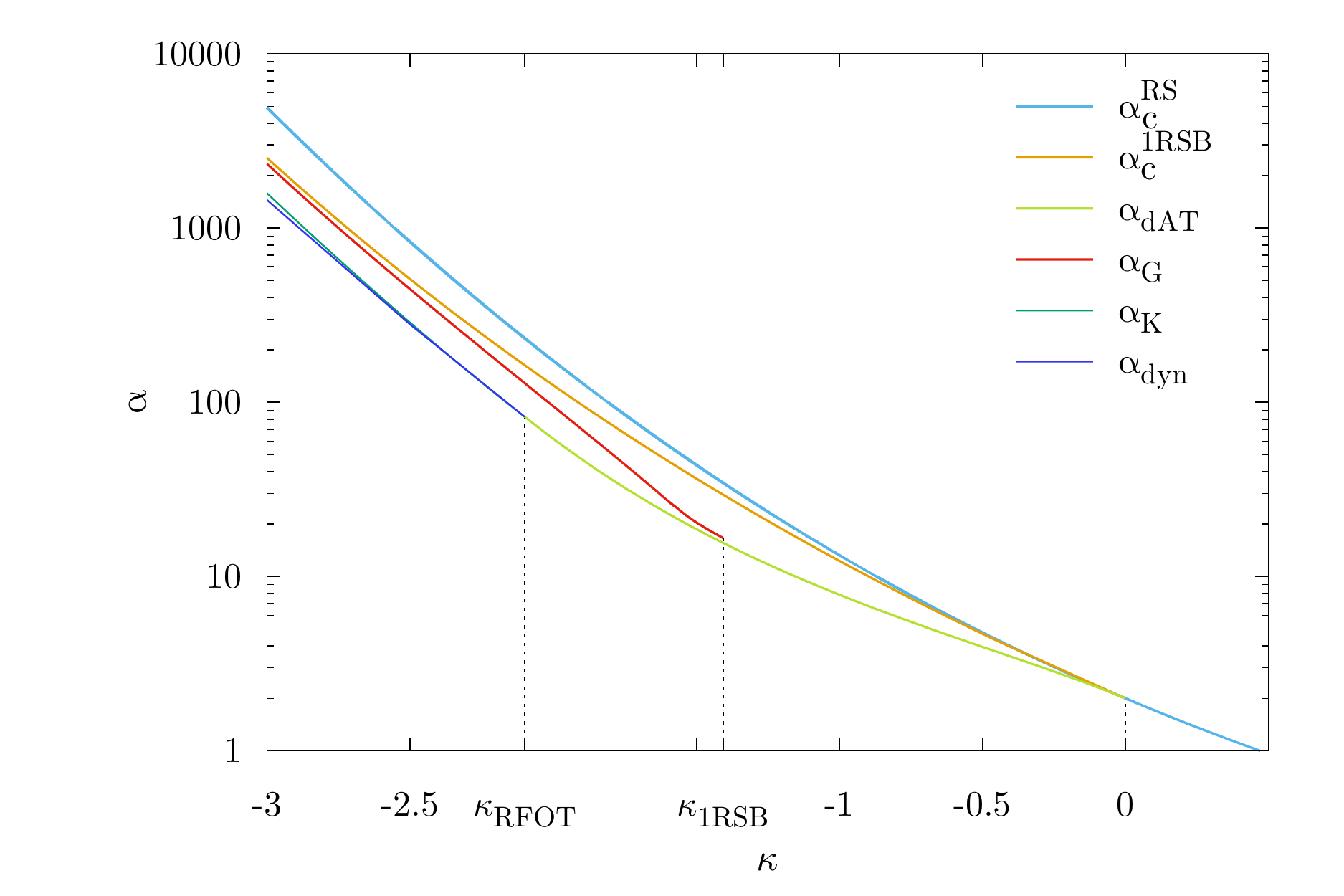}
	\end{centering}
	\caption{\label{fig:phasediagram}The zero temperature phase diagram of the continuous spherical perceptron showing the density of constraints $\alpha$ as a function of the margin $\kp $ by which the problem is solved. For each $\kp\leq 0$ the problem is non-convex and RSB transition effects appear. Reprinted from \cite{baldassi2023typical}.} 
\end{figure}

Here we briefly recall the phase diagram $\alpha(\kp)$ of the model specialized to the zero temperature limit and obtained via the replica method in the thermodynamic limit ($N,P\to\infty$ while keeping $\alpha=\mathcal{O}(1)$). We refer to a more detailed discussion to the works \cite{scipost2017, baldassi2023typical}. For any fixed value of $\ensuremath{\kp}$, by increasing the density of constraints, there exists a critical value $\alpha_{c}(\kp)$ that separates a satisfiable (SAT) phase where the problem admits an exponential number of solutions from an unsatisfiable (UNSAT) one where no solutions of the problem can be found. In the phase diagram reported in Fig.~\ref{fig:phasediagram}, predictions of the SAT/UNSAT transition are calculated in a Replica-Symmetric (RS) ansatz ($\alpha_{c}^{\text{RS}}$ curve) and by refining the analysis using a one-step Replica-Symmetric-Breaking (1RSB) approximation. Transitions coincide and are exact when the problem is convex for $\kp\geq0$ while they differ as soon as $\kp<0$. The \textit{de Almeida-Thouless} (dAT) transition line ($\alpha_{\text{dAT}}(\kp)$) is reported to identify the instability phase of the RS solution towards RSB effects. When one crosses $\alpha_{\text{dAT}}(\kp)$ for $\kp>\kappa_{\text{1RSB}}$, the dAT instability leads to a fullRSB phase, while for $\kappa_{\text{RFOT}}<\kp<\kappa_{\text{1RSB}}$ the dAT line marks the point where the system transitions from an RS phase to a stable 1RSB one. Finally, when $\kp< \kappa_{\text{RFOT}}$, three different transitions follow one another when increasing $\alpha$. Looking at the diagram from bottom to top, for $\alpha_{\text{dyn}} <\alpha < \alpha_{\text{K}}$ the system is in a dynamical 1RSB phase, where $q_1 > q_0$ and the Parisi block parameter is $m=1$, where the subscripts refer respectively to the \textit{dynamical} and \textit{Kautzmann} phase transitions at the basis of the mean field theory of glasses. For $\alpha_{\text{K}} < \alpha < \alpha_{\text{G}}$ the complexity of the clusters dominating the Gibbs-measure vanishes and the system is left only with 1RSB solutions having $m<1$. Upon further increasing $\alpha$, the 1RSB solution becomes unstable and undergoes a \textit{Gardner} transition towards a fullRSB phase. This sequence of transitions has also been observed in systems of hard-spheres in high dimension and are at the basis of the so called ``Random First Order Theory'' (RFOT)~\cite{Parisi_2010, Berthier_2011, Kirkpatrick2015}. 

\section{Typical overlap between pairs of solutions with different margin}
\subsection{Equilibrium configurations}
In order to study the geometrical properties of the solution space, we consider the free entropy landscape of solution-pairs solving the same classification problem of storing a number $P = \alpha N$ of input-output mappings as defined in Sec.~\ref{sec1}, but considering two independent solutions $\boldsymbol{W}^1$, $\boldsymbol{W}^2$ having two different margins respectively $k_{1}$ and $k_{2}$. 
\begin{equation}
	\phi(k_1, k_2) = \frac{1}{N}\mathbb{E}_{\boldsymbol{\xi}} \log Z_{\boldsymbol{\xi}}(k_1, k_2)
\end{equation}
with $Z_{\boldsymbol{\xi}}(k_1, k_2) = Z_{\boldsymbol{\xi}}(k_1) Z_{\boldsymbol{\xi}}(k_2)$, being
\begin{equation}
	Z_{\boldsymbol{\xi}}(k) \equiv \int d\mu(\boldsymbol{W}) \, \prod_{\mu = 1}^P \Theta\left(\frac{1}{\sqrt{N}} \boldsymbol{W} \cdot \boldsymbol{\xi}^\mu - k\right) \equiv \int d\mu(\boldsymbol{W}) \, \mathbb{X}_{\boldsymbol{\xi}}(\boldsymbol{W}; k)
\end{equation}
$d\mu(\boldsymbol{W})$ being the spherical measure over weights
\begin{equation}
	d\mu (\boldsymbol{W}) =  \prod_{i=1}^N dW_i \, \delta \left(\sum_{i=1}^N W_i^2 - N \right)
\end{equation}
Similarly to what has been done in~\cite{Huang_pairs}, this give us access to the typical overlap value $p = \left\langle \frac{\boldsymbol{W}^1 \cdot \boldsymbol{W}^2}{N} \right\rangle_{k_1, k_2}$ between the two solutions. Using the replica trick \cite{mezard1987spin}, the averaged free entropy can be written as
\begin{equation}
	\phi = \lim_{N \to \infty}  \frac{\langle \ln Z \rangle_{\boldsymbol{\xi}}}{N} = \lim_{N \to \infty}\lim_{n \to 0}  \frac{\langle  Z^n \rangle_{\boldsymbol{\xi}} -1}{n N}
\end{equation}
and the problem becomes that of computing the average of $n$ independent copies of the system with the same disorder realization of the patterns $\boldsymbol{\xi}^{\mu}$
\begin{equation}
	\langle Z^n \rangle = \mathbb{E}_{\boldsymbol{\xi}} \int \prod_{ar} d\mu (\boldsymbol{W}^{ar} ) \, \prod_{\mu a r} \Theta \left(\frac{1}{\sqrt{N}}\sum_i W^{ar}_i \xi^\mu_i -k_r \right) 
\end{equation}
where $r=1,2$ runs over the two solutions with margin $k_r$, $a=1,\ldots,n$ runs over the virtual replicas.

Then, we extract the quantity $\lambda^{\mu a}_r = \frac{1}{\sqrt{N}}\sum_i W^{ar}_i \xi^{\mu}_i$ by using the integral representation of the delta function and this allow us to perform the disorder average on the patterns in the limit of large $N$, obtaining
\begin{equation}
	\begin{split}
		\langle Z^n \rangle &= \mathbb{E}_{\boldsymbol{\xi}}\int \prod_{\mu a r} \frac{d\lambda^{\mu a }_r\,d\hat{\lambda}^{\mu a }_r}{2\pi} \, \int \prod_{ar} d\mu (\boldsymbol{W}^{ar}) \, \prod_{\mu a r} \Theta \left(\lambda^{\mu a}_r - k_r \right)\,e^{i\lambda^{\mu a}_r \hat{\lambda}^{\mu a}_r}\,\prod_{\mu i}  e^{-\frac{i}{\sqrt{N}}\xi^{\mu}_i \sum_{ar}W^{ar}_i \hat{\lambda}^{\mu a}_r}\\
		&= \int \prod_{\mu a r} \frac{d\lambda^{\mu a }_r\,d\hat{\lambda}^{\mu a }_r}{2\pi} \, \int \prod_{ar} d\mu (\boldsymbol{W}^{ar}) \, \prod_{\mu a r} \Theta \left(\lambda^{\mu a}_r - k_r \right)\,e^{i\lambda^{\mu a}_r \hat{\lambda}^{\mu a}_r} \prod_\mu e^{-\frac{1}{2N}\sum_{ar,bs}\left(\sum_i W^{ar}_i W^{bs}_i \right)\hat{\lambda}^{\mu a}_r \hat{\lambda}^{\mu b}_s}.
	\end{split}
\end{equation}
By defining appropriate order parameters 
\begin{equation} \label{eq::order_parameters}
	q^{ab}_{rs} = \frac{1}{N}\sum_i W^{ar}_i W^{bs}_i    
\end{equation}
we can conveniently rewrite the partition function as
\begin{equation}
	\langle Z^n \rangle = \int \prod_{a<b,rs} \frac{dq^{ab}_{rs}\,d\hat{q}^{ab}_{rs}}{2\pi}\int \prod_{ar} \frac{d\hat{h}^{a}_r}{2\pi}\,e^{Nn\phi(q, \hat{q}, \hat{h})}
\end{equation}
where we have defined 
\begin{subequations}\label{entropicandenergetic}
	\begin{align}
		&\phi(q, \hat{q}, \hat{h}) = -\frac{1}{2n}\sum_{a<b, rs}q^{ab}_{rs}\hat{q}^{ab}_{rs} +\frac{1}{n}\sum_{ar}\hat{h}^a_r + \frac{1}{n} G_S + \frac{\alpha}{n} \,G_E \\
		&G_S = \ln \int \prod_{ar} dW^{ar} \, e^{\frac{1}{2}\sum_{a<b, rs}W^{ar}W^{bs} \hat{q}^{ab}_{rs}-\sum_{ar}(W^{ar})^2 \hat{h}^a_r}\\
		&G_E = \ln \int \prod_{ar} \frac{d\lambda^a_r\,d\hat{\lambda}^a_r}{2\pi}\,\prod_{ar} e^{i\lambda^a_r \hat{\lambda}^a_r}\,\Theta\left(\lambda^a_r - k_r \right)\,e^{-\frac{1}{2}\sum_{a<b, rs}q^{ab}_{rs}\,\hat{\lambda}^a_r \hat{\lambda}^b_s}.
	\end{align}
\end{subequations}
\subsection{Replica-Symmetric Ansatz} \label{sec:1b}
The first approximation of the most general solution is obtained by imposing a Replica-Symmetric (RS) ansatz on the order parameters as follows
\begin{subequations}
	\label{ansatz}
	\begin{align}
		q^{ab}_{rr} &=\delta_{ab}+q_r(1-\delta_{ab})\,\,\,\,\,\,\, \forall r=s,\\
		q^{ab}_{rs} &= R \,\delta_{ab} + p\,(1-\delta_{ab})\,\,\,\,\,\,\, \forall r\neq s.
	\end{align}
\end{subequations}
where $q_r$ represents the typical overlap of a solution with $\kappa_r$ margin while $p$ and $R$ are respectively the overlaps between the two reference solutions within the same replicas or not. There is no reason in principle to assume $p$ and $R$ to be different since the system is decoupled, however since we are looking for the value of $p$ which would not appear in the calculation if $p=R$, we use the trick of assuming them different and later perform the limit $R \to p$. The calculation follows standard steps and in the end the free entropy can be written as
\begin{equation}\label{entropy_RS}
	\begin{split}
		\phi = \frac{1}{2} \sum_r q_r \hat{q}_r + \sum_r \hat h_r + p\,\hat{p}- R\,\hat{R} +\, \mathcal{G}_S \,+\, \alpha \mathcal{G}_E 
	\end{split}
\end{equation}
where
\begin{subequations}
	\begin{align}
		\mathcal{G}_S &= \lim_{n \to 0} \frac{G_S}{n} = \frac{1}{2}\ln \frac{4\pi^2}{(2\hat{h}_1+\hat{q}_1)(2\hat{h}_2+\hat{q}_2)-(\hat{p}-\hat{R})^2} + \frac{\hat{h}_2 \, \hat{q}_1 +(\hat{h}_1 +\hat{q}_1)\,\hat{q}_2 +\hat{p}\,(\hat{R}-\hat{p})}{(2\hat{h}_1+\hat{q}_1)(2\hat{h}_2+\hat{q}_2)-(\hat{p}-\hat{R})^2}\\
		\mathcal{G}_E &= \lim_{n \to 0} \frac{G_E}{n} = \int \mathcal{D}z_1 \, \mathcal{D}z_2 \ln \int \mathcal{D}x \, H\left(\frac{k_1 +\sqrt{R-p}\,x+\sqrt{q_1}\,z_1}{\sqrt{1-q_1-R+p}} \right)\, H\left(\frac{k_2 +\sqrt{R-p}\,x + \frac{p}{\sqrt{q_1}}\,z_1 +\sqrt{q_2 -\frac{p^2}{q_1}}\,z_2}{\sqrt{1-q_2-R+p}} \right)
	\end{align}
\end{subequations}
and $\mathcal{D}x \equiv dx \, \frac{e^{-x^2/2}}{\sqrt{2\pi}}$, $H(x) = \frac{1}{2} \text{erfc}\left(\frac{x}{\sqrt{2}} \right)$. By differentiating Eq.\eqref{entropy_RS} with respect to the order parameters $q_1,q_2,p,R,\hat{q}_1,\hat{q}_2,\hat{h}_1,\hat{h}_2,\hat{p},\hat{R}$, the saddle point equations take the form of 
\begin{equation}
	\label{eq::RS_SP_equations}
	\begin{aligned}
		q_1 &=  -2 \frac{\partial \mathcal{G}_S}{\partial \hat{q}_1}\,, & q_2 &= -2 \frac{\partial \mathcal{G}_S}{\partial \hat{q}_2}\,, & p &=-\frac{\partial \mathcal{G}_S}{\partial \hat{p}}\,, & R &=\frac{\partial \mathcal{G}_S}{\partial \hat{R}} \,, \\
		\hat q_1 &= -2\alpha \frac{\partial \mathcal{G}_E}{\partial q_1}\,, & \hat q_2 &=-2\alpha \frac{\partial \mathcal{G}_E}{\partial q_2}\,, & \hat p &=-\alpha \frac{\partial \mathcal{G}_E}{\partial p} \,, & \hat R &=\alpha \frac{\partial \mathcal{G}_E}{\partial R} \, , &1+  \frac{\partial \mathcal{G}_S}{\partial \hat{h}_1} &= 0\,,  &1+  \frac{\partial \mathcal{G}_S}{\partial \hat{h}_2} &=0.
	\end{aligned}
\end{equation}
The conjugated parameters can be solved as functions of the non-conjugated ones. Moreover, if the two solutions sampled with margin $\kappa_1$ and $\kappa_2$ are free to arrange in the most probable position, we have $p=R$ as already mentioned. This simplifies the conjugated parameters to
\begin{equation}
	\begin{aligned}
		\hat q_1 &= \frac{q_1}{(1-q_1)^2}\,, & \hat q_2 &=\frac{q_2}{(1-q_2)^2}\,, & \label{eq::ph}\hat p &= \frac{p}{(1-q_1)(1-q_2)}\,, \\
		\hat R &=\frac{p}{(1-q_1)(1-q_2)} \, , &\hat h_1 &= \frac{1-2q_1}{2(1-q_1)^2}\,,  &\hat h_2 &= \frac{1-2q_2}{2(1-q_2)^2}.
	\end{aligned}
\end{equation}
By looking at the expression of the entropy, it is also easy to see that, since the two solutions are sampled independently, in the limit $R \to p$ the expression reduces the sum of the two entropies of single configuration.
\subsection{$R \to p$ limit and simplification of the saddle point equations}
To actually compute the value of $p$ we resort to the saddle point equation $\hat{R} = \alpha \frac{\partial G_E}{\partial R}$ in the limit $R \to p$. Defining 
\begin{equation}
	\begin{split}
		\mathcal{Z} &\equiv \int \mathcal{D}x \, H\left(\frac{k_1 +\sqrt{R-p}\,x+\sqrt{q_1}\,z_1}{\sqrt{1-q_1-R+p}} \right)\, H\left(\frac{k_2 +\sqrt{R-p}\,x + \frac{p}{\sqrt{q_1}}\,z_1 +\sqrt{q_2 -\frac{p^2}{q_1}}\,z_2}{\sqrt{1-q_2-R+p}} \right) \\
		&= \int \mathcal{D}x \, H(a + b x) H(c + d x)
	\end{split}
\end{equation}
and integrating by parts once the limit has been replaced in order to remove divergences, we have
\begin{equation}
	\frac{\partial G_E}{\partial R} = \frac{1}{\sqrt{(1-q_1)(1-q_2)}} \int \mathcal{D}z_1 \mathcal{D}z_2 \frac{G(a) G(c)}{\mathcal{Z}}
\end{equation}
since when $R \to p$, then $b,d \to 0$ as
\begin{subequations}
	\begin{align}
		\lim\limits_{R \to p} \frac{b}{\sqrt{R-p}} &= \frac{1}{\sqrt{1-q_1}}\\
		\lim\limits_{R \to p} \frac{d}{\sqrt{R-p}} &= \frac{1}{\sqrt{1-q_2}}.
	\end{align}
\end{subequations}
Using equation~\eqref{eq::ph} for $\hat{R}$ we therefore get $p$ by solving the following implicit equation
\begin{equation}
	\begin{split}
		p &= \alpha \sqrt{(1-q_1)(1-q_2)} \int \mathcal{D}z_1  \mathcal{D}z_2 \, GH(a) \, GH(c) \\
		&= \alpha \sqrt{(1-q_1)(1-q_2)} \int \mathcal{D}z_1  \mathcal{D}z_2 \, GH\left(\frac{k_1 +\sqrt{q_1}\,z_1}{\sqrt{1-q_1}}\right) \, GH\left(\frac{k_2 + \frac{p}{\sqrt{q_1}}\,z_1 +\sqrt{q_2 -\frac{p^2}{q_1}}\,z_2}{\sqrt{1-q_2}} \right)
	\end{split}
\end{equation}
where $GH(x) \equiv \frac{G(x)}{H(x)}$.
\subsubsection{The $\kappa_{\text{max}}$ limit for the overlap of solution-pairs}
Using the identity
\begin{equation}
	GH(a x) \simeq a x \, \Theta(x) \,, \qquad \text{for} \; a \to \infty
\end{equation}
it is easy to perform the limit $k_2 \to \kappa_{\text{max}}$, i.e. specializing to the case where one of the two configurations solves the problem with the maximum possible margin $\kappa_{\text{max}}$ at a fixed $\alpha$ value (i.e. its internal entropy goes to zero as $q_2 \to 1$; for a more detailed description see \cite{baldassi2021unveiling, baldassi2021learning}).  The corresponding equation for $p$ in this limit becomes
\begin{equation}
	\begin{split}
		p &= \alpha \sqrt{(1-q_1)} \int \mathcal{D}z_1  \mathcal{D}z_2 \, GH\left(\frac{k_1 +\sqrt{q_1}\,z_1}{\sqrt{1-q_1}}\right) \, \Theta\left(\kappa_{\text{max}} + p\,z_1 +\sqrt{1 -\frac{p^2}{q_1}}\,z_2\right) \left(\kappa_{\text{max}} + p\,z_1 +\sqrt{1 -\frac{p^2}{q_1}}\,z_2\right).
	\end{split}
\end{equation}
Integrating in $z_2$ we get the desired expression
\begin{equation}
	\begin{aligned}
		p = \alpha \sqrt{(1-q_1)} \int \mathcal{D}z_1 \, & GH\left(\frac{k_1 +\sqrt{q_1}\,z_1}{\sqrt{1-q_1}}\right) \, \\
		&\times \left[ \left(\kappa_{\text{max}} + \frac{p}{\sqrt{q_1}}\,z_1\right) H\left(- \frac{\kappa_{\text{max}} + \frac{p}{\sqrt{q_1}}z_1 }{\sqrt{1-\frac{p^2}{q_1}}}\right) + \sqrt{1-\frac{p^2}{q_1}} G\left( \frac{\kappa_{\text{max}} + \frac{p}{\sqrt{q_1}}z_1 }{\sqrt{1-\frac{p^2}{q_1}}}\right)\right]. 
	\end{aligned}
\end{equation}


\subsection{Nested overlap structure}\label{sec:p}
\begin{figure}[h]
	\begin{centering}
		\includegraphics[width=0.60\columnwidth]{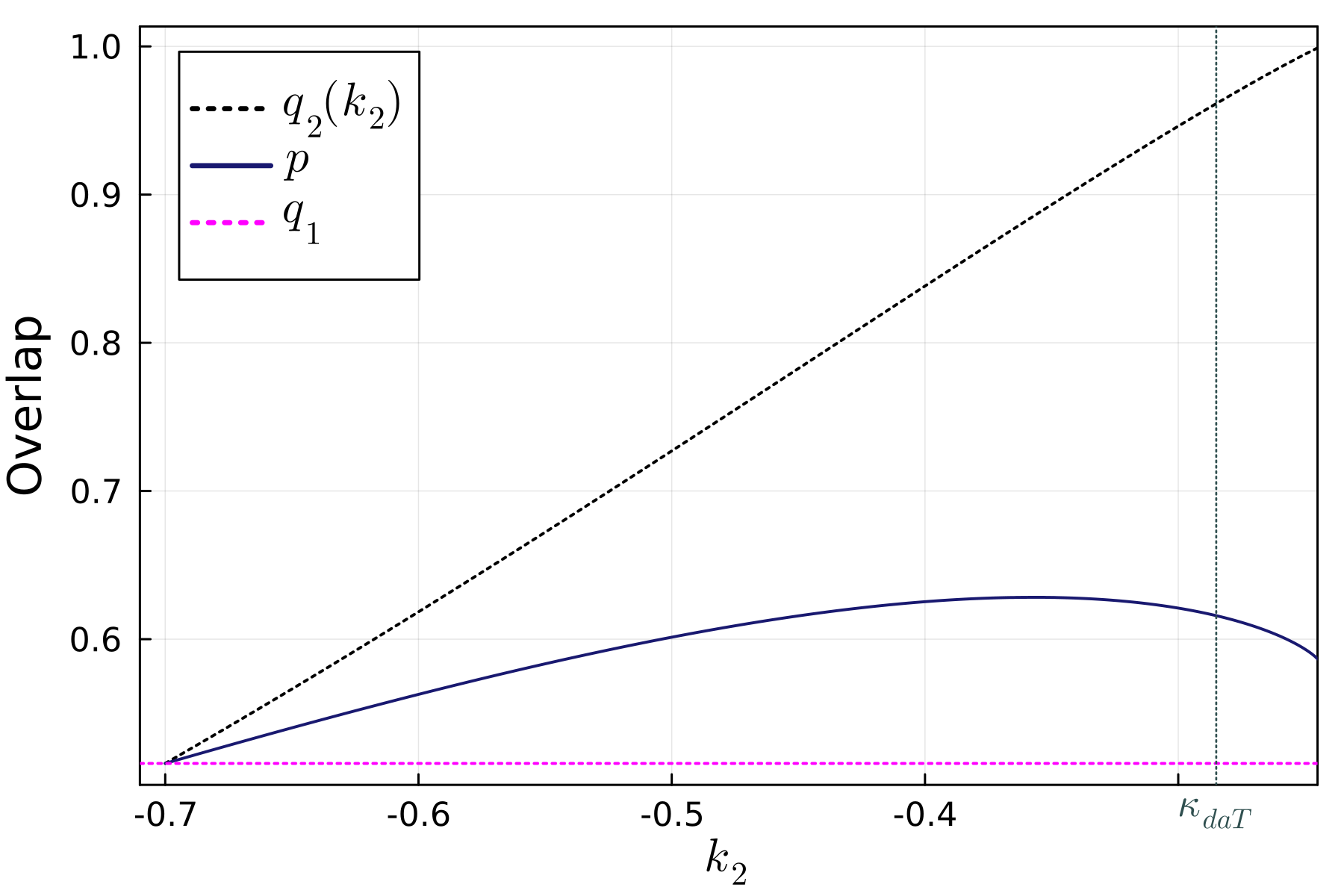}
	\end{centering}
	\caption{\label{fig::heatmap} 
		As a function of the margin $k_2$ from bottom to above: typical overlap $q_1$ between solutions having margin $k_1 = -0.7$ (dashed pink line), typical overlap $p$ between two solutions having respectively margin $k_1$ and $k_2$ (full blue line) and typical overlap $q_2$ between two solutions with margin $k_2$ (dashed black line). Here $\alpha =3$. This plot shows the nested overlap structure of the solutions $q_1< p < q_2$.} 
\end{figure}

In order to understand how configurations with different margin are organized in the solution space at a fixed value of $\alpha$, we study how the order parameters $q_1, \,q_2, \,p$ evolve as a functions of $k_1$ and $k_2$. As already explained in Sec.~\ref{sec1}, since the RS ansatz from which we derive the typical overlap of each solution ($q_1$ and $q_2$) amounts to assuming that the space form a unique connected component, we can trust our predictions as long as the RS solution is stable with respect to RSB effects. Referring to the phase diagram in Fig.~\ref{fig:phasediagram}, this implies that for each value $k_r <0 $ and low enough $|k_r |$ there is a range of $\alpha(k_r) \in [0,\alpha_{\text{dAT}}(k_r))$ for which the RS approximation is reliable (green line). 
\\As discussed in \cite{baldassi2021learning}, we find that for every $\alpha$ the typical overlaps $q_r (k_r)$ of solutions are an increasing function of $k_r$, meaning that the robust solutions are indeed the most sparse (i.e. the external entropy is a decreasing function of $k_r$)  but also the closest to each other. In addition, looking at the inset in Fig.~\ref{fig::heatmap}, for each value of $k_1 < k_2$, the overlap between two solutions is such that $q_1 < p < q_2$.  This suggests that solutions with a given margin are arranged in an hierarchical fashion: the most robust solutions are gradually surrounded by solutions with smaller and smaller margin. Another indication towards this picture is the following (see the inset of Fig.~\ref{fig::heatmap}): for fixed values of $\alpha$ and $k_1$, if we increase $k_2$ then $p \,(k_2; \alpha, k_1)$ is a non-monotonic function of $k_2$. It increases until a certain value $k_2^{\star}(\alpha) $ and then it starts to decrease, showing that the space of these rearranged solutions is not isotropic with respect to high-margin solutions. 

\section{Manifold spanned by $y$ solutions}
In this section we refer to the general computation of the manifold spanned by $y$ solutions sampled with different margins. Our general setting is choosing a non-convex problem solved with margin $\kp$ and in this space considering the training error manifold obtained when sampling one solution with margin $k_1$ and $y-1$ solution with margin $k_2$, with $k_1 \neq k_2 \neq \kp$. In the following subsections we report case studies providing a picture of how connectivity changes when considering typical and atypical solutions.

Consider $y$ weight configurations $\boldsymbol{W}^{r}\in\mathbb{R}^{N}$
on the sphere $\lVert\boldsymbol{W}^{r}\rVert^{2}=N$ and a vector
of interpolation coefficients $\boldsymbol{\gamma}=(\gamma_{1},\dots,\gamma_{y})$
with $\gamma_{r}\geq0$ and such
that $\sum_{r=1}^{y}\gamma_{r}=1$. We call $\boldsymbol{W}_{\boldsymbol{\gamma}}$ the interpolation 
\begin{equation}
	\boldsymbol{W}_{\boldsymbol{\gamma}}=\sqrt{N}\frac{\sum_{r}\gamma_{r}\boldsymbol{W}^{r}}{\rVert\sum_{r}\gamma_{r}\boldsymbol{W}^{r}\rVert}
\end{equation}

We want to compute the corresponding expected asymptotic training error, given by
\begin{equation}
	E_{\boldsymbol{\gamma}} = \lim_{N\to+\infty}\ \mathbb{E}_{\boldsymbol{\xi}}\,\big\langle \Theta\big( - \boldsymbol{W}_{\boldsymbol{\gamma}} \cdot \boldsymbol{\xi}^\mu + \kp\sqrt{N} \big)\big\rangle_{k_1,\dots,k_y}.
\end{equation}

where the expectation $\langle\bullet\rangle$ is over the factorized
measure
\begin{eqnarray}
	\langle\bullet\rangle & = & \frac{\int\prod_{r=1}^{y}d\mu(\boldsymbol{W}^{r})\prod_{\mu r}\Theta\left(\frac{1}{\sqrt{N}}\sum_{i}W^r_{i}\xi_{i}^{\mu}-k_r \right)\ \bullet}{\prod_{r=1}^yZ (k_r)}
\end{eqnarray}

Using the replica trick $Z^{-1}=\lim_{n\to0}Z^{n-1}$ we have
\begin{eqnarray}
	E_{\boldsymbol{\gamma}} & = & \lim_{N\to\infty}\lim_{n\to0}\mathbb{E}_{\boldsymbol{\xi}}\,\int\prod_{a=1}^{n}\prod_{r=1}^y d\mu(\boldsymbol{W}^{ar})\ \prod_{\mu a r}\Theta\left(\frac{1}{\sqrt{N}}\sum_{i}W_{i}^{ar}\xi_{i}^{\mu}-k_r\right)\ \Theta \left(-\sum_i W_{\bgamma,i} ^{1}\xi^{1}_i + \kp\right)\\
	& = & \lim_{N\to\infty}\lim_{n\to0}\mathbb{E}_{\boldsymbol{\xi}}\,\int\prod_{ar}d\mu(\boldsymbol{W}^{ar})\ \prod_{\mu ar}\frac{d\lambda_{r}^{\mu a}d\hat{\lambda}_{r}^{\mu a}}{2\pi}\prod_{\mu ar}\Theta(\lambda_{r}^{\mu a}-k_r)\ \Theta\left(-\frac{\sqrt{N}\sum_{r=1}^{y}\gamma_{r}\lambda_{r}^{11}}{\rVert\sum_{r}\gamma_{r}\boldsymbol{W}^{1r}\rVert}+\kp\right)\\
	&  & \times e^{i\sum_{\mu ar}\hat{\lambda}_{r}^{\mu a}\lambda_{r}^{\mu a}-i\sum_{\mu ari}\frac{1}{\sqrt{N}}\hat{\lambda}_{r}^{\mu a}W_{ri}^a\xi_{i}^{\mu}}\\
	& = & \lim_{N\to\infty}\lim_{n\to0}\int \prod_{a<b,rs} \frac{dq^{ab}_{rs}\,d\hat{q}^{ab}_{rs}}{2\pi}\int \prod_{ar} \frac{d\hat{h}^{a}_r}{2\pi}\,e^{Nn\phi(q, \hat{q}, \hat{h})} g(q)
\end{eqnarray}
where $\phi(q, \hat{q}, \hat{h})$ and $q_{rs}^{ab}$ have been defined respectively in equations~\eqref{entropicandenergetic} and~\ref{eq::order_parameters}. We have also defined the quantity
\begin{eqnarray}
	\label{eq::g(q)}
	g(q) & = & \frac{\int\prod_{ar}\frac{d\lambda_{r}^ad\hat{\lambda}_{r}^a}{2\pi}\prod_{ar}\Theta(\lambda_{r}^a-k_r)\ \Theta\left(-\frac{\sum_{r=1}^{y}\gamma_{r}\lambda_{r}^1}{\sqrt{\sum_{rs} \gamma_r \gamma_s q_{rs}^{a=b}}}+\kp\right) e^{i\sum_{ar}\hat{\lambda}_{r}^a\lambda_{r}^a-\frac{1}{2}\sum_{abrs}q_{rs}^{ab}\hat{\lambda}_{r}^a \hat{\lambda}_{s}^b}}{\int\prod_{ar}\frac{d\lambda_{r}^ad\hat{\lambda}_{r}^a}{2\pi}\prod_{ar}\Theta(\lambda_{r}^a-k_r)\ e^{i\sum_{ar}\hat{\lambda}_{r}^a\lambda_{r}^a-\frac{1}{2}\sum_{abrs}q_{rs}^{ab}\hat{\lambda}_{r}^a\hat{\lambda}_{s}^b}}
\end{eqnarray}
Calling $q_{\star}$ the matrix of overlaps at the saddle point for $\phi$, the small $n$ limit will
lead to 
\begin{equation}
	E_{\boldsymbol{\gamma}}=\lim_{n\to0}\ g(q_{\star})
\end{equation}

\subsection{General structure of the result in the RS ansatz}
In the RS ansatz the tensor $q_{rs}^{ab}$ is identified by two matrices $q_{rs}^{a = b}$ and $q_{rs}^{a\ne b}$. They have the following structure
\begin{subequations}
	\begin{align}
		q^{aa}_{rs} &= \delta_{rs} + (1-\delta_{rs}) \, p_{rs} \\
		q^{a\ne b}_{rs} &\equiv t_{rs} = q_r \delta_{rs} + (1-\delta_{rs}) \, p_{rs} \label{eq::trs}
	\end{align}
\end{subequations}
where $\delta_{rs}$ is the Kronecker delta function. We have denoted by $q_r$, $r\in [y]$ the typical overlap between two solutions having the \emph{same} margin $k_r$, $r\in [y]$ and by $p_{rs}$ the typical overlap between a solution having margin $k_r$ and another one having margin $k_s$, with $r, s \in [y]$ and $r \ne s$. The computation can be carried out by noticing that the term in~\eqref{eq::g(q)} that depends on $q_{rs}^{ab}$ can be written as
\begin{equation}
	\begin{split}
		-\frac{1}{2} \sum_{abrs} q_{rs}^{ab} \hat{\lambda}_r^a \hat{\lambda}_s^b &= -\frac{1}{2} \sum_{r} (1-q_{r}) \sum_a \left(\hat{\lambda}_r^a\right)^2 - \frac{1}{2} \sum_{rs} t_{rs} \sum_{ab} \hat{\lambda}_r^a \hat{\lambda}_s^b \\
		&= -\frac{1}{2} \sum_{r} (1-q_{r}) \sum_a \left(\hat{\lambda}_r^a\right)^2 - \frac{1}{2} \sum_{r} \left( \sum_{as} \mathcal{T}_{rs} \hat{\lambda}^a_s \right)^2
	\end{split}
\end{equation}
where $\mathcal{T}_{rs}$ is $(r,s)$ the element of the \emph{square root} of the matrix $t_{rs}$ defined in~\eqref{eq::trs}. The computation proceeds in a standard way by using a Hubbard-Stratonovich transformation and integrating over $\hat{\lambda}_r^a$. Since the denominator tends to 1 when $n\to 0$ we get
\begin{equation}
	\begin{split}
		E_{\boldsymbol{\gamma}} &= \lim_{n\to 0}\int \prod_r \mathcal{D} x_r \int \prod_{ar}\frac{d\lambda_{r}^a}{\sqrt{2\pi (1-q_r)}} \, \prod_{ar}\Theta(\lambda_{r}^a-k_r)\ \Theta\left(\kp c_{\boldsymbol{\gamma}}-\sum_{r=1}^{y}\gamma_{r}\lambda_{r}^1\right) e^{-\frac{1}{2} \sum_{ar}\frac{\left(\lambda_r^a - \sum_s \mathcal{T}_{rs} x_s\right)^2}{1-q_r}} \\
		&= \int \prod_r \mathcal{D} x_r \frac{\int \prod_r \mathcal{D}\lambda_r \prod_{r}\Theta(\sqrt{1-q_r}\lambda_{r} + \sum_s \mathcal{T}_{rs} x_s -k_r)\ \Theta\left(\kp c_{\boldsymbol{\gamma}}-\sum_{r=1}^{y}\gamma_{r} \left(\sqrt{1-q_r}\lambda_{r} + \sum_s \mathcal{T}_{rs} x_s\right) \right)}{\prod_r H\left(\frac{k_r - \sum_s \mathcal{T}_{rs} x_s}{\sqrt{1-q_r}}\right)}
	\end{split}
\end{equation}
where we have defined the quantity
\begin{equation}
	c_{\boldsymbol{\gamma}} \equiv \sqrt{\sum_{rs} \gamma_r \gamma_s q_{rs}^{a=b}} = \sqrt{\sum_r \gamma_r^2 + \sum_{r \ne s}  p_{rs} \gamma_r \gamma_s}\,.
\end{equation}
Next we can get rid of the theta functions expressing them as integration boundaries. Starting from the 
\begin{equation}
	\begin{split}
		E_{\boldsymbol{\gamma}} &= \int \prod_r \mathcal{D} x_r 
		\frac{\int \prod_r \mathcal{D}\lambda_r \prod_{r}\Theta(\lambda_{r} - A_r)\ \Theta\left(
			\frac{\kp c_{\boldsymbol{\gamma}}- \sum_{rs} \gamma_r \mathcal{T}_{rs} x_s-\sum_{r=1}^{y-1} \sqrt{1-q_r} \gamma_r\lambda_{r}}{\gamma_y \sqrt{1-q_y}}
			- \lambda_y \right)}{\prod_r H\left(A_r\right)} \\
		&= \int \prod_r \mathcal{D} x_r 
		\frac{\int \prod_{r=1}^{y-1} \mathcal{D}\lambda_r \prod_{r=1}^{y-1}\Theta(\lambda_{r} - A_r)\ \Theta\left(
			B_y - A_y \right) \int_{A_y}^{B_y} \mathcal{D} \lambda_y}{\prod_r H\left(A_r\right)} \\
		&= \int \prod_r \mathcal{D} x_r 
		\frac{\int \prod_{r=1}^{y-2} \mathcal{D}\lambda_r \prod_{r=1}^{y-2}\Theta(\lambda_{r} - A_r)\ \Theta\left(
			B_{y-1} - A_{y-1} \right) \int_{A_{y-1}}^{B_{y-1}} \mathcal{D} \lambda_{y-1} \int_{A_y}^{B_y} \mathcal{D} \lambda_y}{\prod_r H\left(A_r\right)} \\
		&= \int \prod_r \mathcal{D} x_r \, \Theta\left(B_1 - A_1 \right) 
		\frac{\int_{A_1}^{B_1} \mathcal{D}\lambda_1 \dots \int_{A_y}^{B_y} \mathcal{D} \lambda_y}{\prod_r H\left(A_r\right)}
	\end{split}
\end{equation}
where we have defined the following quantities
\begin{subequations}
	\label{eq::general_AB}
	\begin{align}
		A_r &\equiv \frac{k_r - \sum_s \mathcal{T}_{rs} x_s}{\sqrt{1-q_r}}\,, \qquad r\in [y] \label{eq::general_A}\\
		B_l &\equiv \frac{\kp c_{\boldsymbol{\gamma}} - \sum_{r=l+1}^{y} \gamma_r k_r - \sum_{r=1}^l \gamma_r \sum_s \mathcal{T}_{rs} x_s-\sum_{r=1}^{l-1}\sqrt{1-q_r} \gamma_r \lambda_{r}}{\gamma_l \sqrt{1-q_l}}\,, \qquad  l\in [y] \label{eq::general_B}
	\end{align}
\end{subequations}
Notably the term 
\begin{equation}
	\Theta\left(B_1 - A_1\right) = \Theta\left( \kp c_{\boldsymbol{\gamma}} - \sum_r \gamma_r k_r \right)
\end{equation}
does not depend on any of the $x_r$, $r\in [y]$. The innermost integral can be performed analytically, leading to the final expression
\begin{equation} \label{eq::general_formula_path}
	E_{\boldsymbol{\gamma}} = \Theta\left( \kp c_{\boldsymbol{\gamma}} - \sum_r \gamma_r k_r \right) \int \prod_r \mathcal{D} z_r \frac{\int_{A_1}^{B^1}\mathcal{D}\lambda_1 ... \int_{A_{y-1}}^{B^{y-1}}\mathcal{D}\lambda_{y-1} \left( H(A_y) - H(B_y) \right) }{\prod_r H(A_r)} \,.
\end{equation}
Therefore, the quantities $f_{\boldsymbol{\gamma}}$ and $I_{\boldsymbol{\gamma}}$ defined in the main text are
\begin{subequations}
	\begin{align}
		f_{\boldsymbol{\gamma}}(\kp, \left\{ k_r \right\}_r) &\equiv \kp c_{\boldsymbol{\gamma}} - \sum_r \gamma_r k_r \\
		I_{\boldsymbol{\gamma}}\left( \kp, \left\{ k_r \right\}_r\right) &\equiv \int \prod_r \mathcal{D} z_r \frac{\int_{A_1}^{B^1}\mathcal{D}\lambda_1 ... \int_{A_{y-1}}^{B^{y-1}}\mathcal{D}\lambda_{y-1} \left( H(A_y) - H(B_y) \right) }{\prod_r H(A_r)}
	\end{align}
\end{subequations}

\subsection{Sampling solutions with the same margin $k$}

We consider the case in which all the vertices of the simplex have the same margin, i.e. $k_r = k \ge \kp$, for $r\in [y]$. In this case the matrix $t_{rs}$ defined in~\eqref{eq::trs} and its square root $\mathcal{T}_{rs}$ contain all equal elements
\begin{subequations}
	\begin{align}
		t_{rs} &= q \\
		\mathcal{T}_{rs} &= \sqrt{\frac{q}{y}}   
	\end{align}
\end{subequations}
where we remind that $q$ is the typical overlap between solutions having margin $k$.
In equation~\eqref{eq::general_formula_path} one can integrate all $x_r$ except one leading to
\begin{equation} \label{eq::interpolation_same_margin}
	E_{\boldsymbol{\gamma}} = \Theta\left( \kp c_{\boldsymbol{\gamma}} - k \right) \int \mathcal{D} z \frac{\int_{A}^{B_1}\mathcal{D}\lambda_1 ... \int_{A}^{B_{y-1}}\mathcal{D}\lambda_{y-1} \left( H(A) - H(B_y) \right) }{H^y(A)}
\end{equation}
where
\begin{subequations}
	\begin{align}
		A &\equiv \frac{\kp - \sqrt{q} x}{\sqrt{1-q}}\,, \\
		B_l&\equiv \frac{\kp c_{\boldsymbol{\gamma}} - \kp \sum_{r=l+1}^{y} \gamma_r - \sqrt{q} x \sum_{r=1}^l \gamma_r - \sqrt{1-q} \sum_{r=1}^{l-1} \gamma_r \lambda_{r}}{\gamma_l \sqrt{1-q}}\,, \qquad  l\in [y] \\
		c_{\boldsymbol{\gamma}} &= \sqrt{(1-q)\sum_r \gamma_r^2 + q}
	\end{align}
\end{subequations}
Notice that since $0\;\leq\gamma_r \leq1$, we have $\sum_r \gamma_r^{2}\leq\sum_r \gamma_r=1$ and consequently
\begin{equation}
	c_{\boldsymbol{\gamma}}^{2}=\left(1-q\right)\sum_r \gamma_r^{2}+q\;\leq1.
\end{equation}
Strict equality holds only on the vertices, while $c_{\boldsymbol{\gamma}}$ attains
its minimum value on the barycenter $\gamma_r \equiv\frac{1}{y}$,
for which
\begin{eqnarray}
	c_{\text{barycenter}}^{2} & = & \left(1-q \right)\frac{1}{y}+q.
\end{eqnarray}

\subsubsection{Case $k = \kp$}
\begin{figure}[h]
	\begin{centering}
		\includegraphics[width=0.43\columnwidth]{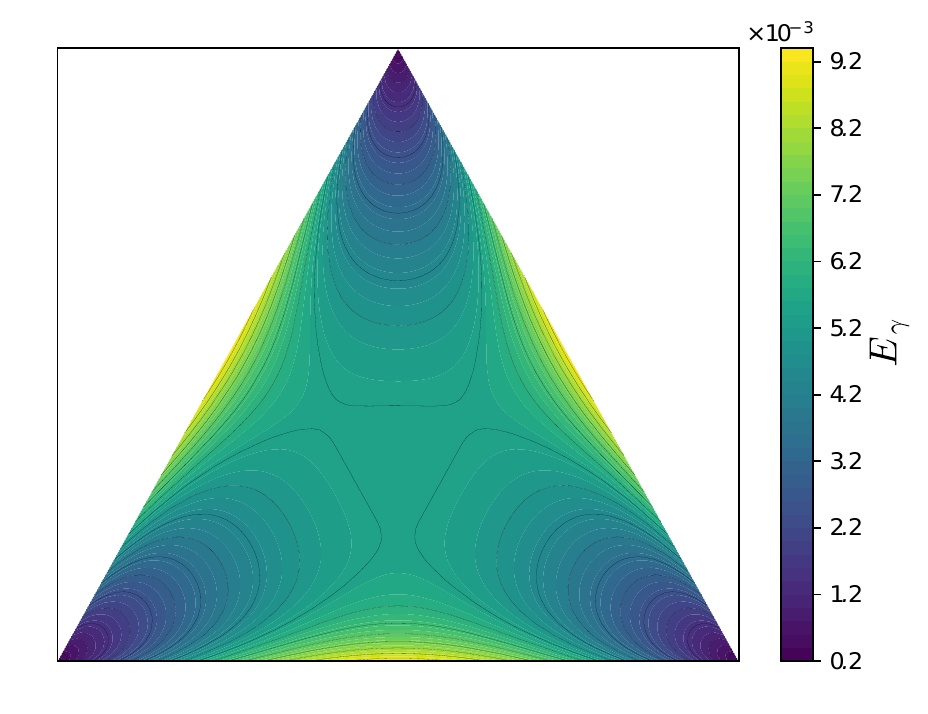}
		\includegraphics[width=.46\columnwidth]{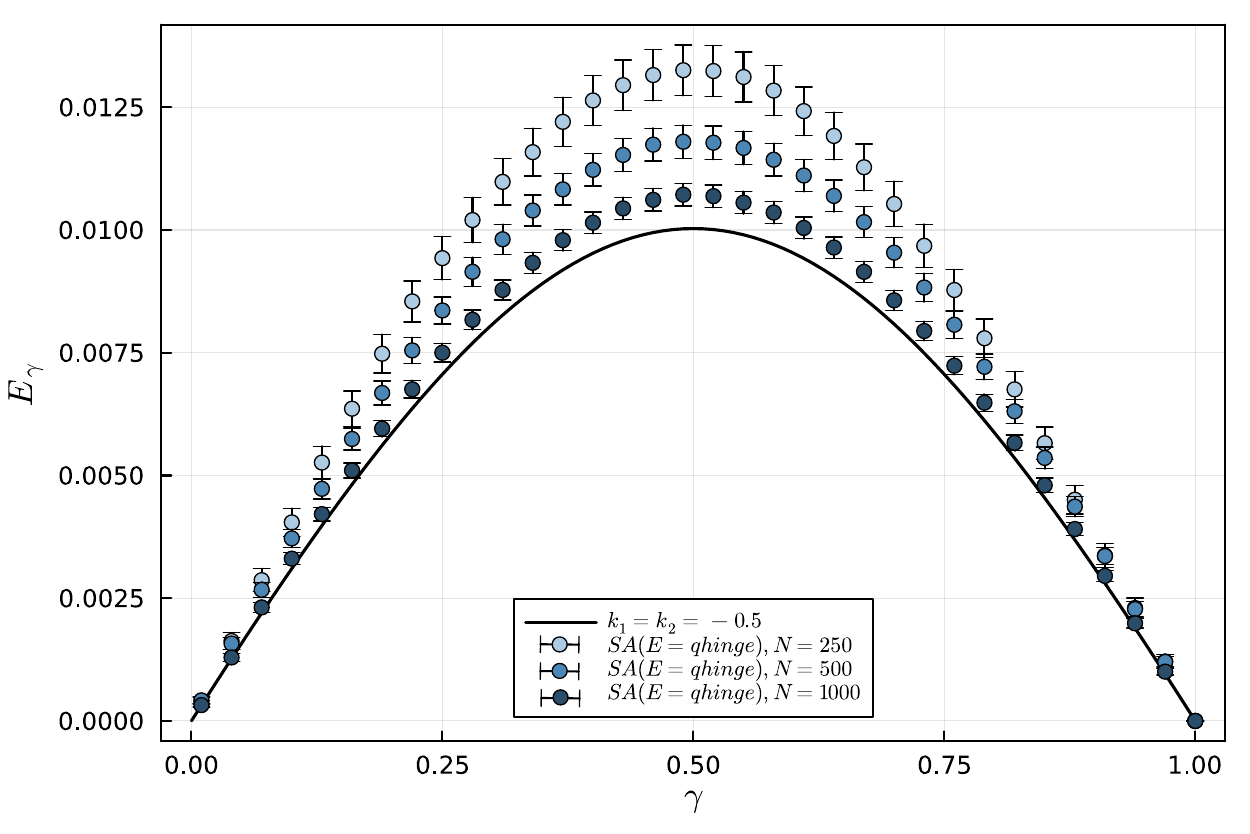}
	\end{centering}
	\caption{\label{fig::typicalsol}(Left panel) Training error manifold spanned by $y=3$ typical solutions where $k = \kp = -0.5$ and $\alpha = 1$. (Right panel) Typical energy barrier between $y=2$ solutions with $k = \kp = -0.5$ and $\alpha = 1$ (corresponding to the edges of the triangle in panel a) compared with SA sampling at different sizes.} 
\end{figure}

Here we start by analyzing the simplest possible case, i.e. when $k = \kp$.  Since $c_{\boldsymbol{\gamma}} \le 1$ the argument of the theta function in~\eqref{eq::interpolation_same_margin}, $\kp c_{\boldsymbol{\gamma}} - \kp$  is always positive when $\kp < 0$ and therefore every point on the simplex has non-vanishing energy. When $\kp > 0$ instead, the argument of the theta function is always negative; when $\kp=0$, the extremes of integration $A = B_1$; in both cases this means that $E_{\boldsymbol{\gamma}} = 0$ on the linear interpolation between two solutions. This is consistent with the fact that for $\kp \ge 0$ the space of solution is convex. 

\begin{figure}[h]
	\begin{centering}
		\includegraphics[width=0.5\columnwidth]{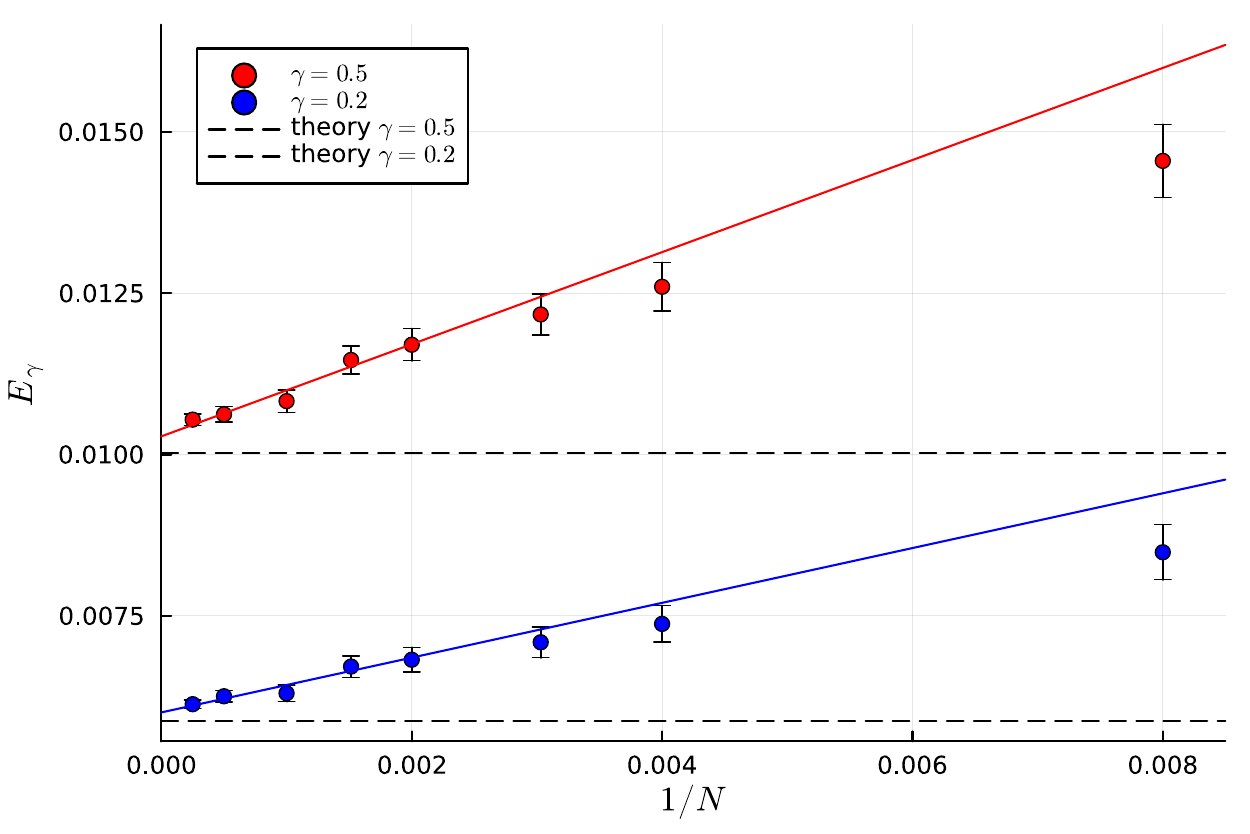}
	\end{centering}
	\caption{\label{fig::extrapolation_energy_barrier}Energy on the linear interpolation between two solutions with margin $k = \kp = -0.5$ found by sampling with SA at various sizes and their extrapolations (full lines) at large $N$. The two colors refer to the case $\gamma = 0.5$ (red) and $\gamma=0.2$ (blue). The horizonthal dashed black lines represents the theoretical predictions. In the extrapolations we have fitted the last 5 points.} 
\end{figure}

In the right panel of Fig.~\ref{fig::typicalsol} we report the training error on the 2-simplex (i.e. a triangle) with solutions having $\kp = -0.5$ placed at its vertices. In the right panel of Fig.~\ref{fig::typicalsol} we report the training error on the line connecting the same $\kp = -0.5$ solutions, i.e. the training error on the edge of the triangle in the left panel of Fig.~\ref{fig::typicalsol}. Notice that the energy is symmetrical by construction with respect to the barycenter of the simplex. 
Note also that the training error of the barycenter obtained imposing $\gamma_{r} = \frac{1}{y}$
\begin{equation}
	\overline{\boldsymbol{W}} = \frac{\sqrt{N} \sum_{r=1}^y \boldsymbol{W}^r}{\lVert \sum_{r=1}^y \boldsymbol{W}^r \rVert}    
\end{equation}
is lower that the maximum of the barrier along the edges. In Sec.~\ref{sec::anisotropy} we also characterize the overlap of the barycenter with the maximum margin solution.
The right panel of Fig.~\ref{fig::typicalsol} shows also the comparison of the energy barrier between the theoretical prediction and the numerical simulations which are obtained by using a Simulated Annealing (SA) protocol at different sizes, see section~\ref{sec::num_sim} for details. Increasing $N$ the energy barrier lowers and improves the agreement with the RS theory presented in the paper. In Fig.~\ref{fig::extrapolation_energy_barrier} we show,  as a function of $1/N$, the behaviour of the training error on two points ($\gamma = 0.5, 0.2$) on linear interpolation between two typical solutions of the problem, together with the extrapolation to large $N$ values.

\subsubsection{Case $k >\kp$}

\begin{figure*}[h]
	\begin{centering}
		\includegraphics[width=.46\columnwidth]{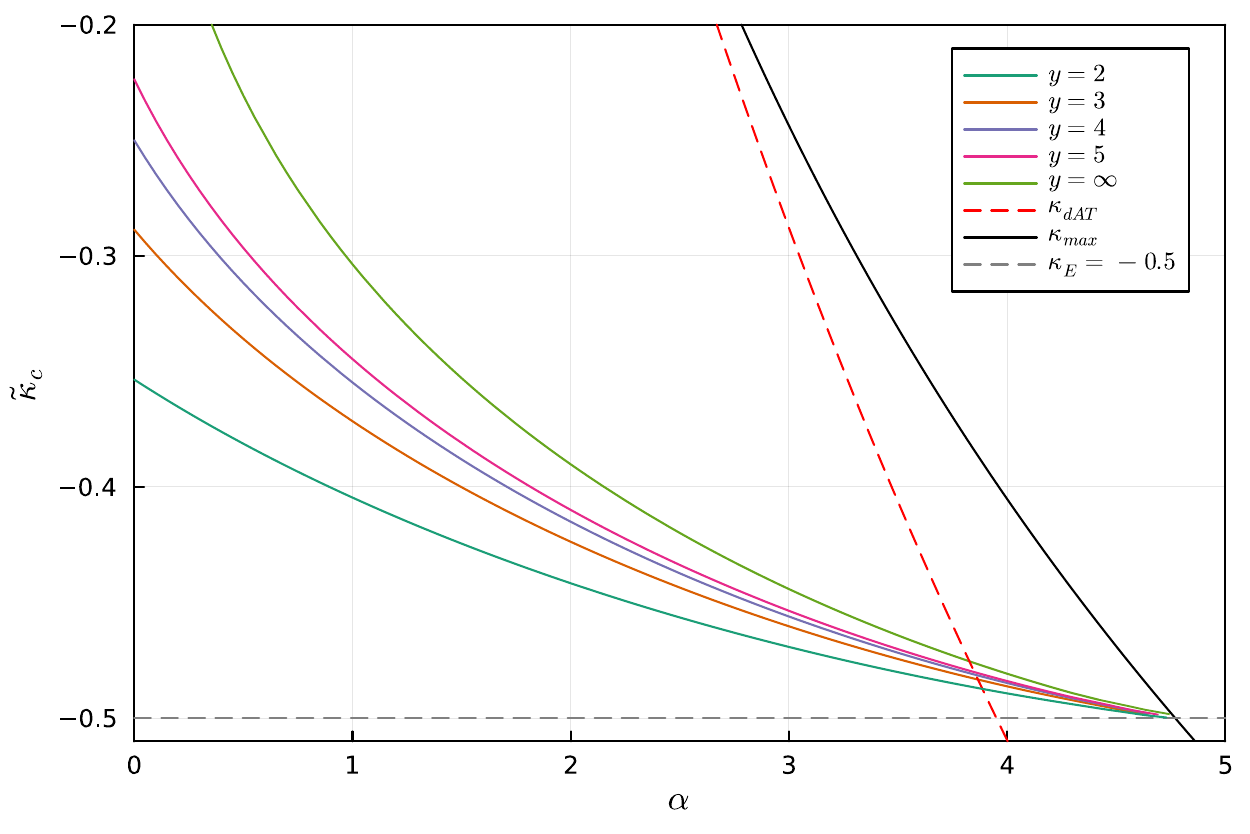}
		\includegraphics[width=.43\columnwidth]{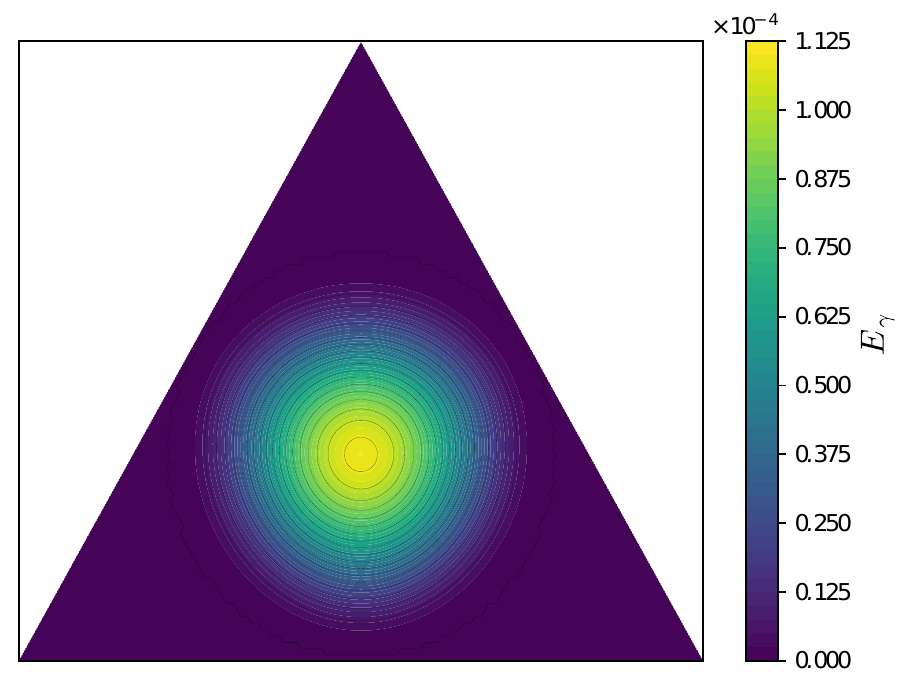}
	\end{centering}
	\caption{\label{fig:phase_diagram_samek} (Left panel) Plot of $\kappa_y^\star$ defined in~\eqref{eq::kappaystar} for $\kp = -0.5$ (grey dashed line) as a function of $\alpha$ for $y=2, 3, 4, 5, \infty$. As showed in the text, the values of $\kappa^\star_y$ for which the whole $(y-1)$ dimensional simplex is at zero energy obey the relations \eqref{eq::kstar_inequalities}. The transition lines are obtained within the RS assumption, which is correct below the dAT transition (red dashed line). (Right panel) Training error on the 2-simplex with vertices all having margin $k$ such that $\kappa_2^\star < k < \kappa_3^\star$. As showed in the text, the training error of the barycenter is indeed at strictly positive energy in this regime.} 
\end{figure*}

\begin{figure*}[h]
	\begin{centering}
		\includegraphics[width=.45\columnwidth]{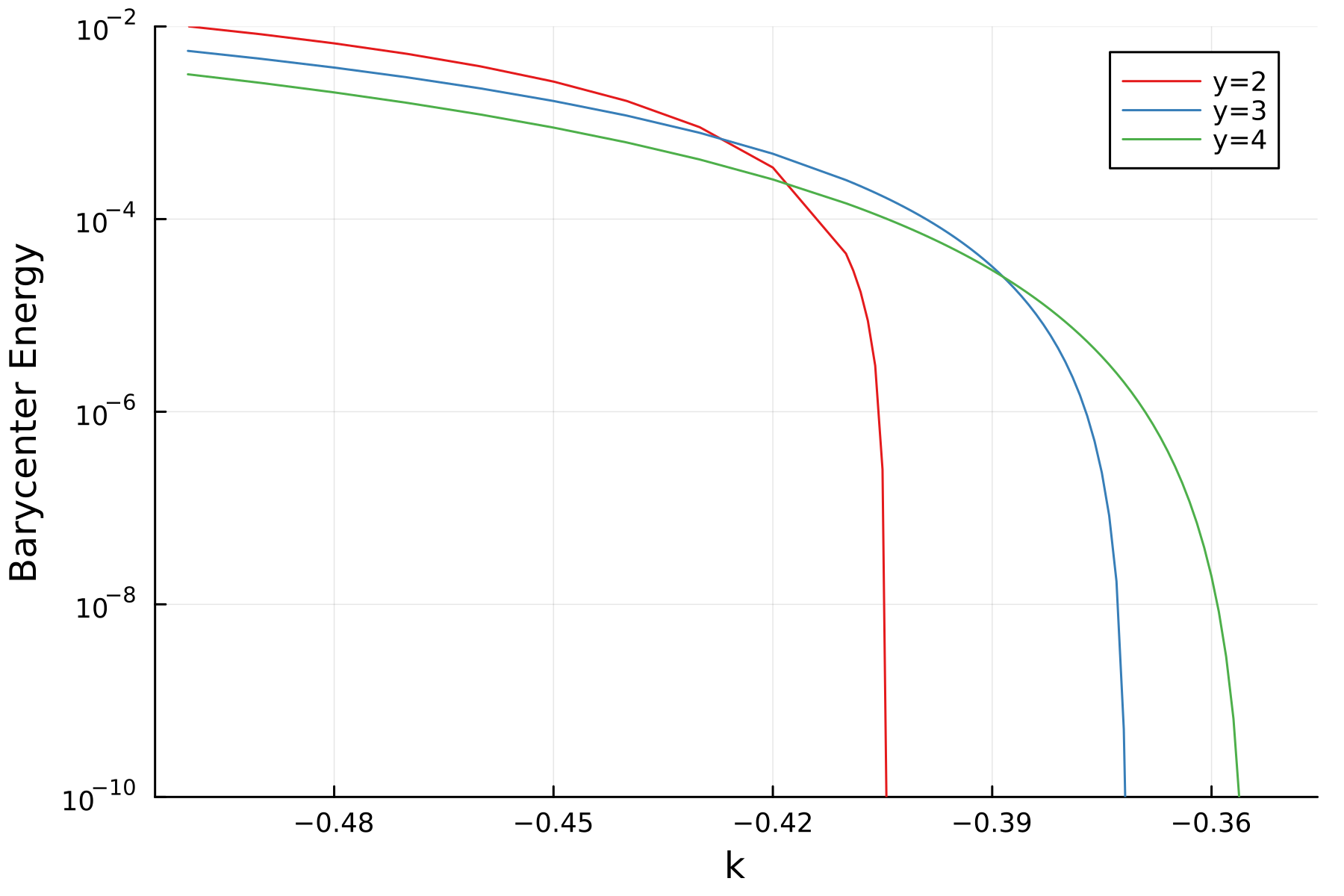}
	\end{centering}
	\caption{\label{fig:center_energy} Training error of the barycenter of the manifold with $3$ different values of $y=2,3,4$, at $\alpha = 1$ and $\kp=-0.5$. Notice the exchange in the curves, due to the ordering in Eq.~\eqref{eq::kstar_inequalities} of the $\kappa^*_y$ thresholds.} 
\end{figure*}
We extend the analysis of the previous subsection to the case $k >\kp$. The condition $\Theta(\kp c_{\boldsymbol{\gamma}}-k)$ allows us to distinguish between three regimes:
\begin{itemize}
	\item for $0\leq\kp\leq k$ the whole convex envelope has
	zero training error (as said before, the space of solutions is convex)
	\item for $\kp < k < \kappa_y^\star(\alpha, \kp) < 0$ an extended region in the proximity of each corner has zero training error. In the case $y=2$ we can extract the extension of this region: the training error is zero if $\gamma \in [0, \gamma_L]$ and $\gamma \in [\gamma_R, 1]$; by symmetry $\frac{1}{2} - \gamma_L  = \gamma_R - \frac{1}{2}$ and
	\begin{equation}
		\gamma_{L, \, R} = \frac{1}{2} \pm \frac{1}{2} \sqrt{1 - 2 \left(\frac{1- \left(\frac{k}{\kp} \right)^2}{1-q} \right)}
	\end{equation}
	\item If the margin is large enough, (i.e. $k >\kappa_y^\star(\alpha, \kp)$) we can find the whole simplex is at zero training energy, meaning that $\kp c_{\boldsymbol{\gamma}} - k < 0$ for any value of $\boldsymbol{\gamma}$. we can find interpolations $\gamma$
	with zero training error if the following relation holds: $\kp\leq\frac{k}{c_{\boldsymbol{\gamma}}}$. In the $y=2$ case it is easy to find the value $\kappa_2^\star$ by imposing that $\gamma_L = \gamma_R$; we find that $\kappa_2^\star$ is found by searching the value of $k$ that satisfies the following implicit equation
	\begin{equation}
		\label{eq::cstar_y=2}
		1 - \left(\frac{k}{\kp}\right)^2 = \frac{1 - q(k)}{2}
	\end{equation}
	where we have explicited for convenience the dependence of the overlap $q$ on $k$.
\end{itemize}
The previous argument to compute the $\kappa_2^\star$ threshold can be easily extended to generic $y$ by reminding that the minimum value of $c_{\boldsymbol{\gamma}}$ is attained on the barycenter $\gamma_r = \frac{1}{y}$; when we increase the value of $k$ by symmetry this is the last point to have non-zero training error. This tells us that if the inequality $\kp c_{\boldsymbol{\gamma}} - k < 0$ is valid for the barycenter, then the whole simplex will have zero training error. Reminding that $\kp$ is negative, we have that the condition for which the whole simplex has zero energy is
\begin{equation}
	\label{eq::kappaystar}
	k > \kp c_{\text{barycenter}} = \kp \sqrt{(1-q) \frac{1}{y} + q } \equiv \kappa_y^\star
\end{equation}
which is consistent with~\eqref{eq::cstar_y=2} for $y=2$. Since the overlap $q$ increases as the margin $k$ increases, then it easy to see that the thresholds are ordered in $y$
\begin{equation}
	\label{eq::kstar_inequalities}
	\kappa_2^\star < \kappa_3^\star < \dots < \kappa_\infty^\star = \kp \sqrt{q}
\end{equation}
In the left panel of Fig.~\ref{fig:phase_diagram_samek} we show these transition curves as function of $\alpha$ on a problem with margin $\kp = -0.5$ together with the dAT line, below which our predictions for $\kappa_y^\star$ are correct. 

In the right panel of Fig.~\ref{fig:phase_diagram_samek} we show the training error on the 2-dimensional simplex with the vertices having $\kappa_2^\star < k < \kappa_3^\star$. In this case, since the solutions are extracted with $k > \kappa_2^*$, the geodesic paths between them (i.e. the edges of the triangle) is at zero energy by definition; however, since $k < \kappa_3^\star $, the 2-simplex presents configurations at non-zero training error near and at its barycenter. 
In Fig.~\ref{fig:center_energy} we show the training error of the barycenter of the $(y-1)$-simplex as a function of $k$ for several values of $y$. We show that, even if for sufficiently small values of $k$ the energy of the barycenter of $y$ solutions has training error larger with respect to the barycenter of $s<y$ solutions, as predicted by the set of inequalities~\eqref{eq::kstar_inequalities}, the energy goes to zero faster if $y$ is lower.

\subsection{Sampling solutions with two different margins}

In order to further explore the structure of the space of solutions we focus on the case of independent sampling of $y$ solutions with two different margins. For simplicity we consider the first $y-1$ vertices to be sampled all with margin $k_1$ and the last vertex with margin $k_2$. 
In this case the matrix $t_{rs}$ is of the type
\begin{equation}
	q^{a\ne b} = t =
	\begin{pmatrix} 
		q_1 & \dots  & q_1 & p \\
		\vdots & \ddots & \vdots & \vdots \\
		q_1 & \dots & q_1 & p\\
		p & \dots  & p & q_2
	\end{pmatrix}
\end{equation}
where $q_1$ and $q_y$ are the typical overlaps between two solutions having respectively margin $k_1$, $k_2$ and $p$ is the typical overlap between a solution having margin $k_1$ and another one with $k_2$. The square root of $t_{rs}$ has the same form of $t_{rs}$
\begin{equation}
	\mathcal{T} =
	\begin{pmatrix} 
		\overline{q}_1 & \dots  & \overline{q}_1 & \overline{p} \\
		\vdots & \ddots & \vdots & \vdots \\
		\overline{q}_1 & \dots & \overline{q}_1 & \overline{p}\\
		\overline{p} & \dots  & \overline{p} & \overline{q}_2 
	\end{pmatrix}
\end{equation}
where
\begin{subequations}
	\begin{align}
		\overline{q}_1 &= \frac{(y-1) q_1 + \sqrt{(y-1) \left(q_1 q_2-p^2\right)}}{(y-1) \mathcal{Z}} \\
		\overline{q}_2 &= \frac{q_2+\sqrt{(y-1) \left(q_1 q_2-p^2\right)}}{\mathcal{Z}} \\
		\overline{p} &= \frac{p}{\mathcal{Z}} \\
		\mathcal{Z} &= q_2 + (y-1) q_1 - 2 \sqrt{(y-1) \left(q_1 q_2-p^2\right)}
	\end{align}
\end{subequations}
The term $\sum_s \mathcal{T}_{rs} x_s$ in~\eqref{eq::general_formula_path} can be written as
\begin{equation}
	\sum_s \mathcal{T}_{rs} x_s = \begin{cases}
		\overline{q}_1 \sum_{s=1}^{y-1} x_s + \overline{p} x_y\,, \qquad &\text{if} \; r \in [y-1]\\
		\overline{p} \sum_{s=1}^{y-1} x_s + \overline{q}_2 x_y\,, \qquad &\text{if} \; r=y
	\end{cases}
\end{equation}
By performing some rotations over the $x_r$ variables and using the identities
\begin{subequations}
	\begin{align}
		(y-1) \overline{q}^2_1 + \overline{p}^2 &= q_1 \\
		\overline{p} ((y - 1) \overline{q}_1 + \overline{q}_2)  &= p \\
		\sqrt{y-1}\left(\overline{q}_1 \overline{q}_2 - \overline{p}^2\right) &= \sqrt{q_1 q_2 - p^2} \\
	\end{align}
\end{subequations}
it is possible to perform $y-2$ of the $y$ integrals over $x_r$. The final result is
\begin{equation} \label{eq::2_k_formula_path}
	E_{\boldsymbol{\gamma}} = \Theta\left( \kp c_{\boldsymbol{\gamma}} - k_1 \sum_{r=1}^{y-1} \gamma_r - k_2 \gamma_y \right) \int \prod_r \mathcal{D} x_1 \, \mathcal{D} x_2 \frac{\int_{A_1}^{B^1}\mathcal{D}\lambda_1 ... \int_{A_{y-1}}^{B^{y-1}}\mathcal{D}\lambda_{y-1} \left( H(A_y) - H(B_y) \right) }{\prod_r H(A_r)}
\end{equation}
where
\begin{subequations}
	\label{eq::2_k_AB}
	\begin{align}
		A_r &= \begin{cases}
			\frac{k_r - \sqrt{q_1} x_1}{\sqrt{1-q}}\,, \qquad &r\in [y-1] \\
			\frac{k_r - \frac{p}{\sqrt{q_1}} x_1 - \sqrt{q_2 - \frac{p^2}{q_1}} x_2 }{\sqrt{1-q_2}}\,, \qquad &r = y
		\end{cases} \\
		B_l &\equiv \begin{cases}
			\frac{\kp c_{\boldsymbol{\gamma}} - k_1 \sum_{r=l+1}^{y-1} \gamma_r - k_2 \gamma_y - \sqrt{q} x_1 \sum_{r=1}^l \gamma_r - \sqrt{1-q} \sum_{r=1}^{l-1} \gamma_r \lambda_{r}}{\gamma_l \sqrt{1-q_l}}\,, \qquad  &l\in [y-1] \\
			\frac{\kp c_{\boldsymbol{\gamma}} - \sqrt{q} x_1 \sum_{r=1}^{y-1} \gamma_r - \gamma_y \left[\frac{p}{\sqrt{q_1}} x_2 +  \sqrt{q_2 - \frac{p^2}{q_1}} x_2\right] - \sqrt{1-q_1} \sum_{r=1}^{y-1} \gamma_r \lambda_{r}}{\gamma_y \sqrt{1-q_2}}\,, \qquad  &l = y
		\end{cases}
	\end{align}
\end{subequations}
and $c^2_{\boldsymbol{\gamma}} = \sum_r \gamma_r^2 + 2 \gamma_y p \sum_{r=1}^{y-1} \gamma_r + 2 q_1 \sum_{r<s<y} \gamma_r \gamma_s$.  

\subsubsection{Finding the solution space kernel}
\begin{figure*}[h]
	\begin{centering}
		\includegraphics[width=0.48\columnwidth]{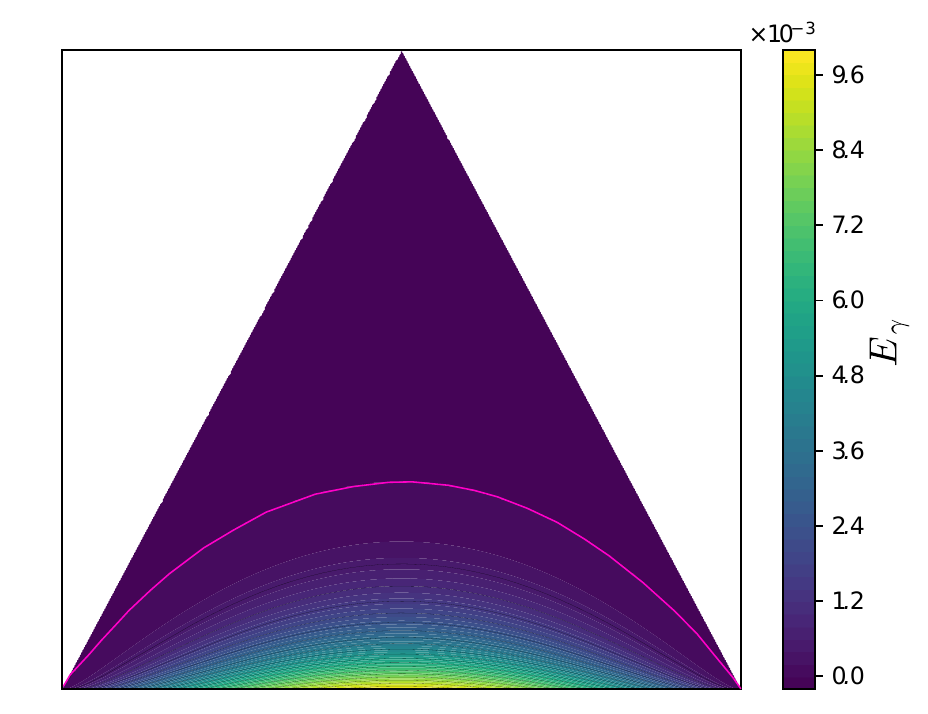}
		\includegraphics[width=.49\columnwidth]{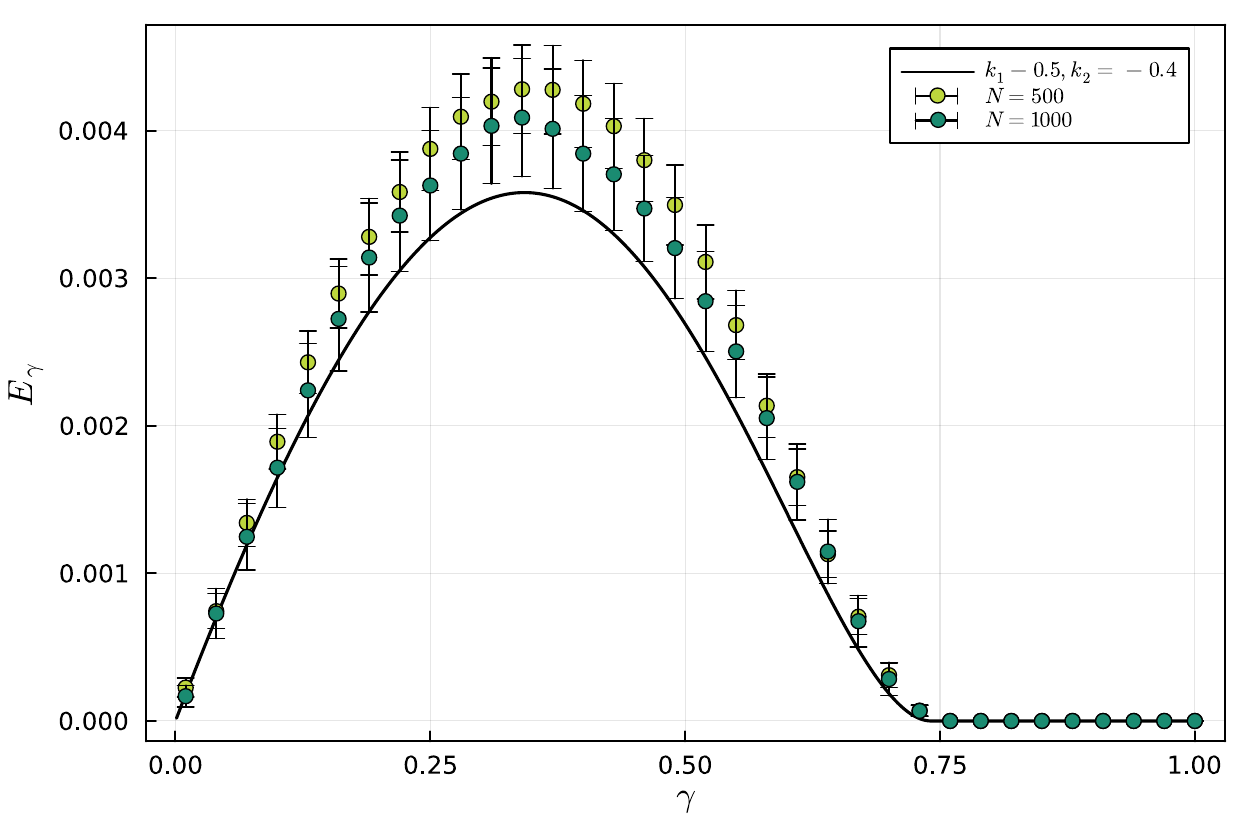}
	\end{centering}
	\caption{\label{fig::diff_margins} (Left panel) Training error on the simplex spanned by $y=3$ solutions with $\alpha=1.0$ and $\kp = -0.5$; the two bottom vertices are two typical solutions to the problem, i.e. have margin margin $k_{1} = \kp$ whereas the top vertex is sampled with $k_2 = -0.1 >\kappa_{\text{krn}}$. The level curve in purple delimits the zero-energy region on the manifold. (Right panel)  Typical energy on the geodesic path between $y=2$ solutions with $k = \kp = -0.5$ and $\alpha = 1$. The solution at $\gamma = 0$ has margin $k_1 = \kp$ whereas the other has $k_2 = -0.4$. The point with errorbars corresponds to the the result of numerical simulations performed by finding the two solutions with SA sampling at different sizes.} 
\end{figure*}

The argument of the theta function in~\eqref{eq::2_k_formula_path} can be inspected analytically in simple cases. We start from the case $y=2$. In this case $c_{\boldsymbol{\gamma}} = \gamma^2 + (1-\gamma)^2 + 2 p \gamma(1-\gamma)$; the training error on a point on a geodesic, which is parameterized by identified by $\gamma \in [0, 1]$, vanishes if
\begin{eqnarray}
	\kp^2 \left[ \gamma^2 + (1-\gamma)^2 + 2p \gamma(1-\gamma) \right] > k_1 \gamma^2 + k_2 (1-\gamma)^2 + 2 k_1 k_2 \gamma (1-\gamma)
\end{eqnarray}
If $k_1 > \kp$ and $k_2 > \kp$ an extended region between $\gamma \in [0, \gamma_L]$ and $\gamma \in [\gamma_R, 1]$ have zero training error with
\begin{equation}
	\gamma_{L, R} = \frac{k_2 (k_2 - k_1)- \kp^2 (1-p) \pm \kp \sqrt{k_1^2 - 2 k_1 k_2 p+k_2^2-\kp^2
			\left(1-p^2\right)}}{(k_1-k_2)^2-2 \kp^2 (1-p)} \,.
\end{equation}
Imposing $\gamma_L = \gamma_r$ we find a condition for $k_2$ that we call $\kappa_{\text{krn}}(k_1)$
\begin{equation}
	\kappa_{\text{krn}}(k_1) = k_1 p - \sqrt{(1-p^2)(\kp^2 - k_1^2)}
\end{equation}
that represents the minimum margin that should be imposed on $\boldsymbol{W}^2$ given the margin $k_1$ on $\boldsymbol{W}^1$ such that the two solutions are geodesically connected. 
Notice that the solution with the plus in front of the square root in the previous equation gives values of $\gamma>1$ so it should be discarded.
Since when $k_1$ increases $p$ increases as well, $\kappa_{\text{krn}}(k_1)$ is a decreasing function of $k_1$. The maximum value of $\kappa_{\text{krn}}$ is obtained therefore when $k_1 = \kp$, for which we find
\begin{equation}
	\kappa_{\text{krn}} = k_1 p
\end{equation}
which was plotted in the main text. Therefore if the margin of a solution is larger than $k_{\text{krn}}$ then we can geodesically connect with every other typical solution with a given margin.

\begin{figure*}[h]
	\begin{centering}
		\includegraphics[width=.32\columnwidth]{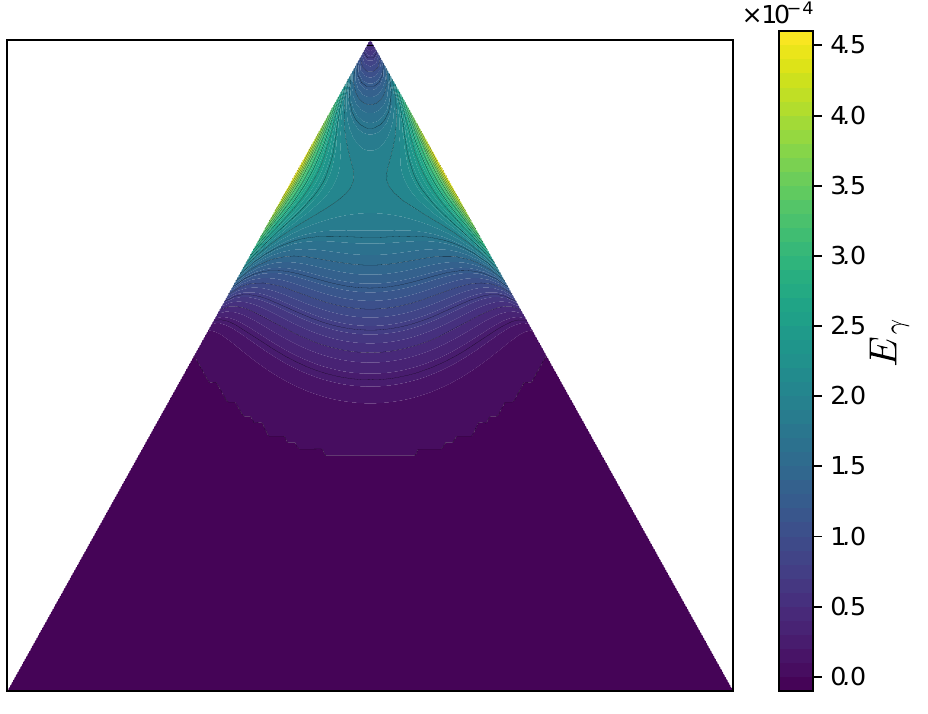}
		\includegraphics[width=.32\columnwidth]{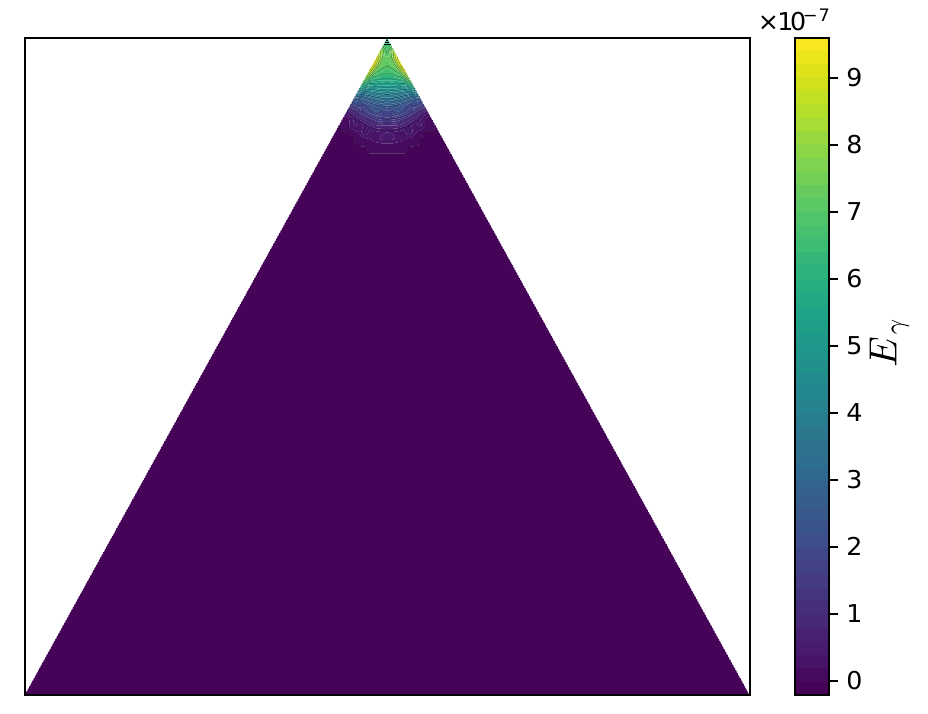}
		\includegraphics[width=.32\columnwidth]{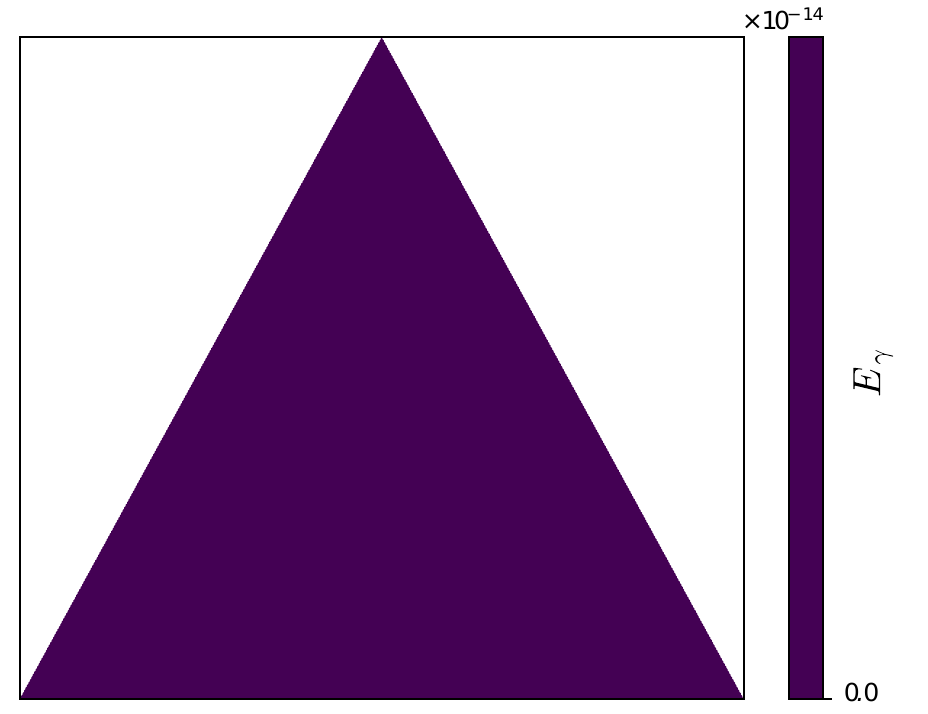}
	\end{centering}
	\caption{\label{fig::diff_margins2} Training error on the simplex spanned by $y=3$ solutions with $\alpha=1.0$ and $\kp = -0.5$; in all plots the top vertex is a typical solutions to the problem, $k_{2} = \kp$. The bottom vertices are sampled with margin $k_1$ respectively -0.3 (left panel), -0.2 (middle) and $k_1 = \kappa_{\text{krn}} \simeq -0.171$ (right). Notice that the left and middle plots are in the case $k_1<\kappa_{\text{krn}} \simeq -0.171$, since the geodesic path between one of the bottom solutions and the top vertex has non-vanishing energy barrier. The lowest level curve in black delimits the zero-energy region on the manifold. The region with positive training error disappears exactly at when $k_1<\kappa_{\text{krn}}$. } 
\end{figure*}

In the left panel of Fig.~\ref{fig::diff_margins} we report the training error on the simplex spanned by $y=3$ solutions in the case where the two vertices have been sampled with the same margin of the problem $k_1 = \kp = -0.5$ and the third one (placed on the top) is a robust solution with margin $k_2>\kp $. In the setting of the left panel of Fig.~\ref{fig::diff_margins}, we have also chosen  $k_2 > \kappa_{\text{krn}}$, i.e. the vertex at the top is geodesically connected to both the two vertices at the base of the triangle. Since the two solutions placed at the bottom are typical solutions to the problem, the energy barrier is non vanishing along the geodesic path; however the two solutions are connected by passing through the center of the interior of the simplex. In the same plot, the red line delimits the region of the manifold at zero-energy (above it) from the one at non-zero energy (below). It also represents the minimum length path needed to pass from a typical solution to the other one. 

In Fig.~\ref{fig::diff_margins2} we show the 2-simplex in the case we sample two solutions with large margin ($k_1>\kp$, bottom vertices) and only one vertex with the margin of the problem ($k_2 = \kp$, top vertex). The three plots corresponds to different values of $k_1$.  We show numerically that the whole simplex goes to zero as soon as the maximum barrier along the geodesic path goes to zero, i.e. when $k_{1} \ge \kappa_{\text{krn}}$.
This shows that $\kappa_{\text{krn}}$ does not have a dependence on the number of replicas $y$ as in the case of $\kappa_y^\star$ and that a robust configuration within the kernel of the solution space are also connected, via a path passing through the simplex, with solutions in the neighborhood of a typical solution. This confirms the overall star-shapedness of the solution space.

\subsection{Convex subsets in a probabilistic setting}

In this SM and in the main paper we have employed a notion of convexity in a probabilistic setting that we clarify here. For a given system size $N$, a load $\alpha$, a margin $\kp$, and a realization of the patterns $\{\bxi^\mu\}_{\mu=1}^{\alpha N}$, we call $S_{\bxi,\kp}$ the solution set of the problem. While the whole random set $S_{\bxi,\kp}$ is with high probability geodesically non-convex, we conjecture that independently sampling $y$ solutions $\{\bW^r\}_{r=1}^y$ with a margin $\kappa > \kappa_y^\star(\alpha,\kp)$, larger than the strictly required one $\kp$, one can span a simplex 
\begin{equation}
	A_y = \left\{\bW_{\boldsymbol{\gamma}}\,:\,\sum_r\gamma_r =1,\ 0\leq\gamma_r\leq1\ \forall r, \boldsymbol{W}_{\boldsymbol{\gamma}}=\sqrt{N}\frac{\sum_{r}\gamma_{r}\boldsymbol{W}^{r}}{\rVert\sum_{r}\gamma_{r}\boldsymbol{W}^{r}\rVert}\right\},
\end{equation}
that with high probability is entirely made of solutions :
\begin{equation}
	\lim_{N\to\infty} \mathbb{P}(A_y \subseteq S_{\bxi,\kp}) = 1.
	\label{eq:strong_convexity}
\end{equation}
Here the randomness is given by both the realization of $\bxi$ and the sampling of the $\bW^r$. While we conjecture Eq. \eqref{eq:strong_convexity} to hold in the $\kappa > \kappa_y^\star(\alpha,\kp)$ regime, we cannot establish it firmly even with our powerful but non-rigorous replica techniques. A much weaker notion of convexity in probability is instead implied by our (again non-rigorous) analysis. In fact, we have that 
\begin{equation}
	\lim_{N\to\infty} \mathbb{P}(\bW_{\bgamma} \in S_{\bxi,\kp}) = 1 \qquad \forall \bgamma.
	\label{eq:weak_convexity}
\end{equation}
This result can be obtained by applying Markov's inequality to the energy of the interpolant configuration: $\mathbb{P}(E(\bW_{\bgamma}) > \epsilon) \leq \frac{\mathbb{E}{E(\bW_{\bgamma})}}{\epsilon}$. The large limit of the expectation $\mathbb{E}{E(\bW_{\bgamma})}$ is in fact the $E_{\bgamma}$ that we compute in the paper. This local result, pointwise in $\bgamma$, could be strengthened by considering the ``total'' energy obtained by integrating the energy over the $\bgamma$ simplex and showing that goes to 0 in expectation for large $N$ and  $\kappa > \kappa_y^\star(\alpha,\kp)$. This leads us to believe that also the maximum energy point on the simplex has asymptotically zero energy by a continuity of the expected energy argument, therefore implying the strong notion of Eq. \eqref{eq:strong_convexity}. More precise characterizations of this stochastic geometry and mathematical proofs of our conjectures remain open problems.

\section{Stability distribution of the interpolated solution}

In this section we want to understand which type of configurations are found along the geodesic path between two solutions. To this end, we derive an analytic expression for the stability distribution on the geodesic path connecting two weights $\boldsymbol{W}_{\boldsymbol{\gamma}} = \sqrt{N} \frac{\gamma \boldsymbol{W}^1 + (1-\gamma)\boldsymbol{W}^2}{\lVert\gamma \boldsymbol{W}^1 + (1-\gamma)\boldsymbol{W}^2\rVert}$, with margin $k_1$ and $k_2$, i.e. 
\begin{equation}
	P(\Delta) = \mathbb{E}_{\boldsymbol{\xi}}\left\langle \delta\Big(\Delta - \frac{\boldsymbol{W}_{\boldsymbol{\gamma}} \cdot \boldsymbol{\xi}^1}{\sqrt{N}}\Big)\right\rangle_{k_1, k_2} = \mathbb{E}_{\boldsymbol{\xi}}\frac{1}{Z_{\boldsymbol{\xi}}(k_1, k_2)} \int d\boldsymbol{W}^1 d\boldsymbol{W}^2\,\mathbb{X}_{\boldsymbol{\xi}}(\boldsymbol{W}^1; k_1) \mathbb{X}_{\boldsymbol{\xi}}(\boldsymbol{W}^2; k_2) \delta\left(\Delta - \frac{\bf{W}_{\boldsymbol{\gamma}} \cdot \boldsymbol{\xi}^1}{\sqrt{N}}\right).
\end{equation}
By writing $Z^{-1} = \lim_{n\to 0} Z^{n-1}$, one can again write this expression as an integral over a set of $n$ replicas, and to subsequently take the $n\to 0 $ limit. Proceeding in a similar fashion to the free entropy computations along the paper, one arrives to the form
\begin{equation} \label{eq:Pdelta}
	P(\Delta) = \lim_{n\to 0} \int \prod_{a<b,rs} \frac{dq^{ab}_{rs}\,d\hat{q}^{ab}_{rs}}{2\pi}\int \prod_{ar} \frac{d\hat{h}^{a}_r}{2\pi}\,e^{-\frac{N}{2}\sum_{a<b, rs}q^{ab}_{rs}\hat{q}^{ab}_{rs} +N\sum_{ae}\hat{h}^a_r +N G_S + N\alpha \,G_E} f_\gamma(\{q^{ab}_{rs}\}, \Delta)
\end{equation}
where 
\begin{equation}
	f_\gamma(\{q^{ab}_{rs}\}, \Delta) = \int\prod_{ar}\frac{d\lambda_{r}^{a}d\hat{\lambda}_{r}^{a}}{2\pi}\prod_{ar}\Theta(\lambda_{r}^{a}-k_{r})e^{i\sum_{ar}\lambda_{r}^{a}\hat{\lambda}_{r}^{a}-\frac{1}{2}\sum_{arbs}\hat{\lambda}_{r}^{a}q_{rs}^{ab}\hat{\lambda}_{s}^{b}}\delta\Big(\Delta-\frac{\gamma\lambda_{1}^{1}+(1-\gamma)\lambda_{2}^{1}}{c_{\boldsymbol{\gamma}}}\Big).
\end{equation}
In the asymptotic $N\to\infty$ limit, by employing Laplace's method we find that the integral in Eq.~\eqref{eq:Pdelta} is given by 
\begin{equation}\label{eq:Pdelta1}
	e^{nN\phi(\{(q_{rs}^{ab})^\star\})}f_\gamma(\{(q^{ab}_{rs})^\star\}, \Delta), 
\end{equation}
where $\{(q_{rs}^{ab})^\star\}$ are the values which extremize the free entropy. In the $n\to 0$ limit the exponential term in Eq.~\eqref{eq:Pdelta1} tends to $1$, and we are thus left with $P(\Delta) = f_\gamma(\{(q^{ab}_{rs})^\star\}, \Delta)$.

By imposing the RS-ansatz given by Eqs.~\ref{ansatz}, we end up with the final formula
\begin{equation}
	\begin{split}
		P(\Delta)&=\sqrt{\frac{\Gamma}{2\pi (1-q_1)}}\frac{c_{\boldsymbol{\gamma}}}{\gamma}\Theta\Big(c_{\boldsymbol{\gamma}}\Delta-\gamma k_{1}-(1-\gamma)k_{2}\Big)\times\\
		&\times\int \mathcal{D}z\mathcal{D}y \,  e^{\frac{1}{2}\Xi^{2}-\frac{\left(b_{1}(y)\gamma+b_{2}(z, y)(1-\gamma)+\Delta W\right){}^{2}}{2(1-q_{1})\gamma^{2}}}\frac{H\left(\frac{\left(b_{2}(z, y)+k_{2}\right)}{\sqrt{\Gamma(1-q_{2})}}-\Xi\right)-H\left(\frac{\left(b_{2}(z, y)(1-\gamma)-\gamma k_{1}+\Delta c_{\boldsymbol{\gamma}}\right)}{\sqrt{\Gamma(1-q_{2})}(1-\gamma)}-\Xi\right)}{\prod_{r}H\Big(\frac{b_{r}+ k_{r}}{\sqrt{1-q_r}}\Big)}
	\end{split}
\end{equation}
where we have introduced the functions
\begin{subequations}
	\begin{align}
		&b_{1}(y)=\sqrt{q_{1}}y\\
		&b_{2}(z, y)=\frac{p}{\sqrt{q_{1}}}y+\sqrt{q_{2}-\frac{p^{2}}{q_{1}}}z
	\end{align}
\end{subequations}
and the constants
\begin{subequations}
	\begin{align}
		&\Gamma=\frac{(1-q_{1})\gamma^{2}}{(1-q_{1})\gamma^{2}+(1-q_{2})(\gamma-1)^{2}}\\
		&\Xi=\frac{(1-\gamma)\sqrt{1-q_{2}}(\lVert W\rVert\Delta+\gamma b_{1}(y)+(1-\gamma)b_{2}(z, y))}{\gamma\sqrt{1-q_{1}}\sqrt{(1-\gamma)^{2}(1-q_{2})+\gamma^{2}(1-q_{1})}}\\
		&c_{\boldsymbol{\gamma}}=\sqrt{\gamma^{2}+(1-\gamma)^{2}+2\gamma(1-\gamma)p}
	\end{align}
\end{subequations}
Similarly to the case of the stability distribution for typical solutions, and since we are dealing with the error counting loss, there is a value of $\Delta$ up to which the distribution is zero. In the case of typical solutions this threshold is simply the margin of the solution $\kp$, while in the interpolated case its value is given by 
\begin{equation}
	\Delta_{eff} = \frac{\gamma k_1 + (1-\gamma) k_2}{c_{\boldsymbol{\gamma}}}
\end{equation}
which at the extremes of the path reduces precisely to the thresholds of the two typical solutions sampled with margins $\Delta_{eff}|_{\gamma=0} = k_2$ and $\Delta_{eff}|_{\gamma = 1} = k_1$, 
while for each $\gamma \in (0,1)$ could be interpreted as the effective margin of solutions along the path. 

\begin{figure}[h]
	\begin{centering}
		\includegraphics[width=0.60\columnwidth]{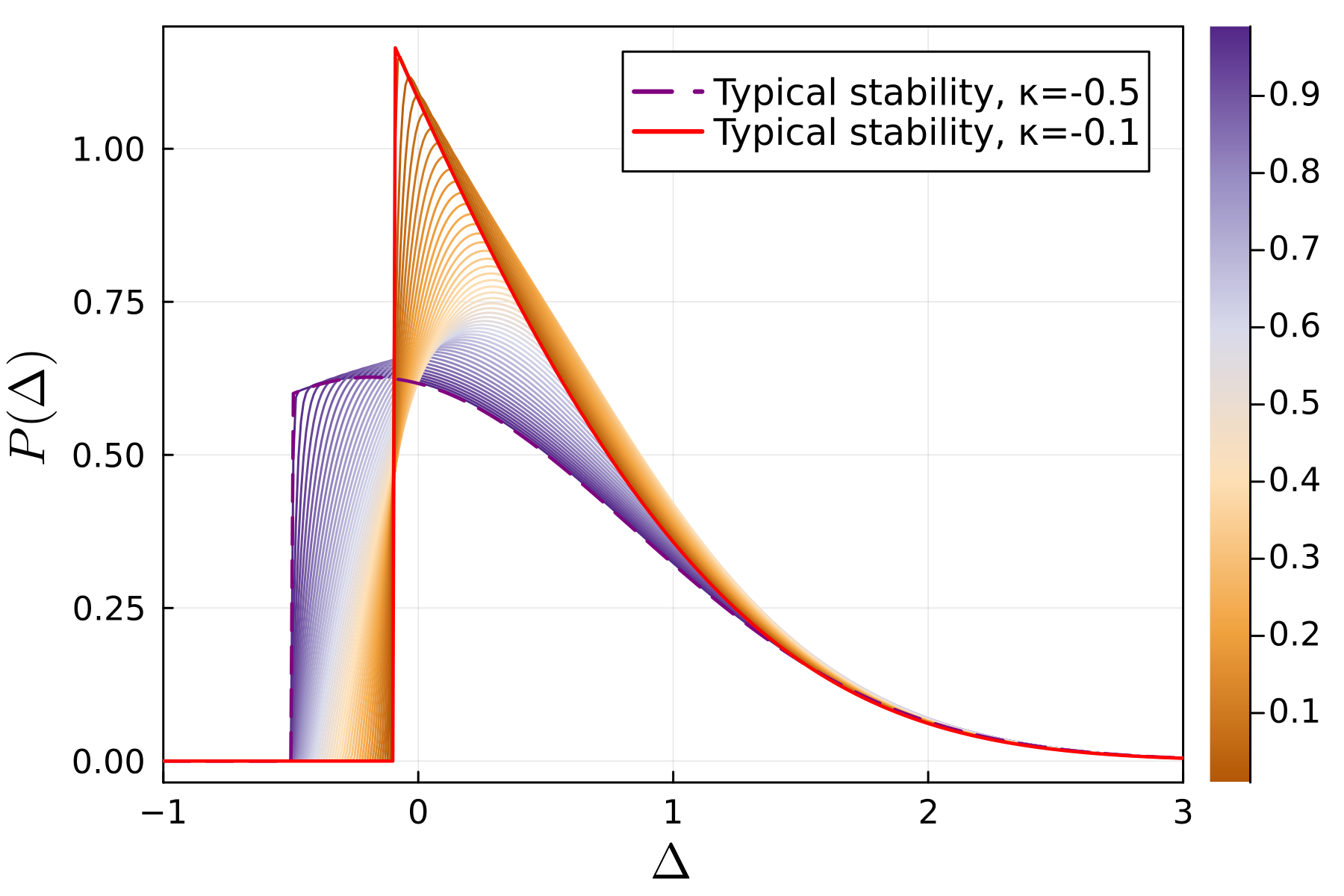}
	\end{centering}
	\caption{\label{fig::stability} Stability distributions of solutions found along the geodesic path between a typical solution at $k_1 = -0.5$ towards a more robust solution at $k_2 = -0.1$, varying the interpolation parameter $\gamma_1$ and for $\alpha=1$. Notice that the geodesic connecting the two solutions in this case is at zero training error. Purple and red lines at the extremes of the curves represent the typical stability distribution for $\kp=-0.5$ and $\kp=-0.1$ respectively.} 
\end{figure}

In Fig.~\ref{fig::stability} we plot the stability distribution of the solutions along the geodesic for $k_2=0.1>\kappa_{\text{krn}}$ (notice that, coherently with the definition of $\kappa_{\text{krn}}$, we find that if $k_2>\kappa_{\text{krn}}$ then $k_1 \leq \Delta_{eff} \leq k_2$ $\forall \gamma\in[0,1]$, implying that all the configurations along the geodesic are solutions). 
Even though the presence of a sharp threshold separating zero from non-zero values of the stability distribution is reminiscent of the typical solutions case, the functional form of the distribution itself beyond this threshold is qualitatively different from the one of typical solutions at a given margin value $k$ (i.e. it does not follow a truncated Gaussian), implying that the solutions found along the geodesic path are not that of the equilibrium measure at that $k$, so we do not have control on their sampling. 


\section{Anisotropy of the solution space}\label{sec::anisotropy}
\begin{figure}[h]
	\begin{centering}
		\includegraphics[width=0.60\columnwidth]{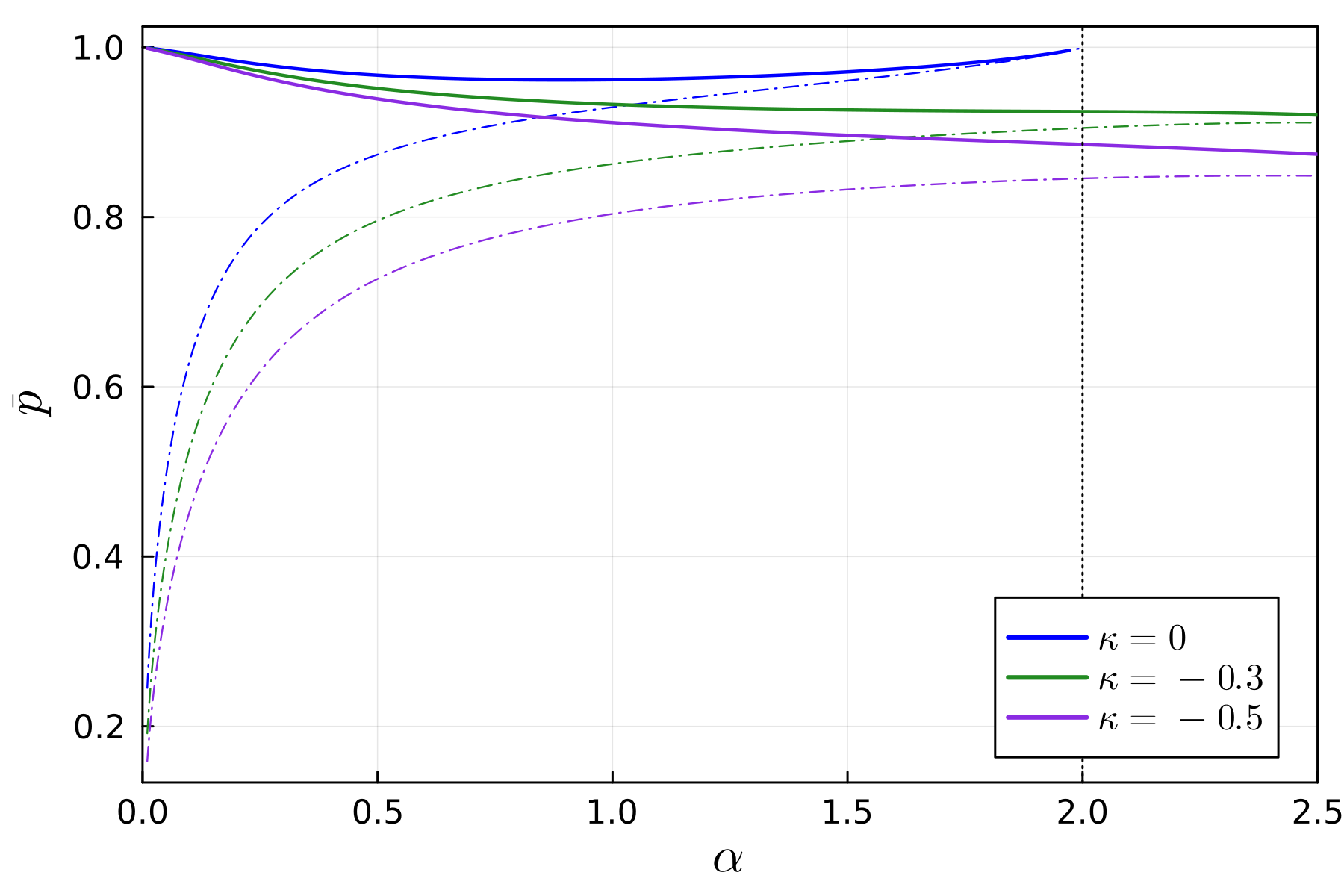}
	\end{centering}
	\caption{\label{fig::anisotropy} Typical overlap $\bar{p}$ between the barycenter of $y$ independently-sampled solutions with margin $\kp$ and the maximum margin solution at $\kappa_{\text{max}}(\alpha).$ The dashed curves refer to the $y=3$ case while the full ones are for $y=\infty$. See the main text for details on the intepretation of the results. After the vertical dashed line ($\alpha_{\text{dAT}}(\kp=0)$) the RS approximation is unstable and the predicted value of $\kappa_{\text{max}}(\alpha)$ is not exact.} 
\end{figure}



We want here to describe the degree of isotropy of the space of solutions. In order to do that, we consider an ensemble of $\kp$-margin solutions and quantify the overlap between their barycenter and the $\kappa_\text{max}(\alpha)$-margin solution, located in the inner region of the nested structure.
The barycenter of $y$ independently-sampled replicas of the system at a given margin $\kp$ can be defined as 
\begin{equation}
	\overline{\boldsymbol{W}} (\kp) = \frac{\sum_{a=1}^y \boldsymbol{W}^a (\kp)}{|| \sum_{a=1}^y \boldsymbol{W}^a (\kp)||}
\end{equation}
such that the vector $ \overline{\boldsymbol{W}} (\kp)$ is normalized to $\lVert\overline{\boldsymbol{W}} (\kp)\rVert^2 = 1 $. The overlap between $ \overline{\boldsymbol{W}} (\kp)$ and the solution with the highest possible margin $\bf{W}(\kappa_{\text{max}})$ is straightforward to define
\begin{equation}
	\overline{p} = \frac{1}{N} \langle \overline{\boldsymbol{W}}(\kp), \boldsymbol{W}(\kappa_{\text{max}}) \rangle = \frac{p\, (\kp, \kappa_{\text{max}})}{\sqrt{\frac{1}{y}+\frac{y-1}{y}\,q(\kp)}} \xrightarrow{y \to \infty} \frac{p \,(\kp, \kappa_{\text{max}})}{\sqrt{q (\kp)}}.
\end{equation}

In Fig.~\ref{fig::anisotropy} we plot $\bar{p}$ in function of $\alpha$ for different values of $\kp$ and for $y=3$ (dashed curves) and $y=\infty$ (full curves). 
When $y$ is small, we do not expect the barycenter of $y$ typical solutions to be in the central region of the solution space in which robust solutions are located. Therefore, we do not expect it to be either a good approximation or close to the most robust solution with margin $\kappa_{\text{max}}$. Indeed, as can be seen from the figure, for $y=3$ and $\alpha=0$ the overlap is such that $\Bar{p}\,(\kp, \kappa_{\text{max}};\alpha=0) \simeq 0$ independently from the margin $\kp$ (as it would happen when randomly sampling two vectors from an $N$-dimensional sphere). Increasing $\alpha$, the overlap is expected to increase as the volume of the solution space and consequently typical distances between solutions decrease. At the opposite, for $\alpha=0$ the barycenter of $y=\infty$ replicas coincides with the $\kappa_{\text{max}}$ solution independently from $\kp$ (the overlap is $1$). Even reducing to the convex setting recovered in the case $\kp=0$ (full blue curve in the figure), we have that $\Bar{p}\,(\kp=0,\kappa_{\text{max}};\alpha >0 )\neq 1$ when increasing $\alpha$. The fact that the overlap $\bar{p}$ is different from zero in this case indicates the anisotropy of the system with respect to the robust solution. In fact, for the $\kp =0$ case, we have that the overlap $\Bar{p}$ is equal to $1$ only at the critical storage capacity $\alpha^c_{\text{RS}}=2$ bounding the SAT/UNSAT transition. 

\section{Numerical simulations}

\subsection{Numerical simulations details}~\label{sec::num_sim}

\textit{Focusing Belief Propagation} 
Focusing Belief Propagation (fBP) is an heuristic modification of BP that targets flat regions in the solution space~\cite{unreasoanable}. In the negative perceptron problem the solution obtained with fBP are a good approximation of the $\kappa_{\textit{max}}$ solutions (see~\cite{baldassi2023typical}).
In the fBP simulations (see~\cite{baldassi2020shaping} for implementation details) we used $y=10$ replicas and an increasing coupling constant $\gamma$ with $15$ exponentially spaced values in the range $[0.1,10]$. For each value of $\gamma$ we initialized fBP with the messages obtained at the previous coupling value. We then run it until the maximum absolute difference between messages was less than $10^{-5}$ or until it reached a maximum number of $500$ messages updates.

\textit{Simulated Annealing} 
In order to sample solutions as uniformly as possible from the solution space, and thus capture the typical solutions, we employ Simulated Annealing (SA) on the quadratic Hinge loss, defined as follows:
\begin{equation}
	\ell(\Delta_{\kp}) \equiv \Delta_{\kp}^2 \Theta(-\Delta_{\kp})
\end{equation}
where $\Delta^\mu_{\kp}= y^\mu (w \cdot \xi^\mu) - \kp \lVert w \lVert$.
Optimizing the Hinge loss instead of the number of errors function makes the sampling much faster as the SA dynamics benefits from the high-energy solutions organization, while the zero-energy manifold is unaltered.
The weight vectors $w$ are initialized on the unit sphere. At each step, we propose a move in a random direction $w^{\prime} = w+\eta$, with $\eta = \epsilon \mathcal{N}(0,1)$, and $\epsilon=5\times 10^{-3}$ in the simulations. 
The weights are updated with probability $p=\mathrm{min}\left(1, e^{-\beta\Delta\ell}\right)$, where $\Delta\ell$ is the loss difference between $w$ and $w^{\prime}$. 
After $N$ move proposals we increase the inverse temperature as $\beta \leftarrow \beta+d\beta$, with $\beta_0=2000$ and $d\beta=2$.
The simulation is interrupted whenever a solution is found or when the maximum number of $200000$ allowed updates per weight is reached.
As it was previously noted in \cite{baldassi2023typical}, the scaling with the instance size $N$ of the convergence times of SA is worst-case exponential, and this fact limits our analysis to small and medium sizes $N\leq 4000$.

\textit{Gradient Descent on cross-entropy}. We employed a standard gradient descent optimization strategy. Notice that the standard cross-entropy/logistic loss can be adapted to the negative perceptron problem by adding the corresponding margin $\kp$ in the exponential term of the sigmoid: 
\begin{equation}
	\ell(\Delta_{\kp}) = \log(1 + e^{-\Delta_{\kp}}).
\end{equation}
The weights are initialized uniformly on the unit sphere, and optimized with a batchsize $200$ and learning rate $1$, with a maximum number of allowed epochs equal to $20000$. Early stopping is implemented, interrupting the trajectory as soon as a zero energy configuration is found.

\textit{Perceptron Algorithm}.
We implemented the standard Perceptron Algotirhm (PA) (see e.g.~\cite{engel-vandenbroek}). For each pattern, if the stability $\Delta^{\mu}$ on the $\mu$~-th pattern is greater than zero, then the weights are updated as $w \leftarrow w+2 \eta \xi^{\mu}$ where $\eta$ is the learning rate. After each weight update, we project the weight vector onto the unit sphere (in order to avoid a huge increase of the norm and a slowing down of the optimization progress). 
Notice that the PA algorithm is equivalent to a gradient descent optimization on the Hinge loss with unit batch size, and similar results can be obtained also with larger batch sizes provided the learning rate is lowered.
By varying the learning rate, we were able to obtain PA solutions with a wide range of stability distribution (see Fig.~\ref{fig:pa_barr} and the discussion in the following Section).

\subsection{Sampling bias of the Perceptron Algorithm as a function of the learning rate}
\begin{figure}[h]
	\begin{centering}
		\includegraphics[width=0.45\columnwidth]{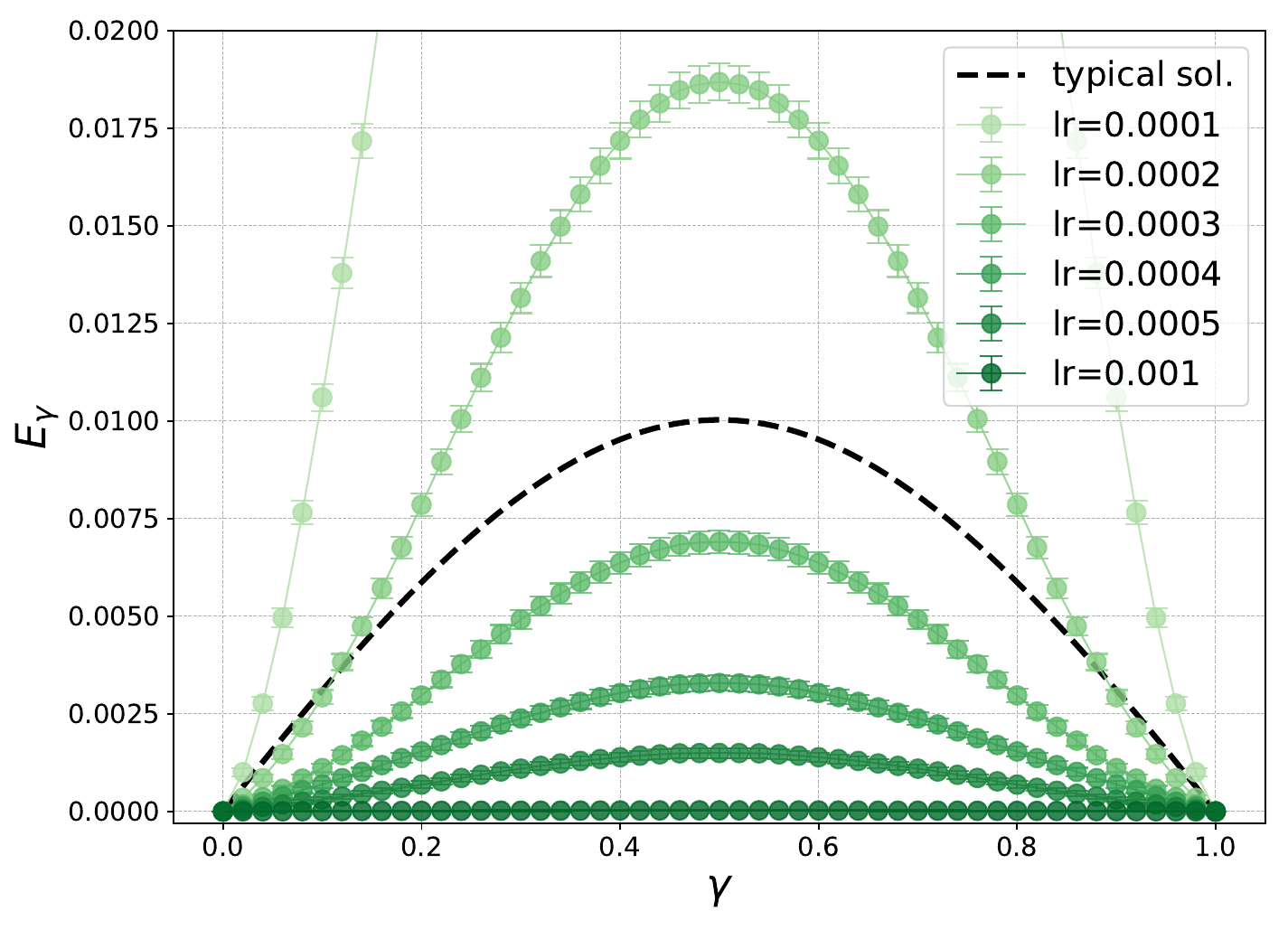}
		\includegraphics[width=0.45\columnwidth]{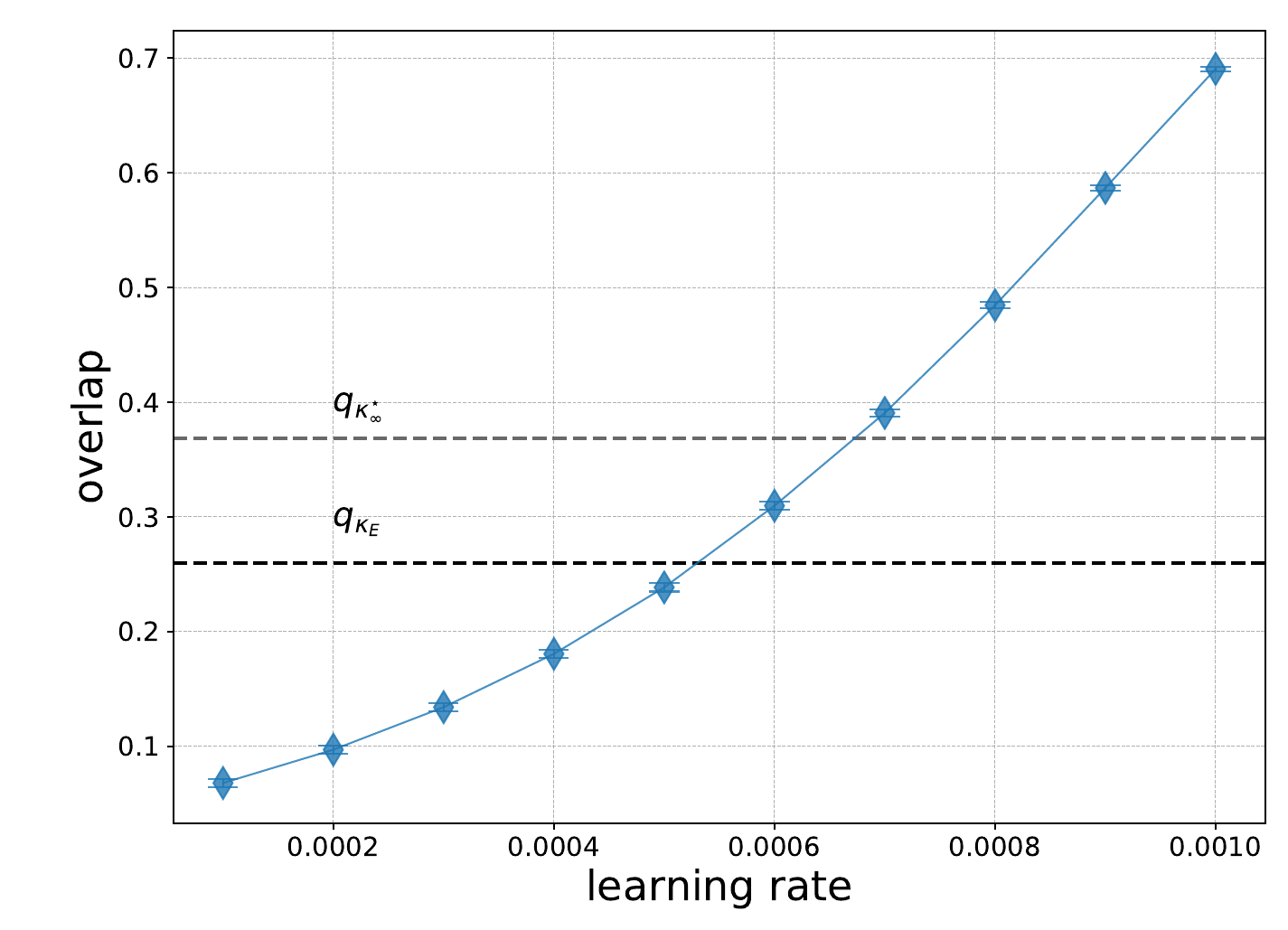}
	\end{centering}
	\caption{\label{fig:pa_barr} (Left)
		Maximum fraction of errors along the geodesic path connecting two solutions obtained with PA at $\alpha=1$ and $\kp=0.5$ using different learning rates at $N=1000$.
		Increasing the learning rate PA is able to find geodesically connected solutions. We also plot as a reference the barrier between typical solutions (dashed line). (Right) Average overlap between PA solutions as a function of the learning rate. Horizontal dashed lines are $q_{\kappa_{E}}$ and $q^{\star}_{\infty}$ at $\alpha=1$.
		When PA finds solutions with higher energy barrier with respect to typical solutions, they are further on average w.r.t. typical solutions (smaller overlap). As soon as the geodesic path connecting PA solutions has zero energy, the solutions have average overlap comparable with the overlap at the coalescence threshold.
	} 
\end{figure}

Finally, we investigate the dependence of the sampling bias of PA as a function of the learning rate employed in the dynamics. As commented above and in the main text, when the learning rate is sufficiently small, this algorithm is able to find solutions that display an atypical distribution of stabilities peaked around the minimum margin allowed in the problem. These solutions are typically found very close to initialization and have an extremely small overlap with one another. This overlap is much smaller than that of the typical solutions of the problem, which numerically dominate the flat measure over solutions. While at low constraint densities these solutions have no barriers with respect to the fBP solutions, they instead show non-zero barriers between each other at any $\alpha$.

Interestingly, the type of solution sampled by PA can be completely altered by increasing the learning rate employed for the optimization. As shown in the left panel of Fig.~\ref{fig:pa_barr}, on the one hand, the quicker optimization trajectories corresponding to higher learning rates find pairs of solutions separated by lower and lower energy barriers, crossing at a certain value the typical energy barrier between the equilibrium solutions (dashed black line). On the other hand, the reciprocal overlap increases up to values that are much larger than that of the equilibrium solutions and comparable with that of the solutions in the kernel regions at margin $\kappa^\star_\infty$, as shown in the right panel of Fig.~\ref{fig:pa_barr}.

\end{document}